%% file: gfossati_MS72579.tex
\documentclass{emulateapj}

\usepackage{apjfonts}
\usepackage{xspace}
\usepackage{calc}
\usepackage{ifthen}

\newcommand{\xray     }{X-ray }%
\newcommand{\xraynosp }{X-ray}%
\newcommand{\xrays    }{X-rays }%
\newcommand{\xraysnosp}{X-rays}%
\newcommand{\gray     }{$\gamma$-ray }%
\newcommand{\graynosp }{$\gamma$-ray}%
\newcommand{\grays    }{$\gamma$-rays }%
\newcommand{\graysnosp}{$\gamma$-rays}%
\newcommand{\mrk      }{Mrk~421 }%
\newcommand{\mrknosp  }{Mrk~421}%
\newcommand{\rxte     }{\textit{Rossi}XTE }%
\newcommand{\rxtenosp }{\textit{Rossi}XTE}%
\newcommand{\sax      }{\textit{Beppo}SAX }%
\newcommand{\saxnosp  }{\textit{Beppo}SAX}%
\newcommand{\ie       }{i.e. }%
\newcommand{\eg       }{e.g. }%
\newcommand{\egnosp   }{e.g.}%
\newcommand{\vs       }{\textit{vs.} }

\newcommand{\etal     }{et al. }%
\newcommand{\etalnosp }{et al.}%

\renewenvironment{itemize}{%
\begin{list}{}
    {
        \setlength{\leftmargin}{12pt}
        \setlength{\itemsep}{0.1\baselineskip}
        \setlength{\topsep}{0.1\baselineskip}
    }
}{%
\end{list}}

\shortauthors{Fossati et al.}
\shorttitle{X-RAY/TeV VARIABILITY OF Mrk 421}
\slugcomment{The Astrophysical Journal, 677:906$-$925, 2008 April 20}

\begin{document}

\title{Multiwavelength Observations of Markarian 421 in March 2001: \\
an Unprecedented View on the X-ray/TeV Correlated Variability.}

\author{%
G.~Fossati,\altaffilmark{1}
J.~H.~Buckley,\altaffilmark{2}      
I.~H.~Bond,\altaffilmark{3}         
S.~M.~Bradbury,\altaffilmark{3}     
D.~A.~Carter-Lewis,\altaffilmark{4} 
Y.~C.~K.~Chow,\altaffilmark{5}      
W.~Cui,\altaffilmark{6}             
A.~D.~Falcone,\altaffilmark{7}      
J.~P.~Finley,\altaffilmark{6}       
J.~A.~Gaidos,\altaffilmark{6}       
J.~Grube,\altaffilmark{3}           
J.~Holder,\altaffilmark{8}          
D.~Horan,\altaffilmark{9,10}        
D.~Horns,\altaffilmark{11}          
M.~M.~Jordan,\altaffilmark{12}      
D.~B.~Kieda,\altaffilmark{13}       
J.~Kildea,\altaffilmark{9}          
H.~Krawczynski,\altaffilmark{2}     
F.~Krennrich,\altaffilmark{4}       
M.~J.~Lang,\altaffilmark{14}        
S.~LeBohec,\altaffilmark{13}        
K.~Lee,\altaffilmark{2}             
P.~Moriarty,\altaffilmark{15}       
R.~A.~Ong,\altaffilmark{5}          
D.~Petry,\altaffilmark{4,16}        
J.~Quinn,\altaffilmark{17}          
G.~H.~Sembroski,\altaffilmark{6}    
S.~P.~Wakely,\altaffilmark{18}      
T.~C.~Weekes,\altaffilmark{9}       
}

\altaffiltext{1}{Department of Physics \& Astronomy, Rice University, Houston, TX 77005, USA}%
\altaffiltext{2}{Department of Physics, Washington University, St. Louis, MO 63130, USA}%
\altaffiltext{3}{School of Physics \& Astronomy, University of Leeds, Leeds, LS2 9JT, UK}%
\altaffiltext{4}{Department of Physics \& Astronomy, Iowa State University, Ames, IA 50011, USA}%
\altaffiltext{5}{Department of Physics \& Astronomy, University of California, Los Angeles, CA 90095, USA}%
\altaffiltext{6}{Department of Physics, Purdue University, West Lafayette, IN 47907, USA}%
\altaffiltext{7}{Department of Astronomy \& Astrophysics, Pennsylvania State University, University Park, PA 16802, USA}%
\altaffiltext{8}{Department of Physics \& Astronomy, University of Delaware, Newark, DE 19716, USA}
\altaffiltext{9}{Fred Lawrence Whipple Observatory, Harvard--Smithsonian CfA, P.O. Box 97, Amado, AZ 85645, USA}%
\altaffiltext{10}{Argonne National Lab, Argonne, IL 60439, USA}%
\altaffiltext{11}{Institut f\"ur Experimentalphysik Univ. Hamburg, Germany}%
\altaffiltext{12}{Raytheon, Tucson, AZ 85706, USA}%
\altaffiltext{13}{Department of Physics, University of Utah, Salt Lake City, UT 84112, USA}%
\altaffiltext{14}{Department of Physics, National University of Ireland, Galway, Ireland}%
\altaffiltext{15}{School of Science, Galway-Mayo Institute of Technology, Galway, Ireland}%
\altaffiltext{16}{Max-Planck-Institut f\"ur Extraterrestrische Physik, D-85741 Garching, Germany}%
\altaffiltext{17}{School of Physics, University College Dublin, Belfield, Dublin 4, Ireland}%
\altaffiltext{18}{University of Chicago, Enrico Fermi Institute, Chicago, IL 60637, USA}%

\email{\small Corresponding authors: Giovanni Fossati <gfossati@rice.edu>
and Jim Buckley <buckley@wuphys.wustl.edu>}

\begin{abstract}
We present a detailed analysis of week-long simultaneous observations of the blazar
\mrk at 2--60~keV \xrays (\rxtenosp) and TeV \grays (Whipple and HEGRA) in 2001.
Accompanying optical monitoring was performed with the Mt. Hopkins 48" telescope.
The unprecedented quality of this dataset enables us to establish
the existence of the correlation between the TeV and \xray luminosities,
and also to start unveiling some of its characteristics, in particular its
energy dependence, and time variability.
The source shows strong variations in both \xray and \gray bands, which are highly
correlated.  No evidence of a \xraynosp/\gray interband lag $\tau$ is found
on the full week dataset, with $\tau\lesssim3$\,ks.
A detailed analysis of the March 19 flare, however, reveals that data are
\textit{not} consistent with the peak of the outburst in the 2--4\,keV
\xray and TeV band being simultaneous.  We estimate a $2.1\pm0.7$\,ks TeV lag.
The amplitudes of the \xray and \gray variations are also highly
correlated, and the TeV luminosity increases more than linearly with
respect to the \xray one.
The high degree of correlation lends further support to the standard model
in which a unique electrons population produces the \xrays by synchrotron
radiation and the \gray component by inverse Compton scattering.
However, the finding that for the individual best observed flares the \gray flux scales
approximately quadratically with respect
to the \xray flux, poses a serious challenge to emission models for TeV blazars,
as it requires rather special conditions and/or fine tuning of the temporal
evolution of the physical parameters of the emission region.
We briefly discuss the astrophysical consequences of these new findings in the
context of the competing models for the jet emission in blazars.
\end{abstract}


\section{Introduction}
\label{sec:intro}

\object[Mrk 421]{\mrk} is the brightest BL Lac object in the \xray and UV sky and the first
extragalactic source detected at TeV energies \citep{punch92_tev_mkn421}.
Like most blazars, its spectral energy distribution shows two smooth broad
band components (\eg \citealp*{sambruna96_sed,umu97_review};
\citealp{fossati98_sed}).
The first one extends from radio to \xrays with a peak in the soft to
medium \xray range; the second one extends up to the GeV to TeV energies,
with a peak presumed to be around 100~GeV.
The emission up to \xrays is thought to be due to synchrotron radiation
from high-energy electrons, while the origin of the luminous \gray radiation
is more uncertain.
Possibilities include inverse Compton scattering of synchrotron (synchro
self-Compton, SSC) or ambient photons (external Compton, EC) off a single
electron population thus accounting for the spectral ``similarity'' of the
two components  
(\eg \citealp{macomb95_mkn421,mastichiadis_kirk97,tavecchio_tev_98,%
mgc92_3c279,sbr94_ec,dermer_etal92}).
Alternative ``hadronic'' models produce \gray from protons, either directly
(proton synchrotron) or indirectly (\eg synchrotron from a second electron
population produced by a cascade induced by the interaction of high-energy
protons with ambient photons) \citep{mucke_etal_2003_SPB_model,bottcher_reimer_2004}.
The synchrotron proton scenario may be more favorable for objects
like \mrk \citep{mucke_etal_2003_SPB_model}, because of the lower density of the
diffuse photon fields necessary for processes like pion photoproduction
to be effective.
Moreover, it is generally true that ``hadronic'' models need a higher level
of tuning in order to reproduce the observed highly correlated \xraynosp/\gray 
variability.
Hence, in this paper we have not addressed this class of models, and instead
focused our limited modeling effort on the pure SSC model.

All of the above models of the \gray emission from blazars have all had some
degree of success in reproducing both single-epoch spectral energy
distributions and their relative epoch-to-epoch changes
(\citealp{vonmontigny95_egret}, \citealp{gg98_sedt}).
These favor the SSC model in \mrk because it is a BL Lac object for which
the ratio of thermal (accretion disk and broad line region) and
synchrotron photons is $\sim0.1$, indicating that the EC mechanism is not
important. 
Detailed modeling of ``blue'' BL~Lacs finds that one-component
SSC model can generally account for the time-averaged spectral energy
distributions \citep{gg98_sedt}.
Some data sets seem to require modifications of the simple model,
introducing either multiple SSC components or additional external seed
photons (\eg \citealp{blazejowski_etal_2005}).

We can, however, further decrease the degeneracy among proposed physical
models by taking advantage of blazars' rapid, large-scale time variability
with simultaneous \xraynosp/TeV monitoring (\eg 
\citealp{tavecchio_tev_98,maraschi99_sax_mkn421,henric01_mrk421_x_tev}).
Different models produce emission at a given frequency with particles of
different energies, cooling times, and cross sections for different
processes (\eg \citealp{blumenthal_gould70,coppi_blandford90}), and thus
are in principle  distinguishable \citep*{henric02_timedependent}.
For example, the SSC model predicts nearly simultaneous variations in both
the synchrotron and Compton components (see however
\S\ref{sec:conclusions}),
while other models predict more complicated timing (\eg \citealp{umu97_review}).

With the possible exception of the \xray and \gray data taken on Mrk\,501, the
multiwavelength observations on which we base our inferences, have often
undersampled the intrinsic variability timescales
\citep{buckley96_mrk421,petry00_mrk501,2001ApJ...563..569T,maraschi99_sax_mkn421}
and lack a sufficiently long baseline to make a quantitative
assertion about the statistical significance of a correlation.  
Recently, there has even been evidence of an ``orphan'' TeV flare for
the Blazar 1ES\,1959+650 \citep{krawczynski_etal_2004_pks1959}; a transient
$\gamma$-ray event that was not accompanied by an obvious \xray flare
in simultaneous data.

For what concerns \mrknosp, there have been regular multiwavelength campaigns
in the last several years, planned with an observing strategy focusing
on month-long timescales \citep{blazejowski_etal_2005,rebillot_etal_2006},
and in turn a relatively sparse time sampling (typically one \rxte snapshot
per night, plus binned \rxtenosp/ASM light curves).
These campaigns showed that \xray and \gray brightnesses vary ``in step''
and there is certainly a ``loose'' correlation, and also raised some
questions about the need of considering additional components to account
for the spectra \citep{blazejowski_etal_2005}, or very
high Doppler factors \citep{rebillot_etal_2006}.

The March 2001 campaign remains the experiment with the
highest density coverage at both \xray and TeV energies, and thus
the best dataset to address questions concerning the characteristics of
the variability of the two spectral energy distribution (SED) components.
Moreover, the brightness state achieved during the week of the observations
was unprecedented and it remains unparalleled.
Preliminary results were presented in \citet{jordan01_icrc,fossati03_veritas03}.

In this paper we present the summary of the multiwavelength
observations, with a particular focus on the correlated \xraynosp/TeV
variability, also including the TeV data taken by the HEGRA (High Energy
Gamma Ray Astronomy) telescope.
An account of the HEGRA March 2001 observations was published
by the HEGRA collaboration \citep{aharonian02_mrk421_spectral_var}.
\citet{giebels_etal_2007_mkn421} report on additional TeV observations
with the CAT observatory (Cerenkok Array at Themis), simultaneous with the
\rxte data, not included in this paper.

The \rxte observations, timing and spectral properties are fully
presented by Fossati \etal (2008, in preparation; hereafter F08).

The paper is organized as follows:  the relevant information about the \xray and
TeV observations, and data reduction is given in \S\ref{sec:data}.
The observational findings are presented in \S\ref{sec:analysis}. and
discussed in \S\ref{sec:conclusions} in the context of the synchrotron self--Compton
model.  Section~\ref{sec:conclusions} summarizes the conclusions.

\section{Observations and Data Reduction}
\label{sec:data}

\subsection{\rxtenosp}
\label{sec:rxte}

\rxte observed \mrk in the spring of 2001, during Cycle\,6 for an approved
exposure time of 350\,ks (\dataset[ADS/Sa.RXTE\#P/60145-01]{ObsID 60145}). 
Observations started on March 18$^\mathrm{th}$, 2001, and lasted until
April 1$^\mathrm{st}$, 2001, yielding a total of 48 pointings.  
The \rxte sampling was very dense until March 25.
During the second week, \rxte observed \mrk only during the visibility
times for Whipple, whose visibility windows were also drastically
shortening.  In this paper we only present and discuss the data obtained
between March 18 and March 25.  The journal of this subset of \rxte
observations is reported in Table~\ref{tab:journal}.
A complete account of the \xray campaign is presented in F08.

There are two pointed instruments on-board \rxtenosp, the Proportional Counter
Array (PCA) \citep{jahoda_SPIE_1996}, and the High Energy X--ray Timing
Experiment (HEXTE) \citep{rothschild_hexte_1998}.

The PCA consists of a set of 5 co-aligned xenon/methane (with an upper
propane layer) Proportional Counter Units (PCUs) with a total effective
area of $\approx$6500 cm$^2$. 
The instrument is sensitive in the energy range from 2~keV to $\approx$100~keV
\citep{jahoda_SPIE_1996}.
The background spectrum in the PCA is modeled by matching the background
conditions of the observation with those in various model files on the basis
of the changing orbital and instrumental parameters.

Since the source was very bright throughout the whole campaign, we used
``bright source'' selection criteria\footnote{%
\texttt{OFFSET}$<$0.02, \texttt{ELV} $>$10, \texttt{TIME\_SINCE\_SAA} $>$5.}
to select good time intervals (GTI), and all layers of the
proportional counter units.
We did not include the PCU0 data\footnote{In May 2000 the propane layer of
PCU0 was lost, resulting in a significantly higher background rate and
calibration uncertainty.} in this analysis.  

The total exposure time for the PCA was $\sim$263~ks, divided over 98
pointing ``segments'', i.e. intervals with the same PCUs on, with
individual exposure times ranging between 144~s and 3.3~ks (we rejected 4
GTIs lasting only 16~s). 
The number of PCUs operational during each pointing varied between 1 and 2
(excluding PCU0), with the vast majority of cases (97/98) having 2 (for
details please refer to F08).

HEXTE consists of two clusters of four NaI(TI)/CsI(Na) phoswich
scintillation counters that are sensitive in the range 15--250 keV.
Its total effective area is $\approx$1600~cm$^2$. 
The HEXTE modules are alternatingly pointed every 32~s at source and
background (offset) positions, to provide a direct measurement of the
background during the observation, therefore allowing background
subtraction with high sensitivity to time variations in the particle flux
at different positions in the spacecraft orbit.  Thus, no calculated
background model is required.
The GTIs for the HEXTE data have been prepared independently, \ie not
requiring any PCU to be active, thus resulting in slightly different
on--source times (see Table~\ref{tab:journal}). 
The total HEXTE on--source time was $\sim$282~ks.

The standard data from both instruments were used for the accumulation of
spectra and light curves with 16\,s time resolution. 
Spectra and light curves were extracted with {\scshape FTOOLS v5.1}.
PCA background models were generated with the tool \textsl{pcabackest}, from
\rxte GOF calibration files for bright sources.

The analysis described in the following is mostly based on the 2--15 keV
data from the Proportional Counter Array.
However, the exceptional brightness of \mrk during this campaign makes it 
possible to accumulate a light curve with good time resolution also for the
higher energy data from the HEXTE detector. HEXTE data between 20~keV and
60\,keV were used.

Light curves for intervals with different active PCUs have been rescaled to
the same count/s/PCU units by using the relative weight for the different
effective areas as discussed in the \rxte GOF website.

The very high count rate in the \rxtenosp/PCA enables us to obtain high S/N
light curves for several different energy bands, and this to study the
energy dependence of the flux variability and \xraynosp/TeV relationship.
Based on the distribution of counts (averaged over the range of observed
spectral variability), and the relative background contribution at
different energies, we selected the following four bands (expressed in
terms of the ``hardware'' channels) that have approximately the same
statistics: ch.\,5--8 (labeled 2--4\,keV), ch.\,10--12 (4--6\,keV),
ch.\,14--18 (6--8\,keV), ch.\,20--37 (9--15\,keV). 
Each band comprises on average $\sim$17--24\% of the PCA counts. 
For more details please refer to F08.

We include here some limited discussion involving HEXTE data (namely
their brightness correlation with the TeV dataset).

A non-variable neutral Hydrogen column density N$_\mathrm{H}$ of
1.61$\times 10^{20}$ cm$^{-2}$ \citep{lockman_savage95_nh} have been used.
However, since the PCA bandpass starts at about 3~keV, the adopted value
for N$_\mathrm{H}$ does not affect significantly our results.

\subsection{Whipple Observatory $\gamma$-ray data}
\label{sec:tev_whipple}

The TeV data on \mrk were taken with the Whipple 10m atmospheric Cherenkov
telescope.  The Whipple telescope detects $\gamma$-rays by imaging the flashes of
atmospheric Cherenkov light emitted by gamma-ray induced electromagnetic showers.
Individual shower images are recorded on a photomultiplier tube (PMT)
camera, and later analyzed to reject the cosmic ray background, select
$\gamma$-ray events, and determine the point of origin and energy of each
detected $\gamma$-ray.  
For the period of the 2001 \mrk observations, the camera consisted of a
hexagonal array of 490 PMTs on a 0.12$^\circ$ pitch PMT camera (only the
inner 379 pixels were used for the present analysis) \citep{finley_2001_icrc}.
Images were characterized by calculating the moments of the angular
distribution of light registered by the PMT camera. 
These moments include the RMS width and length of the shower
images, and orientation of the roughly elliptical images.
Background cosmic-ray showers are rejected using data cuts based on these
parameters according to the procedure described by \citet{delacalleperez_etal_2003}.
The point of origin of each detected $\gamma$-ray can be determined from the
orientation and elongation of the image to with a precision of about 0.12
degree \citep{lessard_etal_2001}. 

The total integrated signal for a shower image (the shower \textsl{size}) is
roughly proportional to the energy of the primary $\gamma$-ray.  To obtain a
better energy estimator, we also correct for the impact parameter of the
shower.  An approximate measure of the impact parameter of each shower is
obtained by measuring the parallax angle between the image centroid (point of
maximum shower development) and the source direction.  Correcting for the
dependence on this centroid distance, one obtains an energy estimator for each
candidate $\gamma$-ray event.  Monte Carlo simulations of $\gamma$-ray showers are
used to determine the relationship between this energy estimator and the actual
energy.  To determine the energy spectrum for a given data set, a histogram of
the energy estimator for candidate $\gamma$-ray events is formed for both the
ON-source and OFF-source data.  The difference in the ON-OFF histograms is then
converted to an energy spectrum by using the effective area function calculated
by detailed Monte Carlo simulations of the air shower and detector.  

For the present analysis, the spectra were reconstructed either using the
method of \citep{mohanty_etal_1998,krennrich01_2001data_cutoff}  
(method\,A) or the forward folding method of \citet{rebillot_etal_2006}  
(method\,B).  
In method\,A, the combined effects of the spectral deconvolution (assuming a
log normal resolution function) and energy-dependent
efficiency are taken into account by dividing the binned fluxes by the
effective area function.  The spectral slope is then determined by fitting a
spectral model to the unfolded spectral data points by $\chi^2$ minimization.
Following \citet{krennrich01_2001data_cutoff}, we assume that the
\mrk spectrum can be characterized by a single power-law with an
exponential energy cutoff at a fixed value of 4.3\,TeV.

In method\,B, instead of fitting to the unfolded spectrum, we directly compared
the energy estimator distribution of a simulated $\gamma$-ray data set to the
measured distribution for the real data.  A grid search was used to find the
parameters (flux normalization and spectral index) that resulted in the best
fit to the data.  To make this computationally tractable, we did not repeat the
simulations for each trial, but drew our random datasets from a weighted
simulation database.  The measured and simulated energy-estimator distributions
were compared to form a likelihood for each trial value of the spectral index.
An adaptive grid search was used to find the best fit value of the flux
normalization and spectral index, and to determine the confidence interval on
the best fit parameter.   

For our spectral analysis, it is important to know the energy scale to get a
reliable value for the flux normalization.  For this season the trigger
condition (which directly impacts the energy threshold) consisted of the
requirement that 3 adjacent PMT signals exceeded a threshold of 32\,mV.
By comparing muon-ring images obtained in the data with those generated
in Monte Carlo simulations, we obtained a gain calibration for the 2001 season.

Even in its relatively active state, the $\gamma$-ray observations of \mrk 
are often background limited.  To adequately characterize the background of
misidentified cosmic-ray events we either used an ON-OFF analysis
or tracking analysis.  For these observations, we employed an
observing strategy that was a compromise between the desire to obtain
continuous coverage for the time-series analysis, and the need for interspersed
OFF-source control runs to adequately characterize systematic variations in
background.

In a typical night, a single ON/OFF pair was taken in which a 28-minute ON-run
run with the 10m telescope tracking \mrk was followed by another
28-minute observation that tracks the same range of azimuth and elevation
angles, but offset by 30 minutes in right ascension.  This mode of data taking
reduces systematic errors from difference in sky brightness, atmospheric
conditions and instrument variations but reduces the duty cycle of observations.
Other runs taken in the same night were typically acquired in a \textsl{staring}
or \textsl{tracking} mode where the telescope continuously tracks the source
without interruption for off-source observation. 

Data selection cuts based on the RMS \textsl{width} and \textsl{length} of the shower
images were used to determine candidate $\gamma$-ray events.  The angle between
the major axis of the Cherenkov image and the line connecting the image
centroid to the angular position of the source (designated $\alpha$),
is used to define the signal and background region in the image
parameter space. In each ON-source run, the events whose major axis points to
the position of the source in the field of view are used to estimate the
signal, and those misaligned events pointing away from the center are
used to determine the background \cite{delacalleperez_etal_2003}. 
The number of
candidate $\gamma$-ray events was determined by subtracting the number of
on-source events from off-source events.  Significances were determined
by the method of \citet{lima_1983_gammaray_significance}.

Since the data were taken over a range of zenith angles, a first order
correction was made to take into account variations in energy threshold
and effective collection area with zenith angle.  For the light curves
presented in this paper we choose
the simple empirical method of normalizing to the flux from the Crab Nebula
at the corresponding zenith angle.  Like relative photometry, this method
cancels systematic errors and should provide a better method for eventually
combining data with data obtained by different detectors calibrated by
different Monte Carlo simulations.  However, this method results in a
systematic (second order) error if the \mrk spectrum differs significantly
from that of the Crab Nebula.

\begin{figure}
\centerline{
\includegraphics[width=0.99\linewidth]{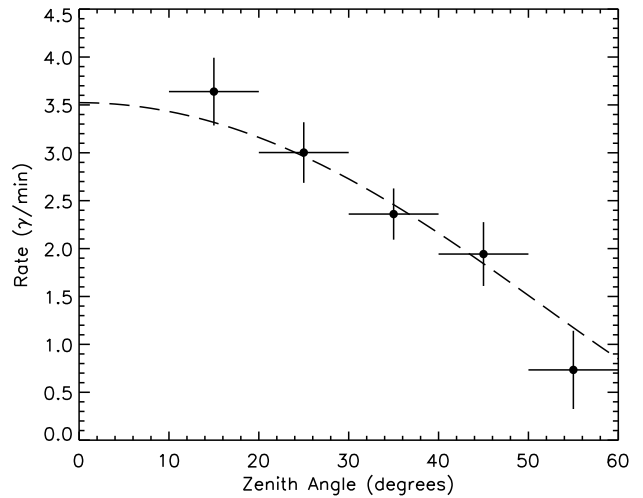}
}
\caption{%
\small
Dependence of the observed rate of gamma rays from the Crab Nebula as a
function of zenith angle, as observed in the 2000--2001 season.
The fit function indicated by the dashed curve was used
to normalize the raw \mrk rates.
\label{fig:crabzen}
}
\end{figure}

The Crab Nebula can be used to empirically quantify the combined effect of
these changes for a source with the same spectrum $N(E) \sim E^{-2.49}$ 
\citep{hillas_etal_1998_crab}.
Figure~\ref{fig:crabzen} shows a fit to the Crab rate as a function of zenith
angle for 50 ON/OFF runs from October 25, 2000 to the end of the 2001
observing season.

The average spectral index of \mrk has been measured to have a similar
value to the Crab nebula in the Whipple energy range, but with evidence for
some spectral variability \citep{piron01_mrk421,krennrich02_mrk421_spectral_var,aharonian02_mrk421_spectral_var}.
Since the energy threshold only varies by a factor of roughly 30\% over the
range of zenith angles of our observations, we estimate a systematic error
in the flux of roughly 10\% if we assume that the spectral index varies by
$\triangle\Gamma\simeq0.3$. 
But we lack the ability to determine the spectral index to this
precision for most individual runs, and thus subsequently admit the
possibility of some systematic errors, 
and normalize our observations to a functional form for the zenith-angle
dependent rate from the Crab Nebula.

\begin{equation} 
R_{\rm Crab}(\theta) = 
\frac{7.423 \; sec^2(\theta)}{\left[\exp\left(0.5 \sec(\theta)) \sec^2(\theta)\right)\right]^{1.49}}
\end{equation}

This function is derived from an empirical fit to data taken on the Crab
Nebula at various zenith angles during the same observing season (see  
Figure~\ref{fig:crabzen}).

Total observations for the March 2001 campaign came to 10.3 hours of ON/OFF
data and 37.7 hours of TRACKING data for a total of 48 hours of data, only
a fraction of which are covered in this paper.

For our multiwavelength studies we chose three arbitrary, a-priori bin widths 
i) one corresponding to a single day of observations, 
ii) one corresponding to a single 28 minutes data run, and
iii) a shorter bin width of $\approx$4 minutes. 
The latter choice is arbitrary, but was derived from past analyses of
strong flares as giving a good compromise between statistics and temporal
resolution.

\subsubsection{HEGRA Observatory $\gamma$-ray data}
\label{sec:tev_hegra}

The HEGRA TeV data, light curve and spectra, utilized in this paper
have been previously published by \citet{aharonian02_mrk421_spectral_var}.
Please refer to the original paper for details concerning the data reduction.

\subsection{Combining Whipple and HEGRA $\gamma$-ray data}
\label{sec:combining_whipple_and_hegra}

\begin{figure*}[t]
\centerline{
\includegraphics[height=0.99\linewidth,angle=270]{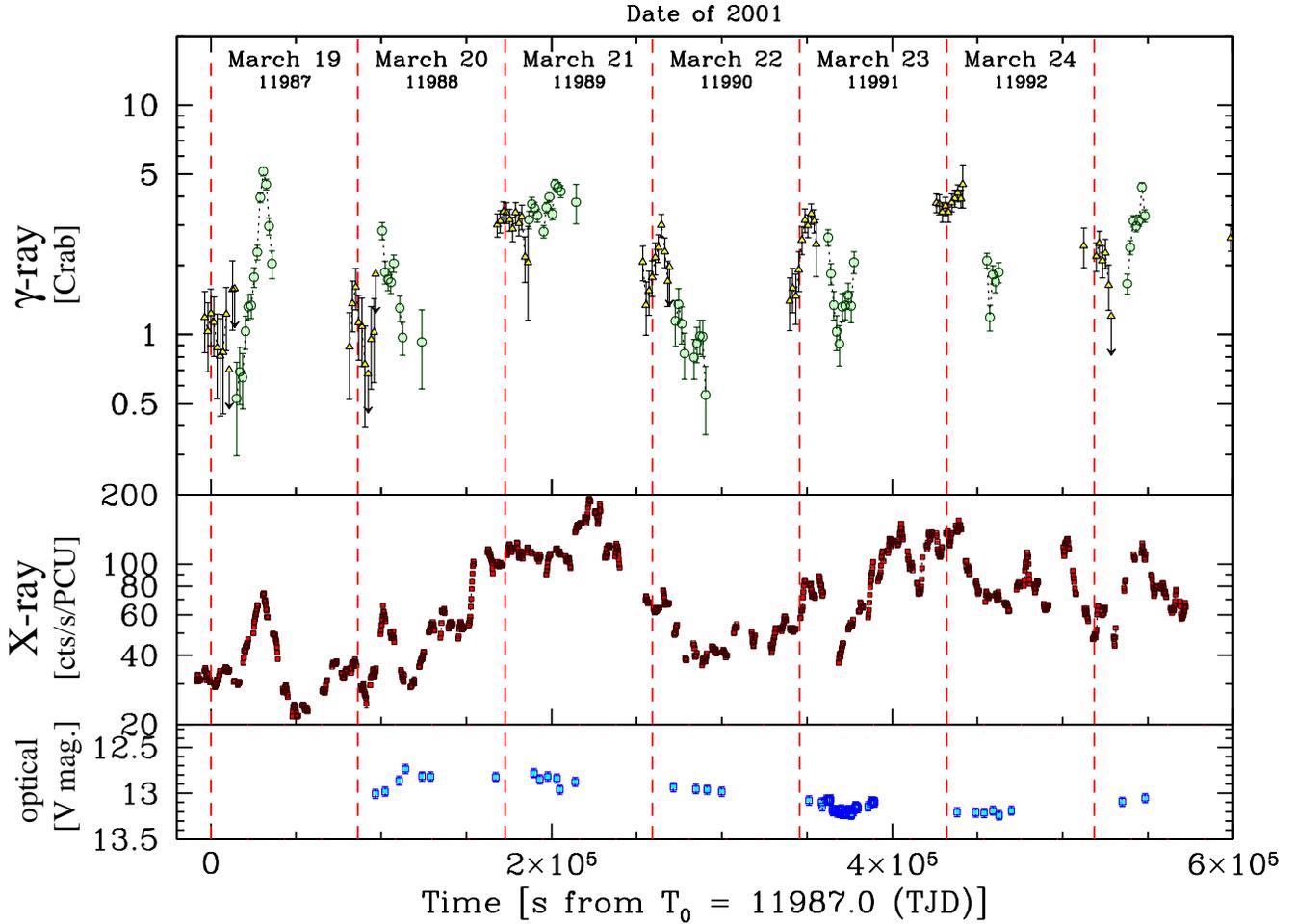}
}
\caption{%
\small
Simultaneous optical (V band, bottom), \xray (2--10~keV, middle) and TeV $\gamma$-ray 
($E>0.4$\,TeV, but see \S\ref{sec:combining_whipple_and_hegra}, top) 
(top) light curves for \mrk for the March 18--March 25 period.
\rxtenosp/PCA data are shown here in 256~s bins.
HEGRA data (dark triangles) are integrated over $1800$~s bins, Whipple data
(white circles) over $1680$~s bins.  HEGRA data precede Whipple's.
The optical data have been rebinned to yield a s/n ratio of at least 8, but with
bin length not exceeding 1500~s.
The logarithmic scales span a factor of $\sqrt{10}$, 10 and 100 for the
optical, \xray and \gray light curves, respectively. 
The length of axes scale accordingly ($\times 2$ between them), so that 
relative amplitude variability can be directly compared.
\label{fig:x_and_g}
\label{fig:x_and_g_and_o}
\label{fig:x_and_g_all_week}
}
\end{figure*}

\begin{figure*}[t]
\centerline{%
\hfill
\includegraphics[width=0.49\linewidth,clip=]{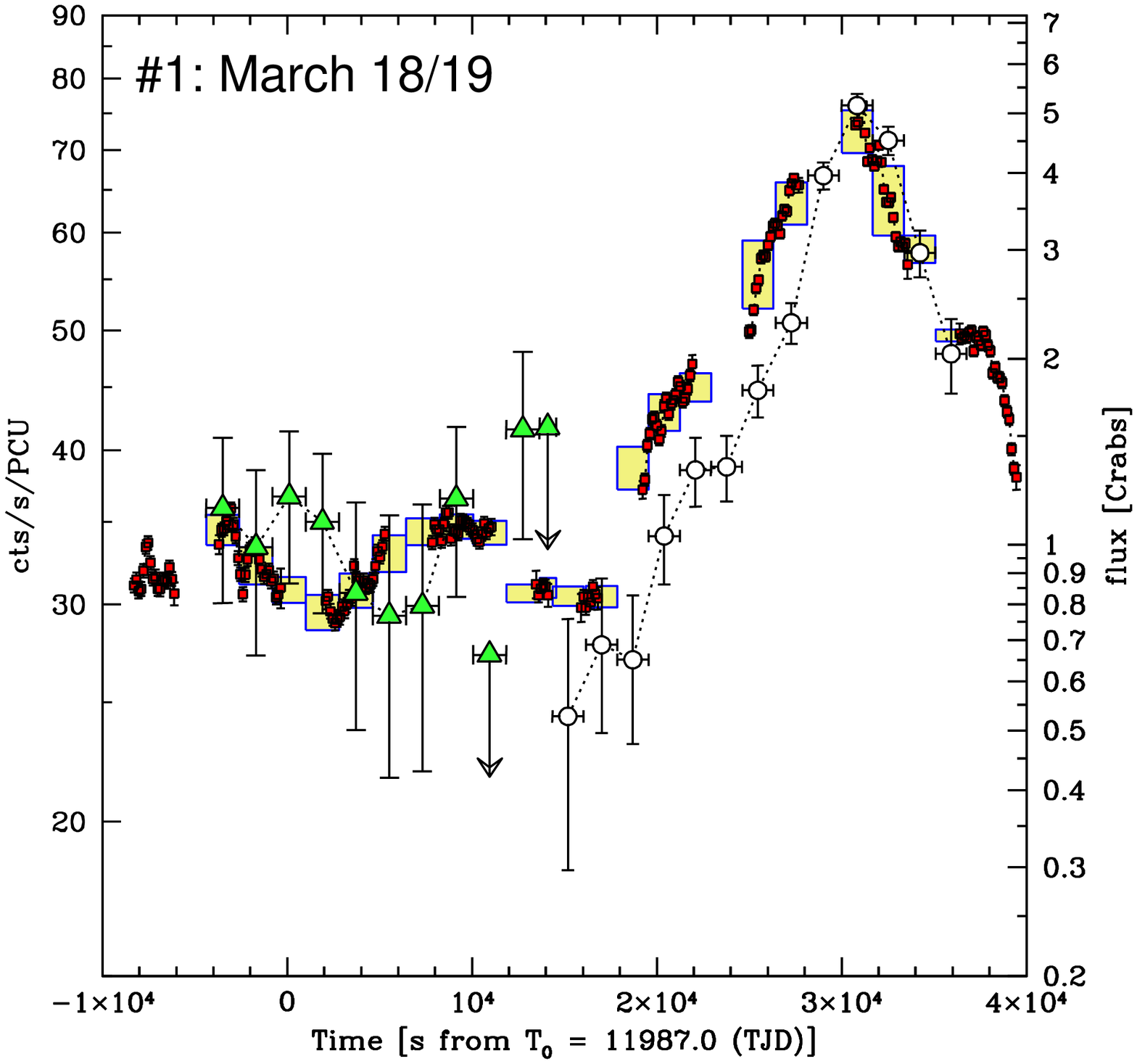}
\hfill
\includegraphics[width=0.49\linewidth,clip=]{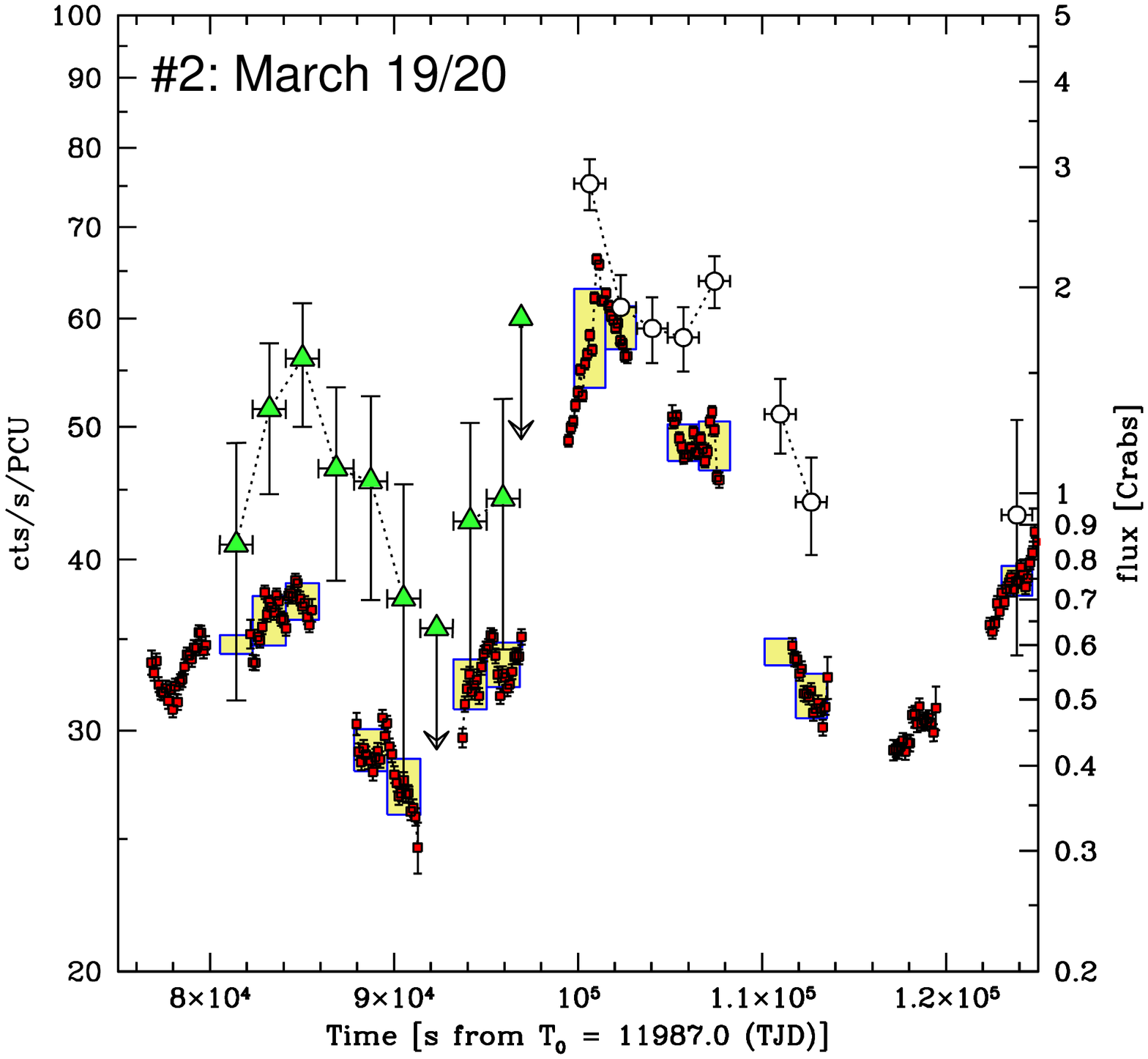}
\hfill
}
\centerline{%
\hfill
\includegraphics[width=0.49\linewidth,clip=]{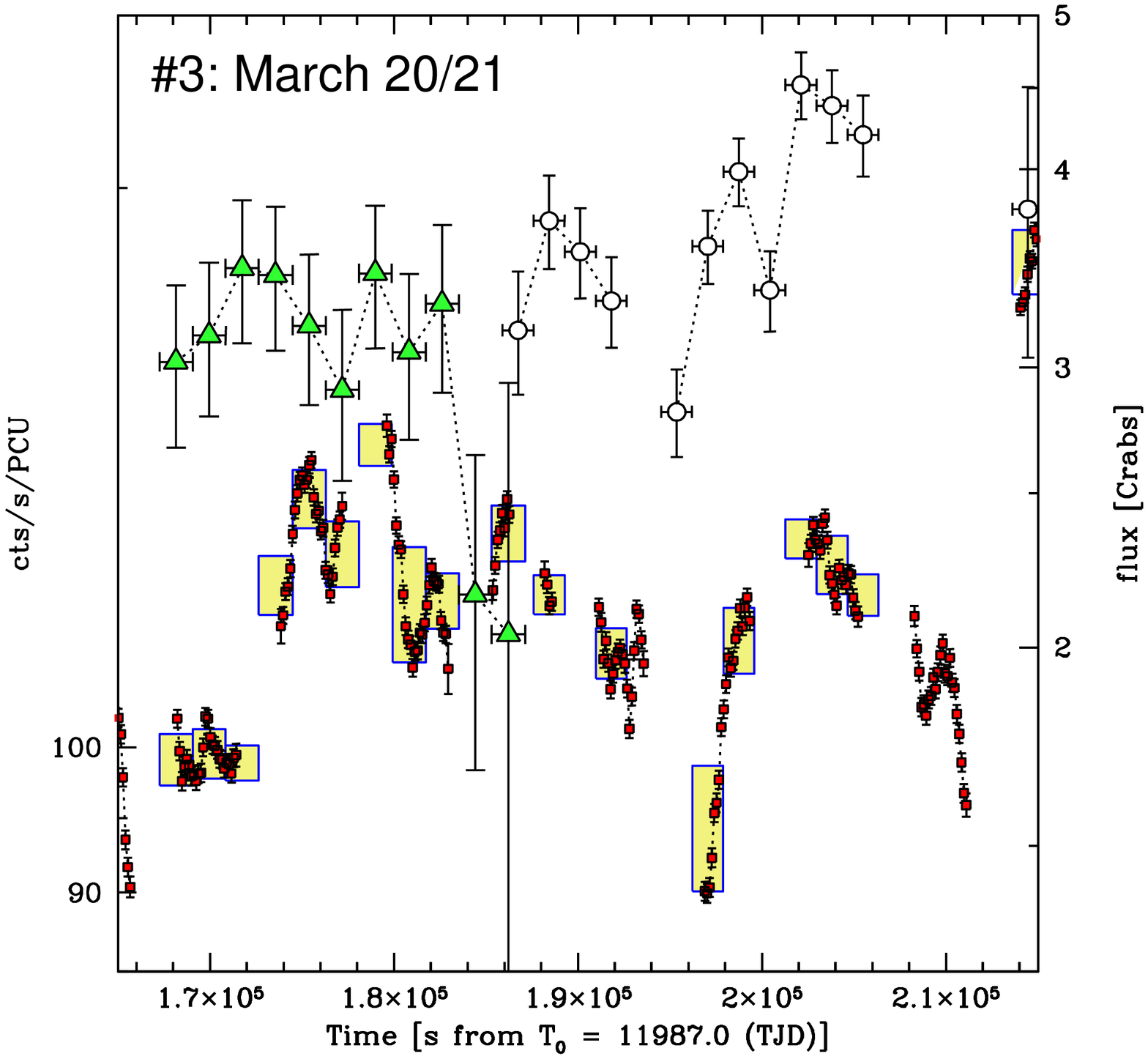}
\hfill
\includegraphics[width=0.49\linewidth,clip=]{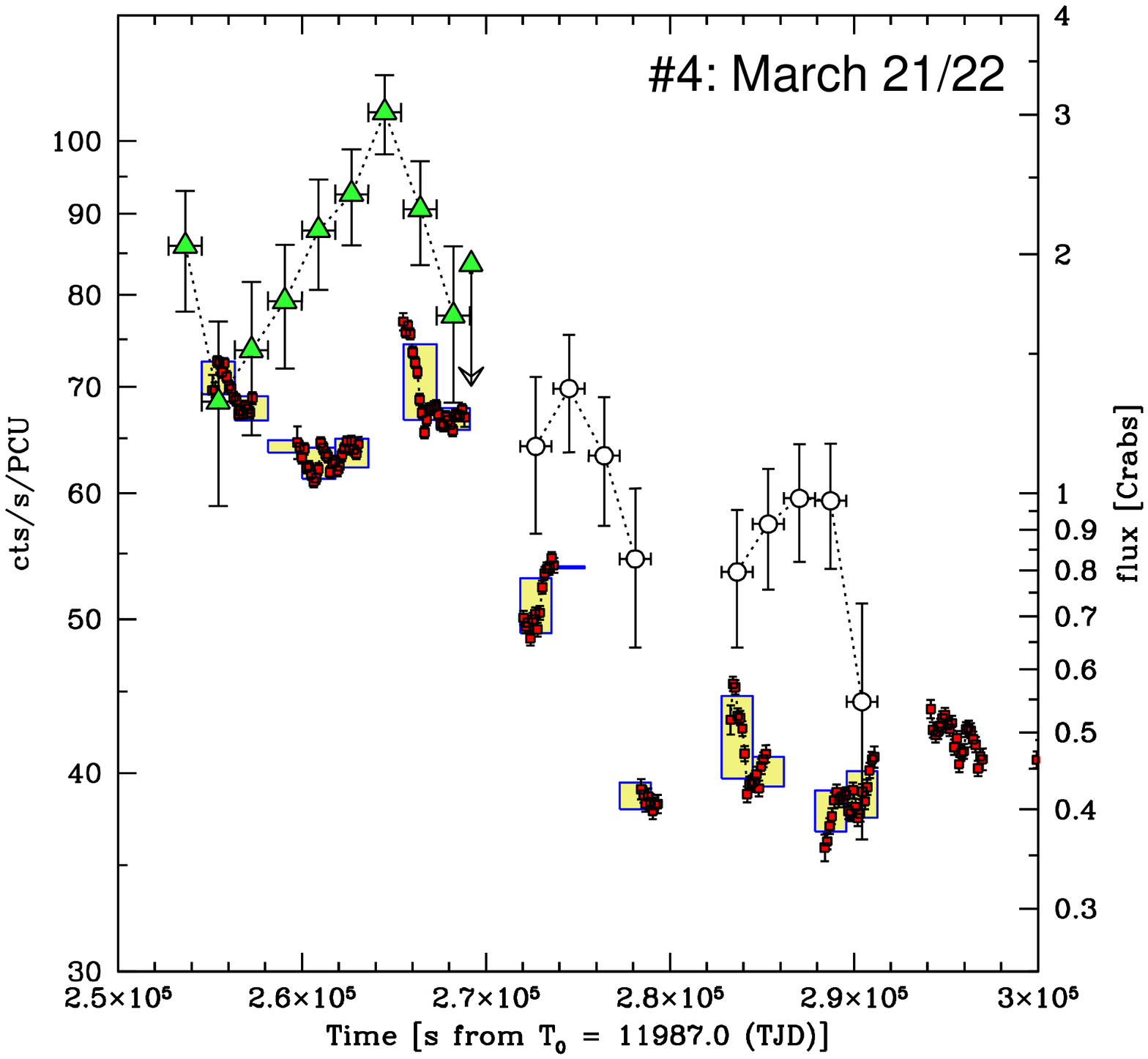}
\hfill
}
\caption{%
\small
Simultaneous 2--10~keV \xray and TeV (see text) $\gamma$-ray light curves for
individual nights. 
Gray triangles are HEGRA data, in $\approx$1800~s bins, 
White circles are Whipple data, integrated over $\approx$1680~s bins.
Dense dark points are \rxtenosp/PCA, in 128~s bins.
The shaded boxes represent the average and variance of the \xray data for
each (longer duration) TeV bin, that are the values used in the
Flux--Flux correlation analyses.
The rate scales for the \xray data are on the left Y-axes, and the flux scales
for the TeV data on the right Y-axes.
The time span is the same for all panels, 50\,ks.
The vertical scales are not the same in all panels, but are adjusted to show
each day in the best possible detail.  The \xray dynamic ranges are (time
ordered) $\times 6$, $\times 5$, $\times 2$, $\times 4$, $\times 4$, $\times 4$, $\times 4$.  
In order to allow for an easier comparison of the relative variability
amplitude, in all panels, the Y--axis range for the \gray light curve is
the square of that used to plot the \xray data.
The source shows strong, highly-correlated variability in both energy
bands, with no evidence for any interband lag (see however \S\ref{sec:lags}).
\label{fig:x_and_g_individual_nights}
\label{fig:x_and_g_nights}
\label{fig:good_nights}
\label{fig:goodnights}
}
\end{figure*}

The locations of the Whipple and HEGRA telescopes, in Arizona and the Canary
islands, respectively, separated by approximately 6 hours, make it possible to
achieve un-interrupted coverage of \mrk during the spring, when the target is
observable at ``small'' zenith angles for up to 7 hours each night from
each site.
The visibility windows thus complement each other very well 
(see Figures~\ref{fig:x_and_g_all_week} and~\ref{fig:x_and_g_nights}),
but don't overlap.
For this campaign we achieved an unprecedented coverage of the target:
as illustrated in Tables~\ref{tab:overlap} and~\ref{tab:overlap_all}  the total
net exposure time for the two TeV observatories was of about 62 hours, over
seven nights.
Ignoring gaps shorter than 1.5 hours, the on-source time was of about 72 hours,
\ie more than 10 hours per night.

The combination of data from different instruments is always a very delicate step,
most robustly addressed by comparing data taken simultaneously, which is not
possible for Whipple and HEGRA.
The next best option is to cross-calibrate the data using a standard candle as
reference, typically the Crab Nebula for high energy emission.
The main issue concerning the combination of the data from the Whipple and
HEGRA telescopes arises from the fact that \textit{i)} they gather data with different
lower energy threshold, 0.4 and 1.0\,TeV respectively, and that \textit{ii)} \mrk TeV
spectrum is in general (significantly) harder than the Crab's.
To illustrate the problem, let's take spectral indices 
$\Gamma=2.5$ for the Crab, 
and $\Gamma=2.2$ for \mrk \citep{krennrich02_mrk421_spectral_var,aharonian02_mrk421_spectral_var}.
Ignoring the fact that a detector response is energy dependent we can compare
the ratio between the \mrk and Crab fluxes computed for different energy
thresholds, namely 0.4 and 1\,TeV.
The flux above an energy $E$ for a simple power law with spectral index $\Gamma$ is:
\begin{equation}
F(E,\Gamma) = \frac{N_0\, E^{1-\Gamma}}{\Gamma-1}
\end{equation}
Therefore
\begin{equation}
\tilde{F}_{mrk}(E) = \frac{F(E,\Gamma_{mrk})}{F(E,\Gamma_{Crab})} = 
\frac{N_{0,mrk}}{N_{0,Crab}}\, \frac{\Gamma_{Crab}-1}{\Gamma_{mrk}-1}\, E^{\Gamma_{Crab}-\Gamma_{mrk}}
\end{equation}
and comparing data taken at two different thresholds yields, 
\begin{equation}
f = \frac{\tilde{F}_{mrk}(E_A)}{\tilde{F}_{mrk}(E_B)} = \left(\frac{E_A}{E_B}\right)^{\Gamma_{Crab}-\Gamma_{mrk}}
\end{equation}
For $E_A=1.0$\,TeV (HEGRA) and $E_B=0.4$\,TeV (Whipple), this ratio is 1.32,
\ie a spectrum yielding a Whipple flux of 1\,Crab will be observed at
$1.32$\,Crab by HEGRA.
Moreover, this effect is a function of the spectral indices.
While we can safely consider the Crab spectrum non variable, significant
variability is observed in \mrk
\citep{krennrich02_mrk421_spectral_var,aharonian02_mrk421_spectral_var}.
The value of $f=1.32$ for the HEGRA/Whipple flux ratio obtained for $\Gamma_{mrk}=2.2$
becomes $f=$1.44, 1.10, 0.83 for $\Gamma_{mrk}=2.1, 2.4, 2.7$, respectively. 
Hence, HEGRA data would also show a larger variance.
Finally, it is worth noting that the ratio between the March 2001 Whipple and HEGRA
mean fluxes, and between their variances, are consistent with what expected
based on the simple arguments just illustrated and the observed spectral indices.


Since it is not possible to derive from ``first principles'' a robust way to
convert Whipple and HEGRA light curve measurements into each other, we
conclude that the most robust procedure at hand is to scale the data of one
telescope to the same mean and variance of those of the other one.
We adopt Whipple as reference and scale the HEGRA fluxes.
We deem that this approach is reliable, primarily because of the large size
of the two datasets: they both span seven days (with alternating visibility
windows), and have comparable variability amplitudes, thus both should be
sampling an equally representative subset of the source phenomenology.
Being a linear conversion, this procedure can not correct for the slight
non-linearity of the Whipple vs. HEGRA flux relationship introduced by the
brightness-hardness correlation of the spectra.
In order to mitigate the effect of outliers, and because of the intrinsic
exponential nature of the source variability, we performed the scaling of mean
and variance on the logarithm of the light curves.

\subsection{Optical data}
\label{sec:optical_data}

UBVRI optical monitoring was performed using the Harvard-Smithsonian
48" telescope on Mt. Hopkins.
Data were analyzed using relative aperture photometry, using comparison
star \#3 
as listed in \citet{villata98_refstars} 
Galaxy background light was subtracted using the simple empirical method
described in \citet{nillson99_galaxy_subtraction}. 
Here we use \citet{nillson99_galaxy_subtraction}'s determination of the
contribution of \mrk galaxy background light in the $R$ band, but extrapolate
this to the $V$ band using the $R-V$ color of the host galaxy as given in
\citet{hickson82_mrk421_photometry}. 
The errors shown on the light curve are the systematic uncertainties, which
dominate the small statistical errors.
These errorbars were determined from the measured variance of the reference
stars with respect to each other, 
and may not include other effects such as the
bleeding of starlight from the bright stars SAO 62387 and SAO 62392 
that lie about 2\,arcmin from \mrknosp, in the 48" telescope field of view.    
Other systematic effects come from the relatively high level of galaxy light
from \mrk and the very nearby satellite elliptical galaxy.
These problems, coupled with the lack of good reference stars 
combine to make \mrk one of the more difficult BL\,Lac objects for optical
monitoring; we estimate that this measurement should be given a systematic
flux uncertainty of about 15\%.
                                                                                                                              
\begin{figure*}
\centerline{%
\hfill
\includegraphics[width=0.49\linewidth,clip=]{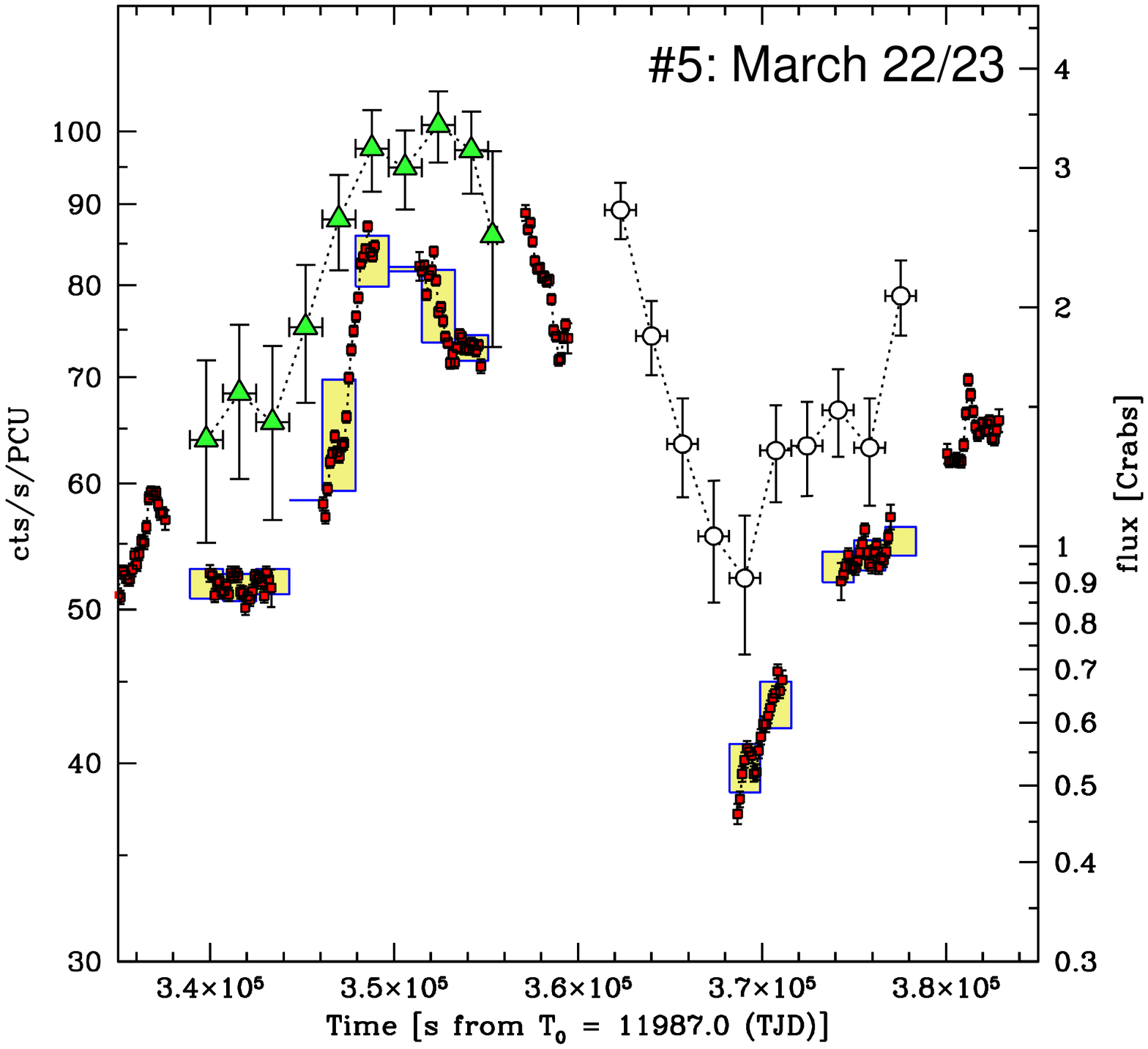}
\hfill
\includegraphics[width=0.49\linewidth,clip=]{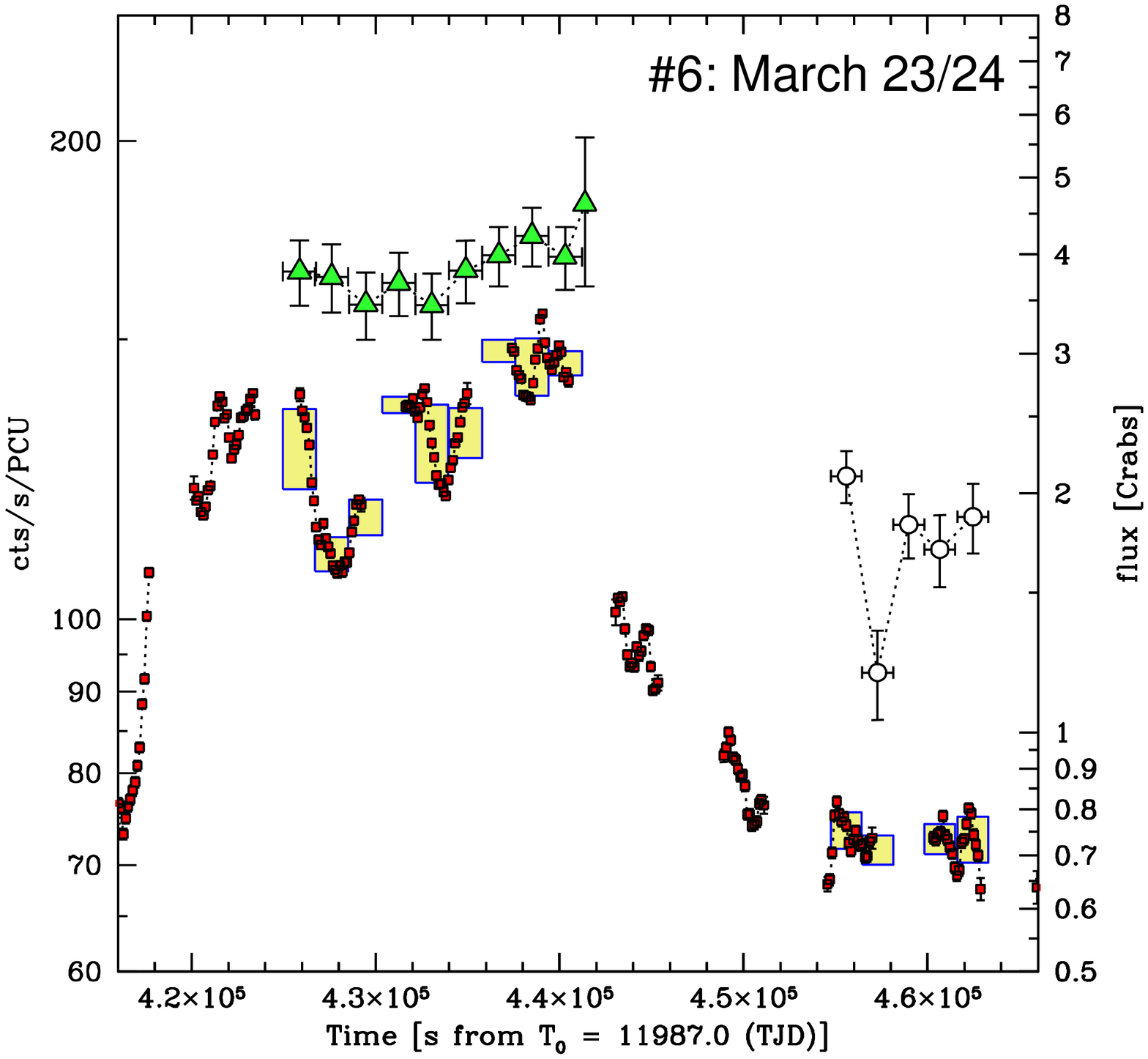}
\hfill
}
\centerline{%
\includegraphics[width=0.49\linewidth,clip=]{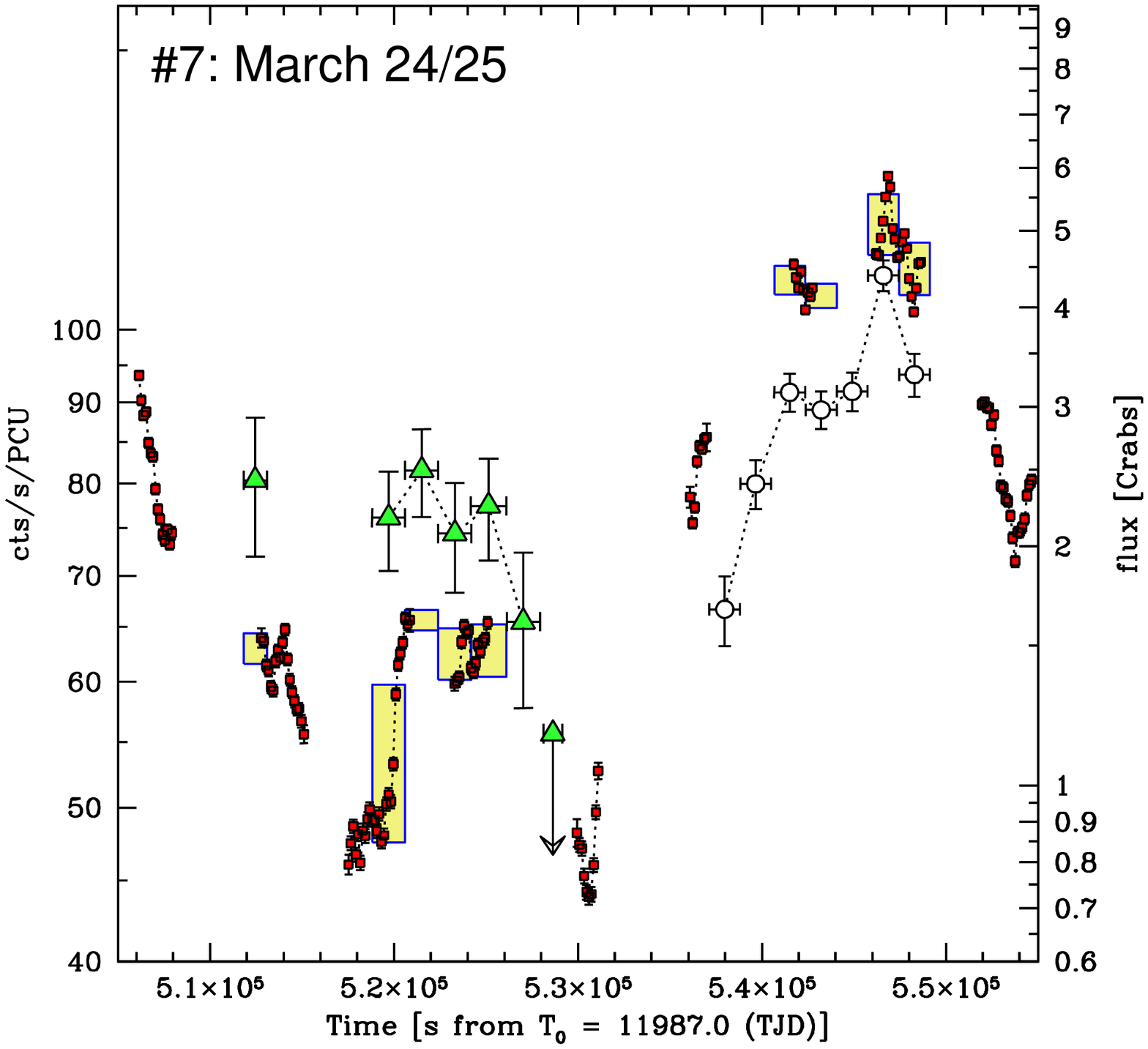}
}
\centerline{Fig. 3. --- Continued.}
\end{figure*}

\section{Temporal Analysis}
\label{sec:analysis}

\subsection{\rxtenosp/Whipple$+$HEGRA data overlap}
\label{sec:overlap}

The principal statistics regarding the quality of the \xraynosp/\gray overlap
are summarized in Table~\ref{tab:overlap} and~\ref{tab:overlap_all}.
There we report two different measures of the overlap between the two telescopes.
The fractions listed in columns 5 and 8 refer to the absolute ``covering
factor'' by \rxte of the on--source time of Whipple or HEGRA. 
These numbers give an idea of how ``representative'' the collected \xray
data are of the actual \xray brightness.
In columns 6 (and 9), instead, we report the fraction of Whipple (and HEGRA)
data points (from the run-by-run light curves) that have a non--zero overlap
with \textit{any fraction} of the \rxtenosp/PCA GTIs: the actual exposed fractions
for the individual runs range between 7\% and 100\% (with average of the order
of the value reported in column 5 and 8).
Although the \rxte schedule was optimized to observe \mrk as much as possible\footnote{%
Between Match 19 and 25 \rxte observed \mrk for 94 out of 102 orbits, and in only
one case there were two consecutive orbits missed.  One missed orbit corresponds
to a $\simeq8.5$\,ks gap.},
we still missed about 1/4 of the Whipple runs, and during
each Whipple/HEGRA $\approx0.5$~hr integration window \xray data were
collected only for about 50\% of the time.
Since the \xray and \gray brightness can vary significantly on timescales
faster than $\approx0.5$~hr, we may expect this to affect the \xraynosp/TeV flux
correlation. 
We carefully inspected the \xray light curves to assess the impact of the
un-even coverage and concluded that the effect is not important.  
The largest \xray variation during any $0.5$~hr interval is $\sim$30\%, with an
average of $\sim$15\%. 

\subsection{Light curves}
\label{sec:lc}

The resulting light curves for the full week are shown in Figure~\ref{fig:x_and_g_all_week}.
The overall brightness ranges are approximately $\times0.5$, $\times10$,
$\times10$ for optical, \xray and \gray respectively.
Although the time coverage was the best possible and the datasets encompass
several days and multiple flares, these ranges can be affected by the lack
of continuous coverage by the ground based telescopes (\eg there are no
optical or \gray data for the time of the lowest \xray flux, on March 19).

Although the large gaps in the \gray light curves may limit the interpretation,
some effects are obvious.
The source shows strong variations at both bands, and these variations are
highly correlated, with features in one band generally showing up in the
other band as well.
This is clearly illustrated in Figure~\ref{fig:goodnights}, where we plot
50~ks sections of each night, centered on  the time window covered by the
TeV observations.

The best example is the March 19 observation, when a well defined, isolated, 
flare was observed both in the \xray and \gray bands, from its onset
through its peak (more in Sections~\ref{sec:lags} and~\ref{sec:ff_correl}).
In several other nights, despite the lack of other major ``isolated'' features
in the light curve(s) that can be provide an unambiguous reference for the
comparison, the variations at \xray and \gray are clearly correlated.
Given the high degree of variability (rapid and high amplitude in both bands)
it is always possible that intrinsically un-correlated flares in the two bands
end up being simultaneous by chance.
However, the unprecedented level of detail (\ie time resolution) and extension
of the simultaneous coverage, allows us to ``match'' several relatively minor
features in the light curves, and effectively confirm the correlation of 
the variations in the \xray and \gray bands.

In particular, within the detail allowed by the coarser TeV sampling, the \xray
and \gray light curves seem to track each other closely in all cases
when the sampling is good, namely for the nights \#1, 2, 4, 5, 7.
For nights \#3 and 6 
there seem to be significant deviations from this general trend.
It is, however worth noting that these two nights correspond to the highest
brightness level in the \xraynosp, and the \xray light curves show several
fast variations (intra-orbit), that may not have been sampled
properly by the TeV observations (\eg in \#3), or may in fact not
have a TeV counterpart (\eg \#6, where the TeV data seem to have good
signal to noise). 
In a broad sense, even in \#3 and \#6, the data are consistent with
correlated variations in the two energy bands.

In the next two subsections (\S\ref{sec:lags}, \S\ref{sec:ff_correl}), we
are going to examine in detail the properties of the \xraynosp/TeV correlation
from two complementary point of views, phase and amplitude.

\subsection{X--ray/TeV interband lags}
\label{sec:lags}

Cross-correlation functions were measured to quantify the degree of
correlation and phase differences (lags) between variations in the \xray
and \gray bands, using the discrete correlation function (DCF) of
\citet{edelson_krolik88}.  
For the \xrays we consider two different bands (2$-$4\,keV and 9$-$15\,keV)
at the usable ends of the PCA bandpass, to explore the possible
energy dependence of the correlation.   
The results for the whole-week and March 19 flare data are shown
in Figures~\ref{fig:ccf}a--d.
It is worth noting that, as discussed in F08, variations in the harder
\xray emission seem overall to lag those in the softer PCA band.  

\subsubsection{Full week-long dataset}
\label{sec:dcf_full_week}

\begin{figure*}[t]
\centerline{%
\hfill
\includegraphics[width=0.35\linewidth,angle=270]{f04a.ps}
\hfill
\includegraphics[width=0.35\linewidth,angle=270]{f04b.ps}
\hfill
}
\centerline{%
\hfill
\includegraphics[width=0.35\linewidth,angle=270]{f04c.ps}
\hfill
\includegraphics[width=0.35\linewidth,angle=270]{f04d.ps}
\hfill
}
\caption{%
\small
Cross correlation between the \xray and the TeV light curves. 
(a) 2--4~keV vs. TeV (Whipple$+$HEGRA) for the whole campaign (computed
over 2048~s bins, from \xray data on 256~s bins, and TeV data on
$\simeq$750--900~s bins).
(b) 9--15~keV vs. TeV (Whipple$+$HEGRA) for the whole campaign (computed
over 2048~s bins, from \xray data on 256~s bins, and TeV data on
$\simeq$750--900~s bins).
(c) 2--4~keV vs. TeV (Whipple) for the night of March 18-19 (the flare of
Figure~\ref{fig:goodnights}a) (computed over 1024~s bins, from \xray data
on 128~s bins, and Whipple data on 256~s bins).
(d) 9--15~keV vs. TeV (Whipple) for the night of March 18-19 (computed over 1024~s
bins, from \xray data on 128~s bins, and Whipple data on 256~s bins).
\label{fig:dcf}
\label{fig:ccf}
}
\centerline{%
\hfill
\includegraphics[width=0.33\linewidth]{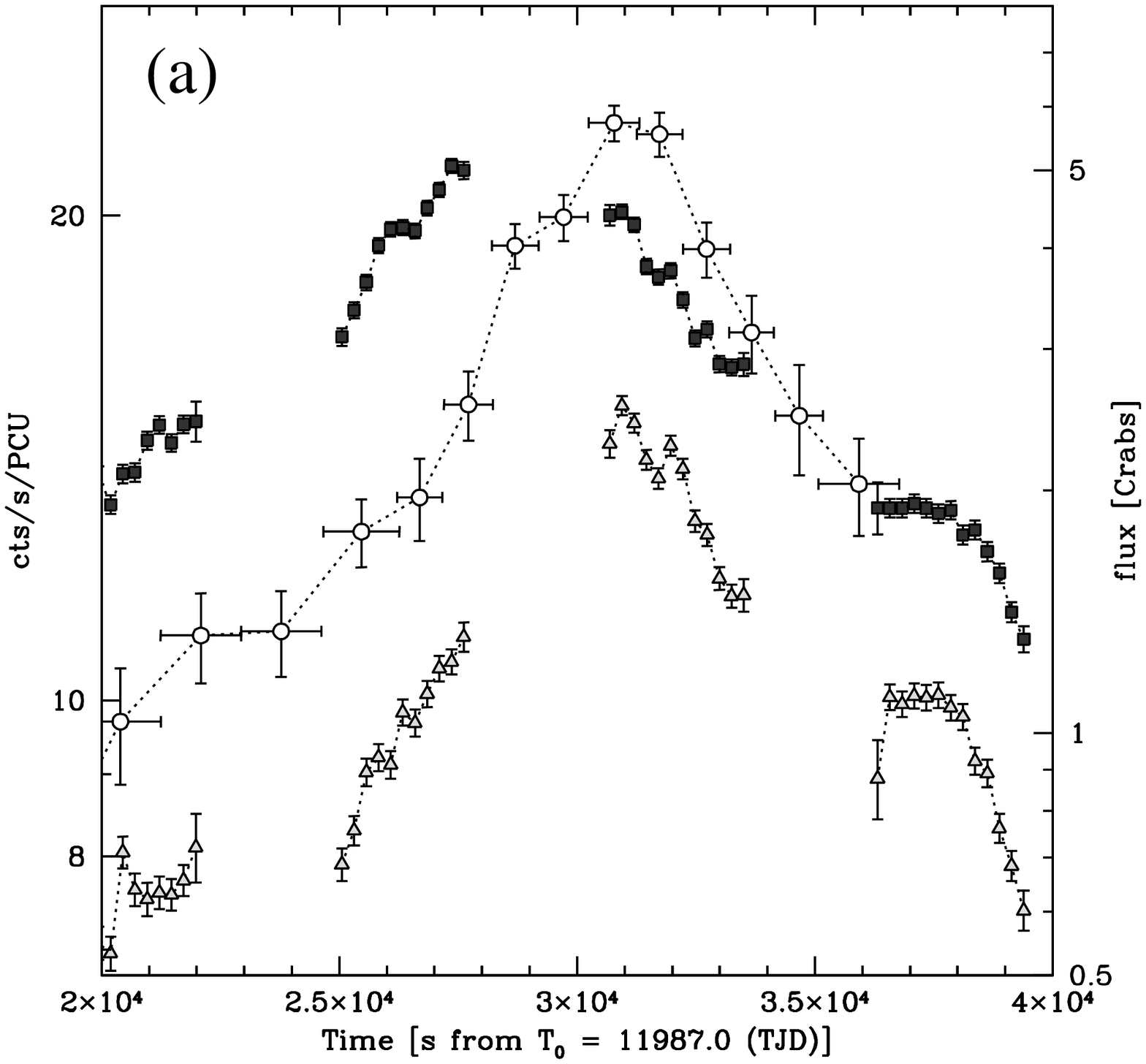}
\hfill
\includegraphics[width=0.33\linewidth]{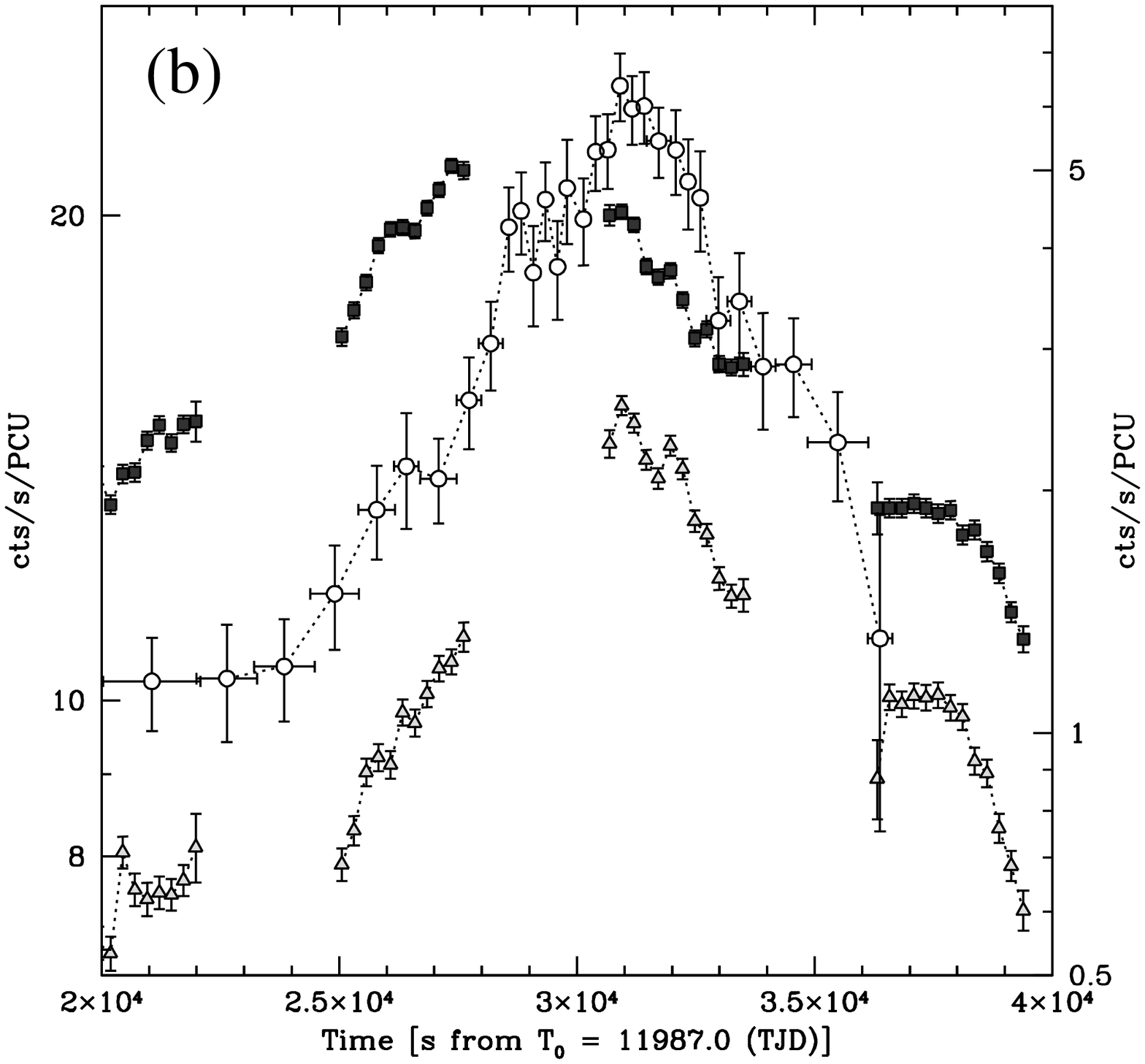}
\hfill
\includegraphics[width=0.33\linewidth]{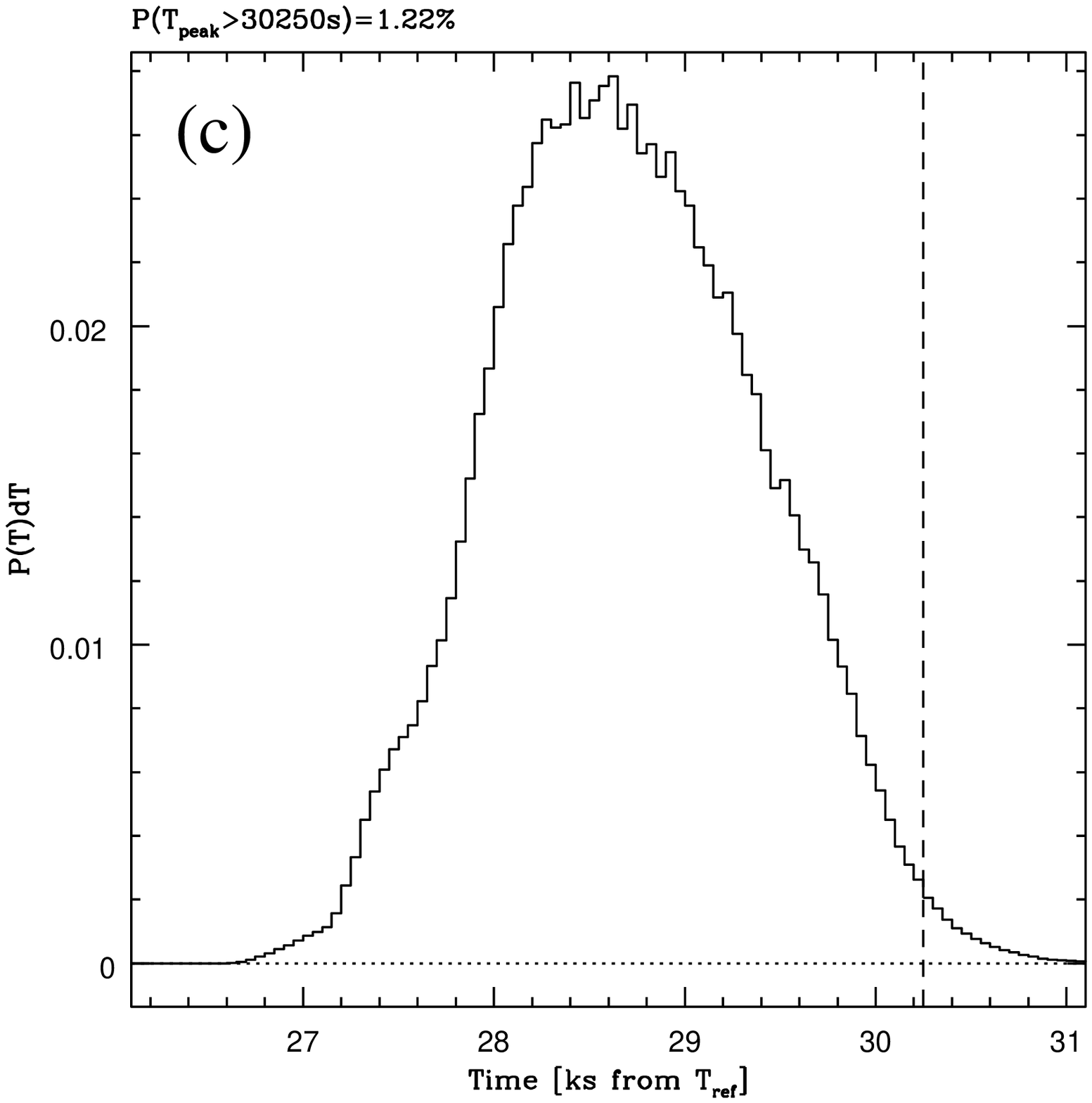}
\hfill
}
\caption{%
\small
a) March 19 light curves for Whipple \gray (connected white circles, 
$\simeq$ 1000 or 1680\,s bins), and two \xray bands (dark squares for
2--4\,keV, gray triangles for 9--15~keV, both in 256\,s bins). 
The rate scale for the \xray data is on the left Y-axis, and the flux scale
for the Whipple data on the right Y-axis.
To allow for an easier comparison of the relative variability amplitude,
the Y--axis range for the \gray light curve ($\times$16) is the square of
that used to plot the \xray data ($\times$4).
b) Same as a) panel, except that here the TeV (Whipple)
light curve data have been adaptively rebinned from the 256\,s data
to a signal-to-noise ratio of at least 6. 
c) Probability distribution of soft \xray flare peak time derived
from the general statistical properties of the short term variability (see
text, \S\ref{sec:lags}).  
The vertical dashed line marks the leftmost boundary of the
Whipple time interval comprising the TeV flare peak.
\label{fig:march19}
\label{fig:histo_t_peak}
}
\end{figure*}

For what concerns the full week dataset (Fig.~\ref{fig:ccf}a,b), we
\textit{do not} find a measurable lag, with either \xray band.
The statistical properties of the DCF are complicated by the presence of
several regular patterns in the data trains (chiefly the diurnal gaps in
the ground based TeV data, and the orbital gaps of the \rxte data).
The DCFs peak at zero-lag, and the correlation coefficients are quite
high, $\simeq0.8$, despite the poor sampling (the combination of the large
diurnal gaps due to the ground based visibility, and the \rxte orbital gaps
yields an efficiency of about $1/2\times3/4 \simeq 40\%$).
By means of simple data-based simulations comprising flux randomization
(\eg \citealp{peterson_etal_1998_lags}), and the effect of introducing a
shift in one of the timeseries, we estimate an upper limit of
$|\tau| \lesssim 3$~ks on the value of the soft-\xraynosp/\gray lag,
possibly smaller for the harder-\xraynosp/\gray case.

\subsubsection{The March 19 flare}
\label{sec:dcf_march19}

Beside the full dataset, we focused our attention on the isolated outburst
of March 18/19 that uniquely comprises many favorable observational
characteristics, namely 
i) the best TeV coverage, 
ii) the least \rxte data gaps, and 
iii) the best \rxtenosp/Whipple overlap, 
iv) the largest brightness excursion (in both bands, $\times10$ in \graysnosp, 
$\times3$ in \xraysnosp).
Because of this indeed rare combination of properties, the March 18/19 flare's DCF
is probably not significantly affected by the sampling.
Because of its relative isolation from other outburst, and its large amplitude,
we can regard this event as a rather clean ``experiment'',
providing us a good view of the variability mechanism at work.
On the other hand, results obtained for this particular event may not
necessarily be representative of all flares.

In this case we used the Whipple data in their shortest available time binning,
256\,s bins (an example of this is the peak region of the March 19 burst
shown in Figure\,\ref{fig:march19}c).
The DCFs for the two different \xray bands, computed over 1024\,s time steps,
are plotted in Figures~\ref{fig:dcf}c and ~\ref{fig:dcf}d.
The DCFs for both \xray bands show a high correlation coefficient, peaking
at 0.84 (at a lag $\tau \approx +2$~ks) for the soft \xraysnosp, and at 0.88 (at
zero lag) for the harder \xraysnosp.

The most remarkable feature is that there seems to be a hint for the \grays 
lagging the softer \xraysnosp, while being ``synchronized'' with the harder \xray
photons.
A thorough analysis and characterization of the properties of the \xray
variability is discussed in F08.
In Figure~\ref{fig:march19}a,c, we just show a 20\,ks section of the
three light curves for March 19, centered around the possible peak position.
Although the peak of the outburst was not directly observed in \xraysnosp, we
can make the following heuristic arguments concerning the possibility,
and value, of a interband lag between the softer \xrays and the TeV data.

\subsubsection{Constraining the lag for the March 19 flare}
\label{sec:constraint_on_lag}

By exploiting the statistical knowledge of the
characteristics of the \xray variability (see F08), during this campaign, 
we can try to assess the probability that a lag at the flare peak in fact
exists between the TeV and the soft \xray light curves.
The idea is to assign a probability distribution for the soft \xray peak to
occur at times $t_\mathrm{peak}$ during the data gap, and then use it to
evaluate the probability that the peak in fact occurred within the time
interval comprising the peak of the TeV outburst, that is at $T - T_{ref}
\ge 30.25$\,ks.

The basic building blocks for these probability estimates $\{{\cal
P}(t_\mathrm{peak})\}$ are the observed distributions of doubling and
halving times, ${\cal P}(\tau_2)$ and ${\cal P}(\tau_{1/2})$, derived from
the entire week long soft \xray dataset. 
With these we assign the probability for the peak of the \xray light
curve to occur at a certain time and brightness
$(t_\mathrm{peak},F_\mathrm{peak})$ within the data gap, by taking the joint
probability of having the $\tau_{2}$ and $\tau_{1/2}$ required to reach
each trial $(t_\mathrm{peak},F_\mathrm{peak})$ position ``moving''
from the left (\ie before, $t_\mathrm{bp},F_\mathrm{bp}$) and right
boundaries of the data gap (\ie after, $t_\mathrm{ap},F_\mathrm{ap}$).
\begin{equation}
{\cal P}(t_\mathrm{peak},F_\mathrm{peak}) \sim 
{\cal P}(\tau_2(t_\mathrm{bp},F_\mathrm{bp};t_\mathrm{peak},F_\mathrm{peak})) \cdot 
{\cal P}(\tau_{1/2}(t_\mathrm{ap},F_\mathrm{ap};t_\mathrm{peak},F_\mathrm{peak})) 
\end{equation}
Since we are not interested on $F_\mathrm{peak}$ this distribution is then
summed over all $F_\mathrm{peak}$ to yield just ${\cal P}(t_\mathrm{peak})$.

The ${\cal P}(\tau_2)$ and ${\cal P}(\tau_{1/2})$ adopted in this analysis
were derived from the doubling and halving times from all data
pairs whose separation in time $\triangle T_{ij}$ is between 0.25 and 5\,ks. 
We restricted our sampling to this subset of data pairs because we wanted
the distribution to be representative of the same type of variations that
could have occurred during the data gap, which spans $\simeq$3\,ks. 
The inclusion of larger pair separations would ``spuriously'' bias the
probability towards large values of $\tau_2$ and $\tau_{1/2}$.  
On the other hand, relaxing the limit on the minimum time separation picks up very
fast variations, which are not relevant for this analysis, because their
influence on where the peak could fall is marginal (given their limited
amplitude), and their overall contribution is already taken into account
(smoothed out) by the ``slopes'' measured on longer timescales.

We have performed this analysis with different choices of i) the allowed
range of $\triangle T_{ij}$, and of ii) the ``starting'' points on both sides of
the gap (namely we checked points up to $\pm$2~ks from the gap).
The results do not change significantly.

The probability distribution for $t_\mathrm{peak}$ resulting from this
analysis is shown in Figure~\ref{fig:march19}c.
The average of the all the different tests yields a probability
of ${\cal P}(t_\mathrm{peak}>30.25)\simeq1-2$\% for the flare peak to occur
later then $T-T_\mathrm{ref}=30.25$\,ks, \ie within the Whipple peak time
interval.  
The most probable $t_\mathrm{peak}$ estimated by this method is
$t_\mathrm{peak} = 28.6 \pm 0.8$~ks (1$\sigma$), two sigma below the ``first
possible time'' for the peak of the TeV flare.

We can push this type of analysis a little further to estimate the most likely
value for the lag between soft \xrays and TeV. 
In order to do this we need to assign a probability for the time of the TeV
peak ${\cal P}(t_\mathrm{peak,TeV})$.
We tried the following simple distributions for ${\cal P}(t_\mathrm{peak,TeV})$:
i) a uniform distribution within the 1680~s integration window, 
ii) a ``tent'' function centered on the top interval and going to zero at its
boundaries, 
iii) a ``tent'' function centered on the top interval, but extending over half
of each of the two adjacent intervals (\ie, $T-T_{ref}\simeq29.7-31.7$~ks).
The convolution of the ${\cal P}(t_\mathrm{peak,X})$ with ${\cal
P}(t_\mathrm{peak,TeV})$ shifted by $\tau$ gives the probability for a
given lag $\tau$.  The result does not change significantly with the
different choices i)--iii), and it is $\tau = 2.06^{+0.69}_{-0.79}$~ks (1$\sigma$).

\begin{figure*}[t]
\centerline{%
\hfill
\includegraphics[width=0.35\linewidth]{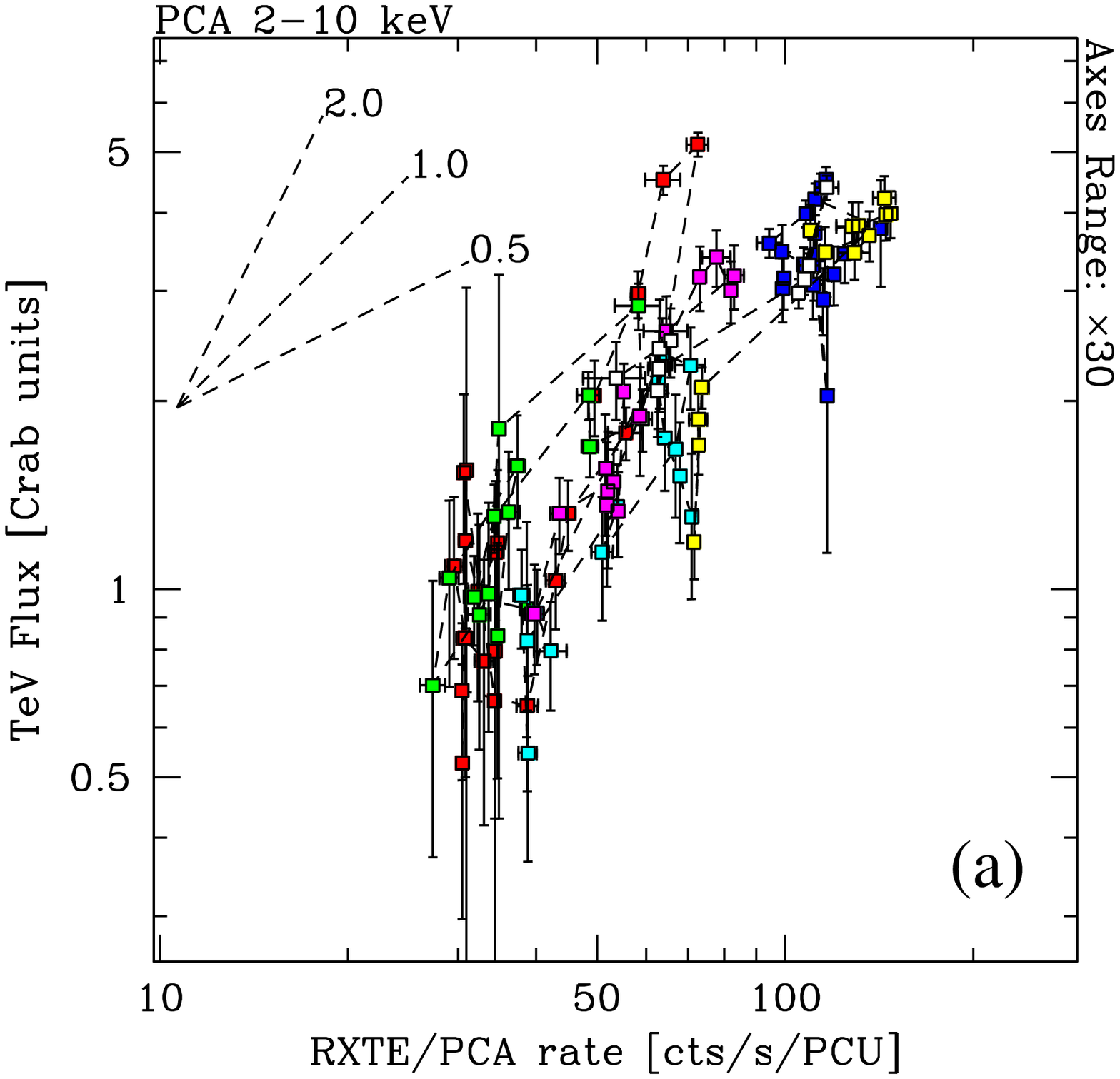}
\hfill
\includegraphics[width=0.35\linewidth]{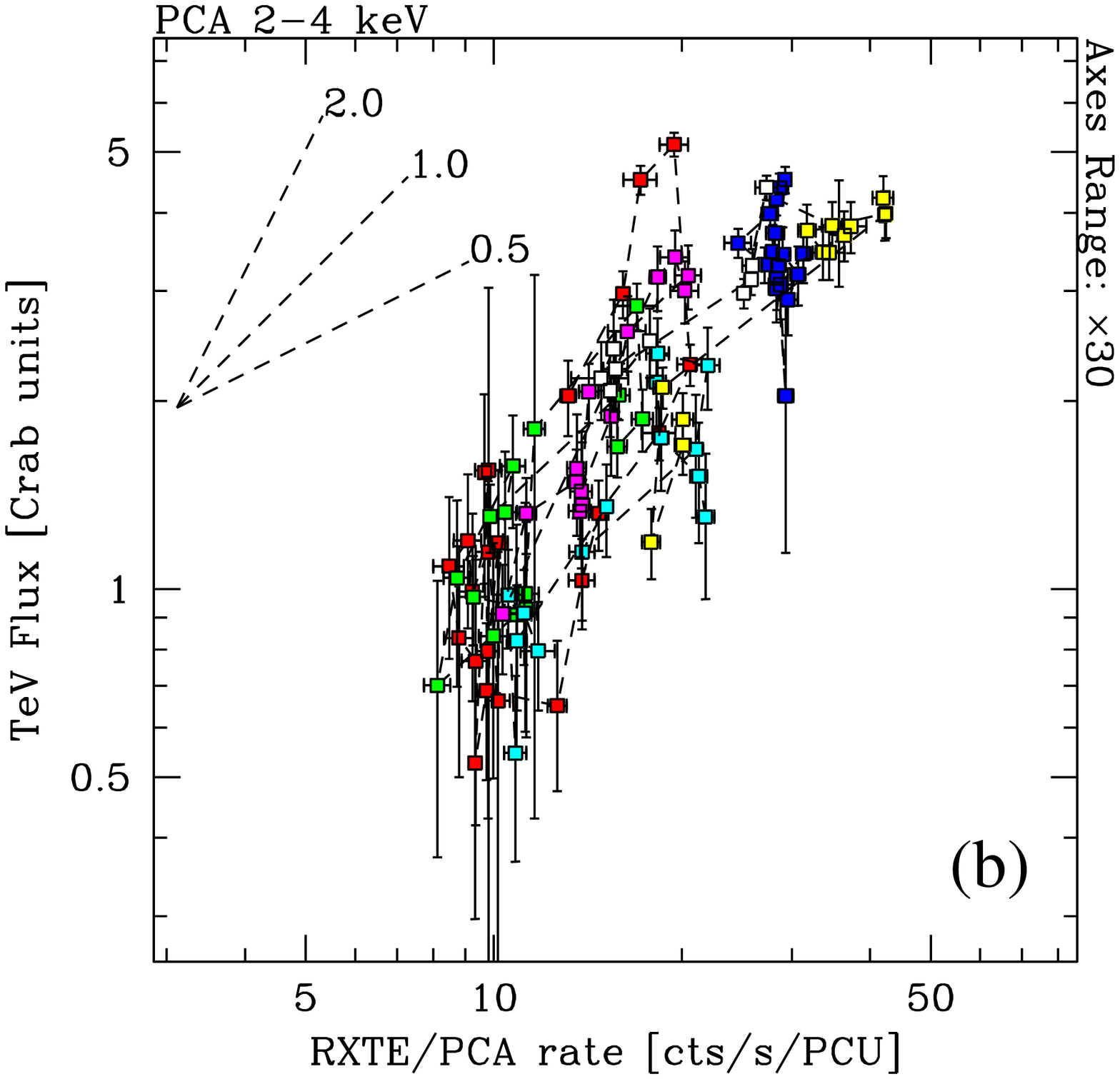}
\hfill
\includegraphics[width=0.35\linewidth]{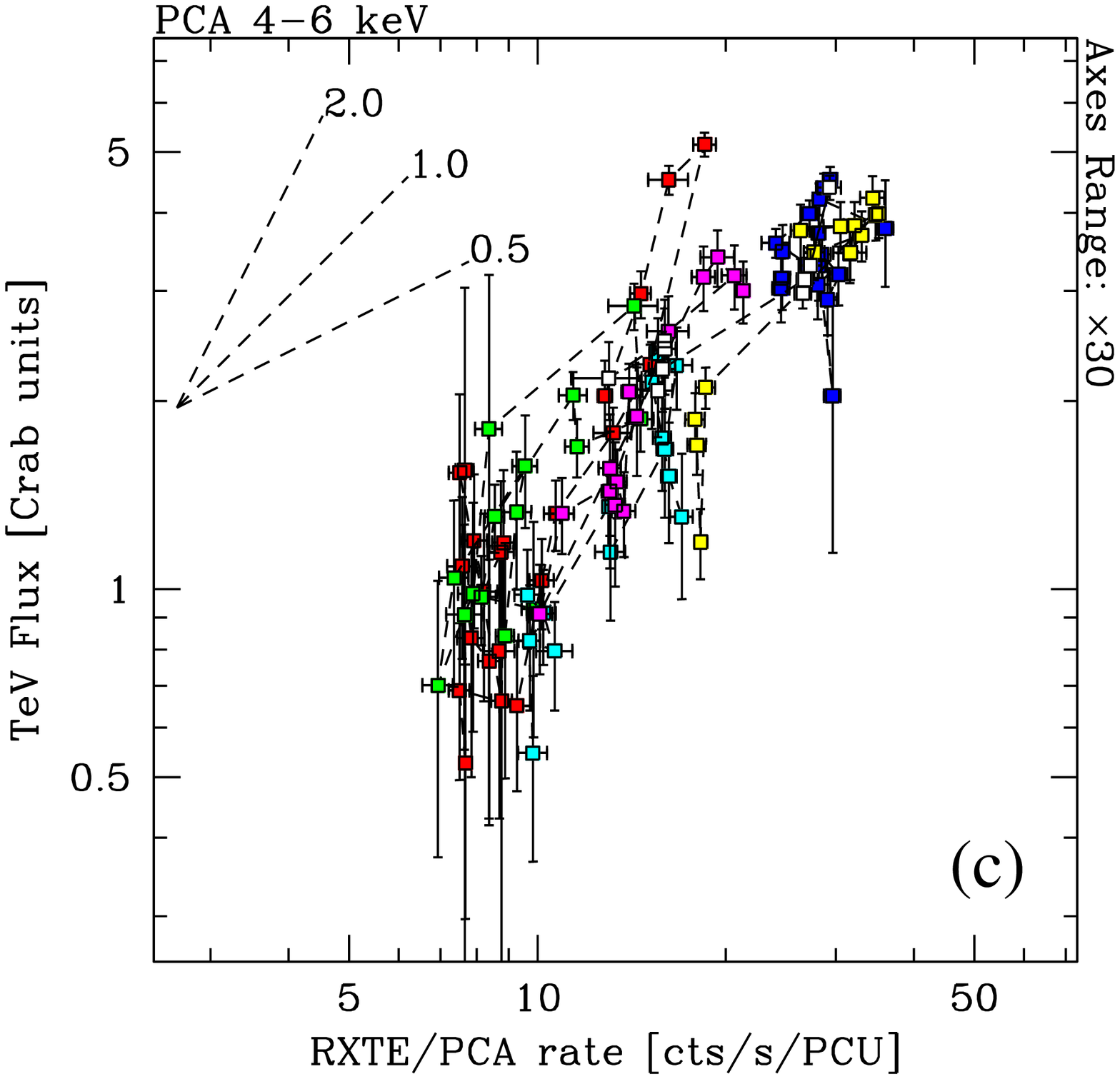}
\hfill
}
\centerline{%
\hfill
\includegraphics[width=0.35\linewidth]{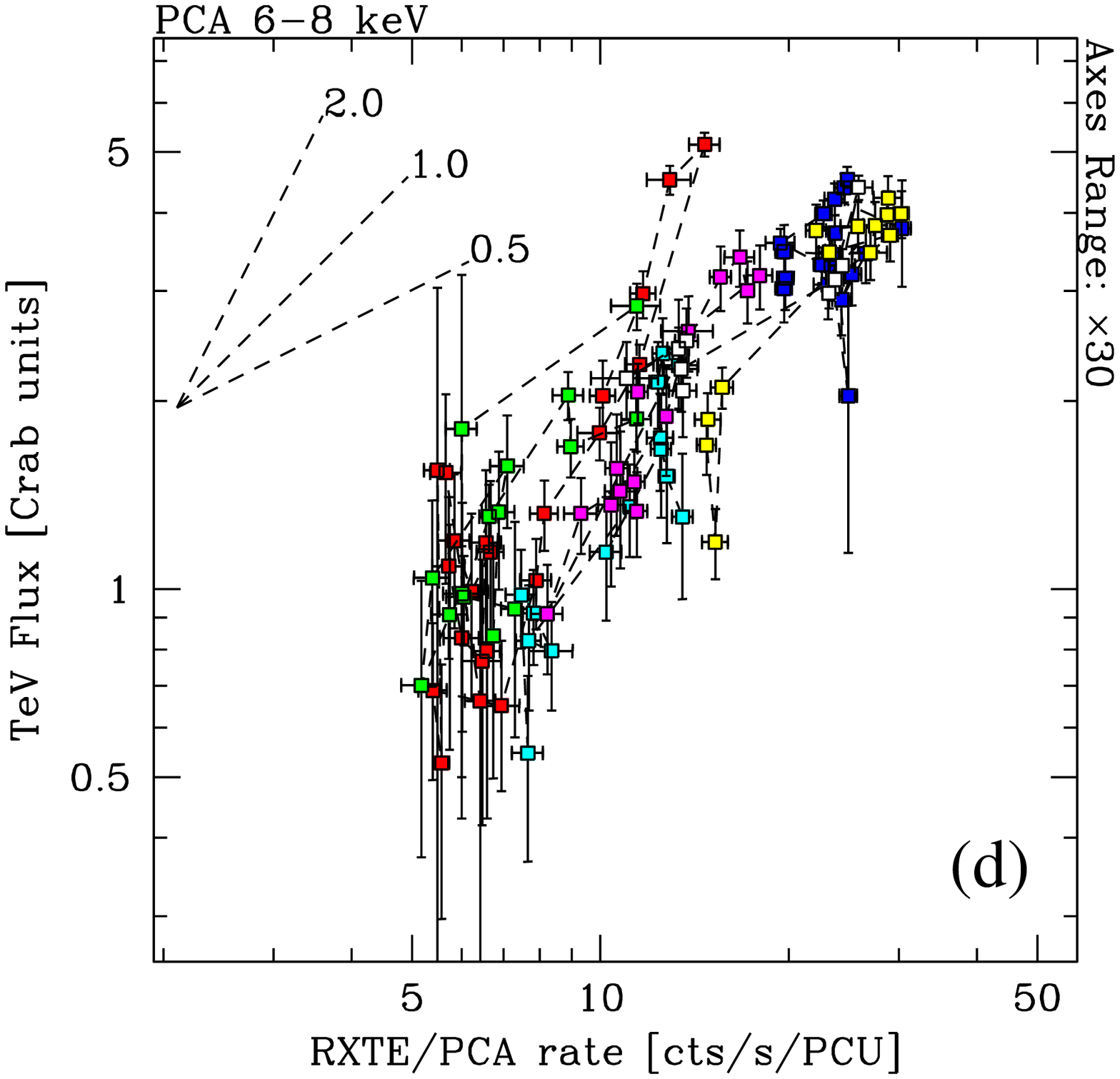}
\hfill
\includegraphics[width=0.35\linewidth]{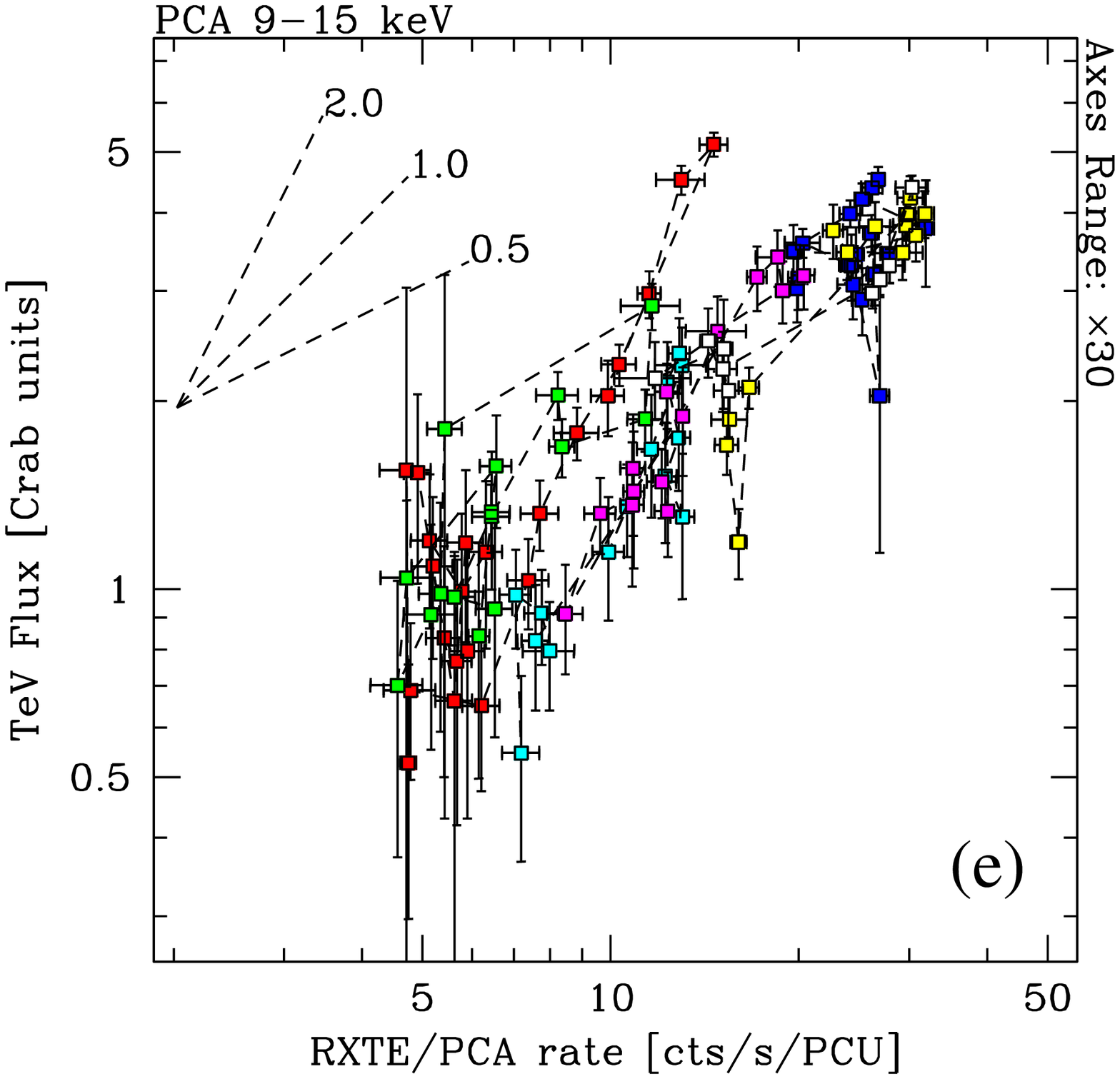}
\hfill
\includegraphics[width=0.35\linewidth]{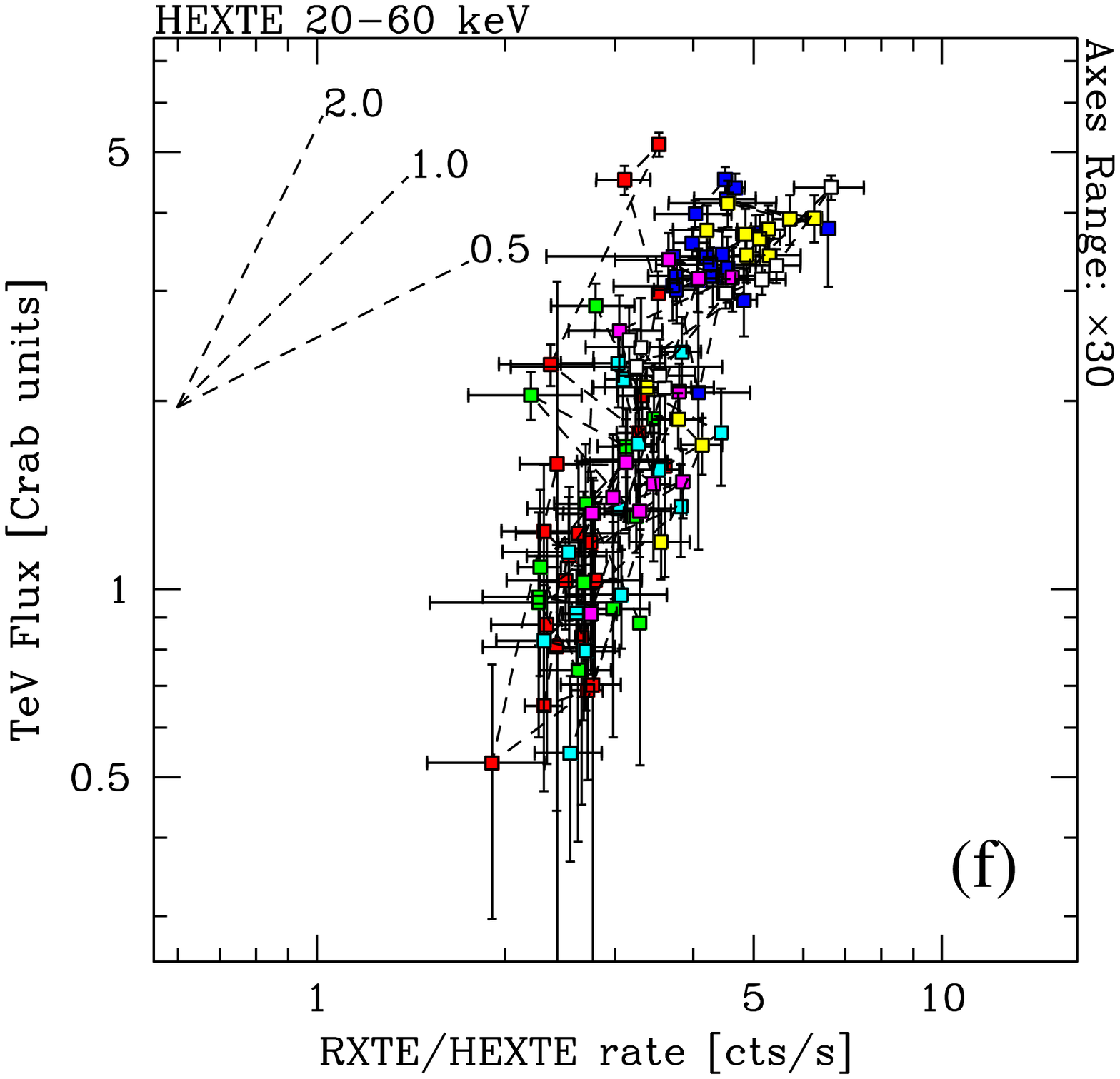}
\hfill
}
\caption{%
\small
Plot of the Whipple$+$HEGRA TeV flux \vs the \xray count rate
in different energy bands (see labels), for the entire week.
Different gray shades correspond to different nights.
For reference, in each panel are shown segments indicating 
different slopes for the relationship between the plotted fluxes.
\label{fig:ff_all_week}
\label{fig:ff_energy_bands}
}
\end{figure*}

It is important to stress that this analysis rests on a few assumptions,
that we deem reasonable, that are here summarized.
\begin{itemize}

\item[$\bullet$] 
The statistical properties of the \xray variability change on a
timescale longer than our experiment.  In this respect we checked that the
distribution of $\tau_{2}$ and $\tau_{1/2}$ for different subsets of the
week-long dataset are consistent with each other.

\item[$\bullet$] 
The power spectrum of the variations in \xrays and TeV is such that
the there is only a negligible probability that the peak of the \xray light
curve occurred before or after of the data gap, and that of the \gray light
curve during one of the earlier or later integration windows.
For instance we rule out that the TeV flare peak could have been reached by
means of a very fast and very large amplitude variation (a spike not resolved,
and smoothed out, by the coarse Whipple binning), during one of the two Whipple
bins falling during the gap in the \rxte data.  
Moreover, higher sampling Whipple light curves (see Figure\,\ref{fig:march19}b) 
provide a further constraint on the probability, and characteristics, of
this type of extreme event.
For what concerns the \xraysnosp, we have the possibility of investigating in
more detail the properties of the variability on fast(er) timescales (see
F08). 
A broad assessment of the reliability of our assumption can be made
by considering the likelihood of a large amplitude variation on a timescale
shorter than \eg 250~s, the cut-off we applied to our sampling of
$\tau_{2}$ and $\tau_{1/2}$.
The analysis of the fractional rate variability for $\triangle t$ between
$32-250$~s shows that the probability for a $\triangle F/F \ge 20$\% is
only $\sim6$\%.

\item[$\bullet$]
We would also like to point out that it would be desirable to use not
simply the probability distribution for the $\tau$'s, but the probability
for a given change in rate $\triangle F/F$ for each given $\tau$.  
However, despite the size of the \rxte dataset, it is not possible to have
a good enough sampling for ${\cal P}(\frac{\triangle F}{F},\tau)$, to
constitute a significant improvement over the uncertainty inherent in the
assumption that all $\triangle F/F$ are equally probable for a given $\tau$.
\end{itemize}

The same analysis performed for the 9--15~keV light curve yields a
$t_\mathrm{peak} = 30.0\pm0.7$~ks (1$\sigma$), a ${\cal
P}(t_\mathrm{peak}>30.25\,\mathrm{ks}) \simeq 39$\%, and an estimate of the
lag of the TeV peak of $\tau = 0.73\pm0.80$~ks, \ie no measurable lag.

\begin{figure*}[t]
\centerline{%
\includegraphics[width=0.33\linewidth]{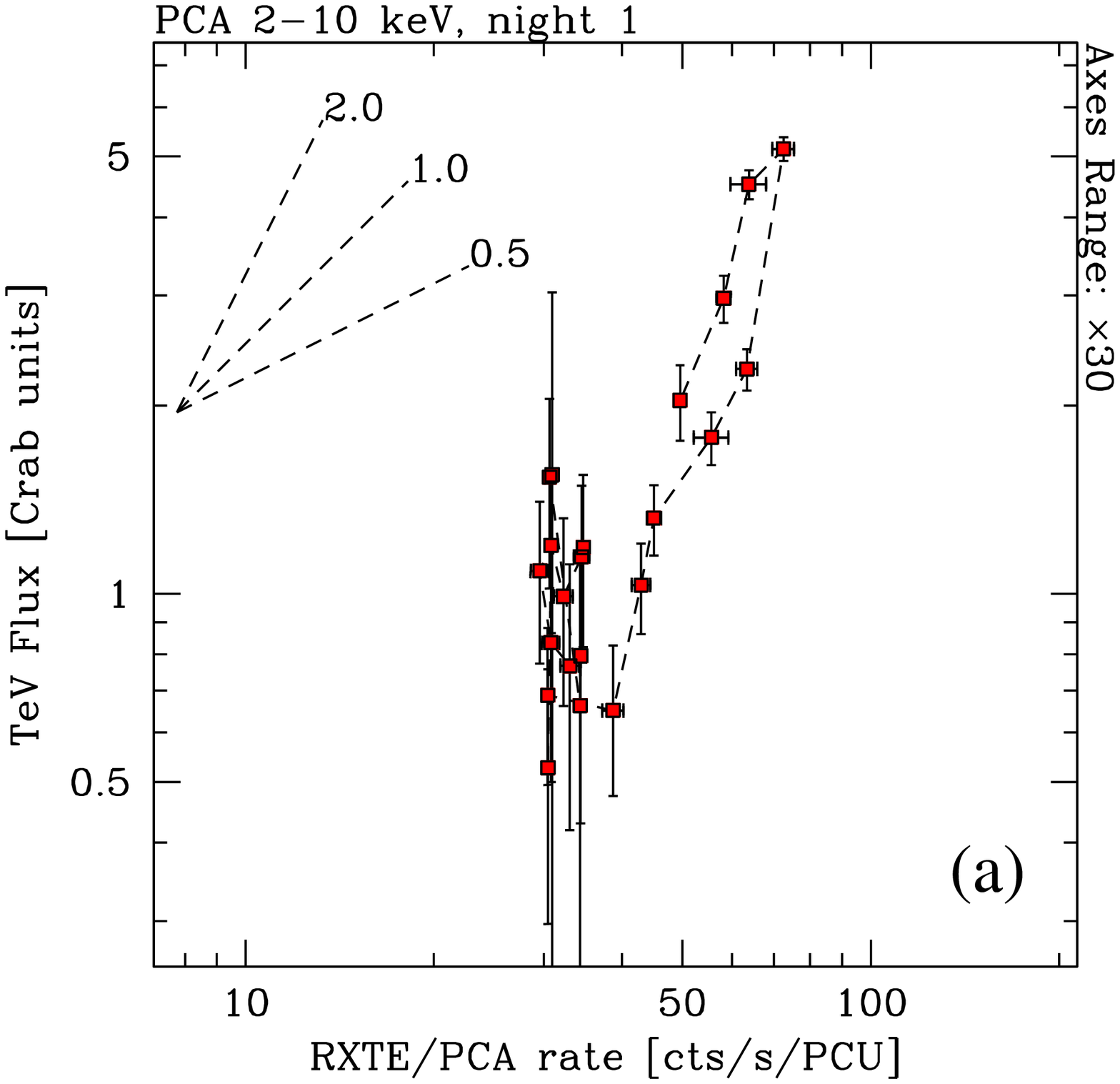}
\hfill
\includegraphics[width=0.33\linewidth]{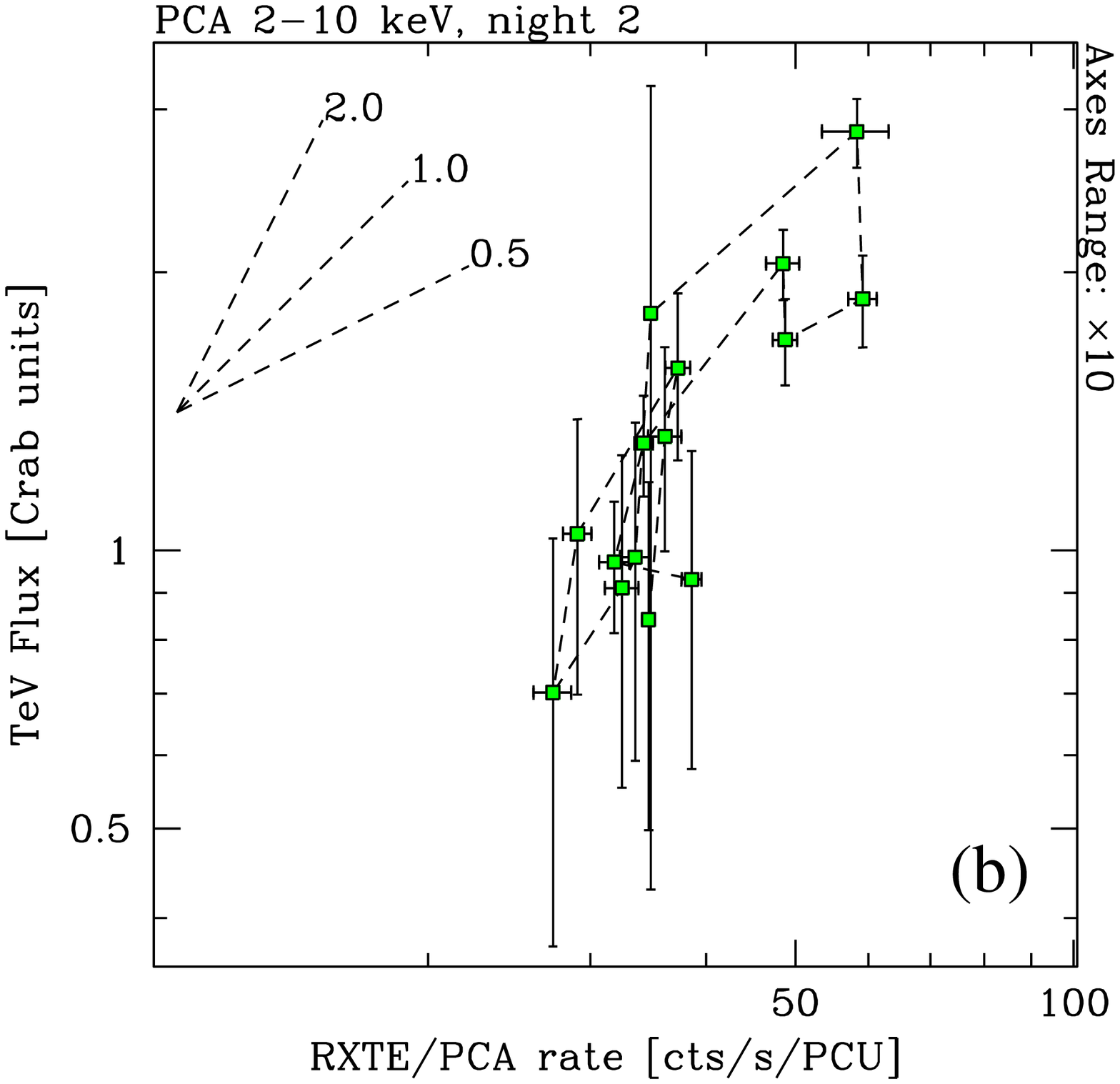}
\hfill
\includegraphics[width=0.33\linewidth]{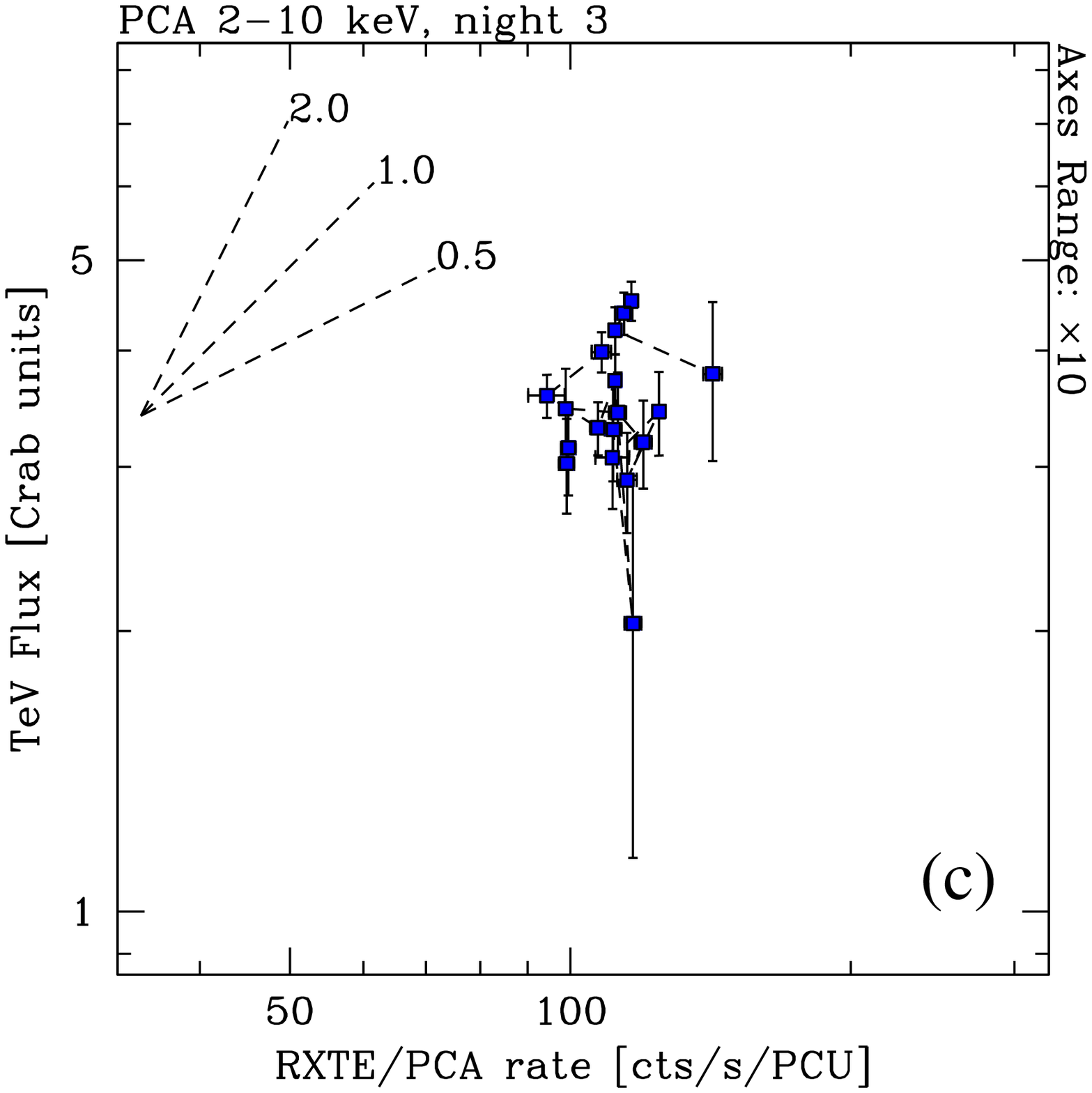}
}
\centerline{%
\includegraphics[width=0.33\linewidth]{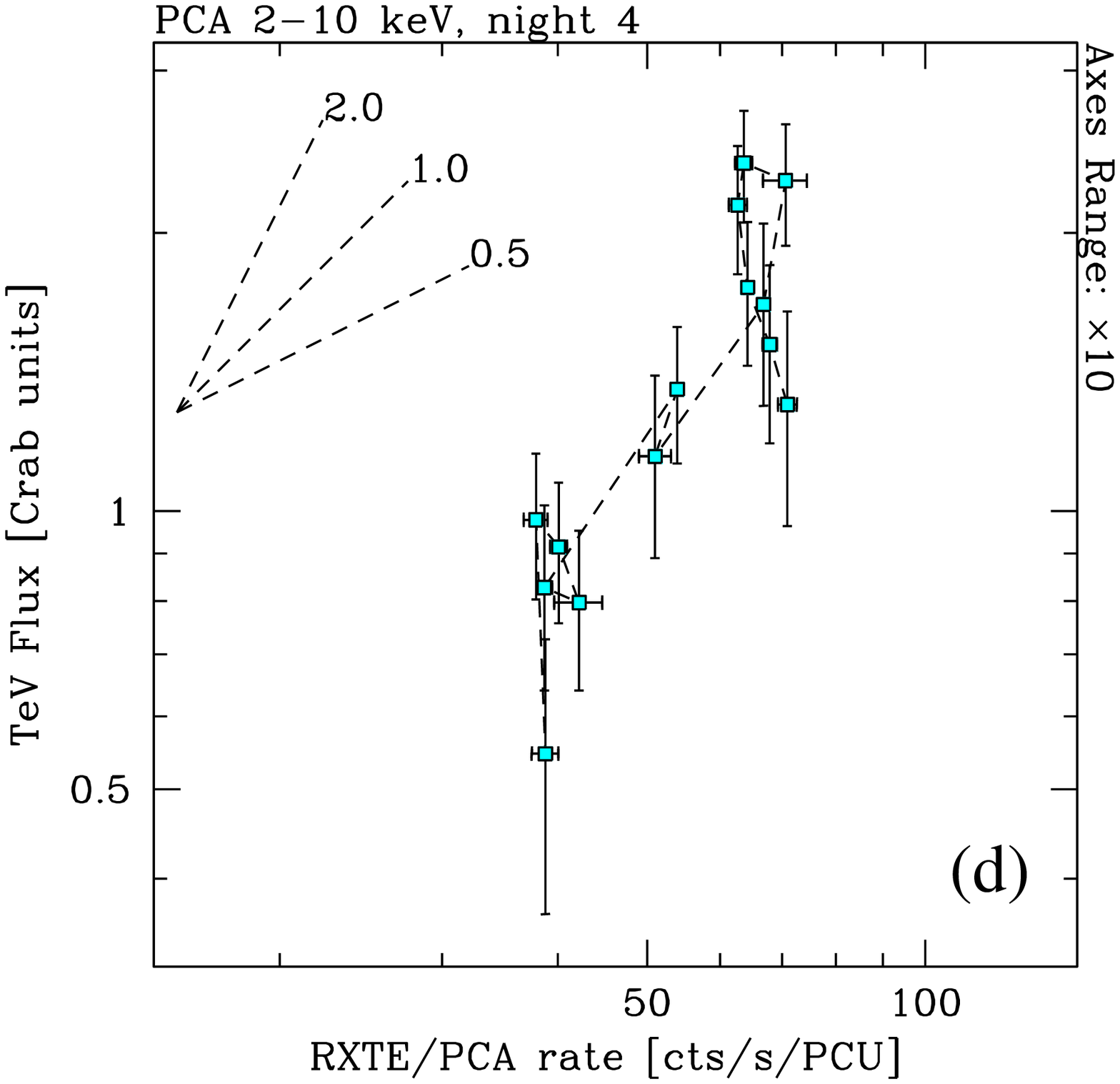}
\hfill
\includegraphics[width=0.33\linewidth]{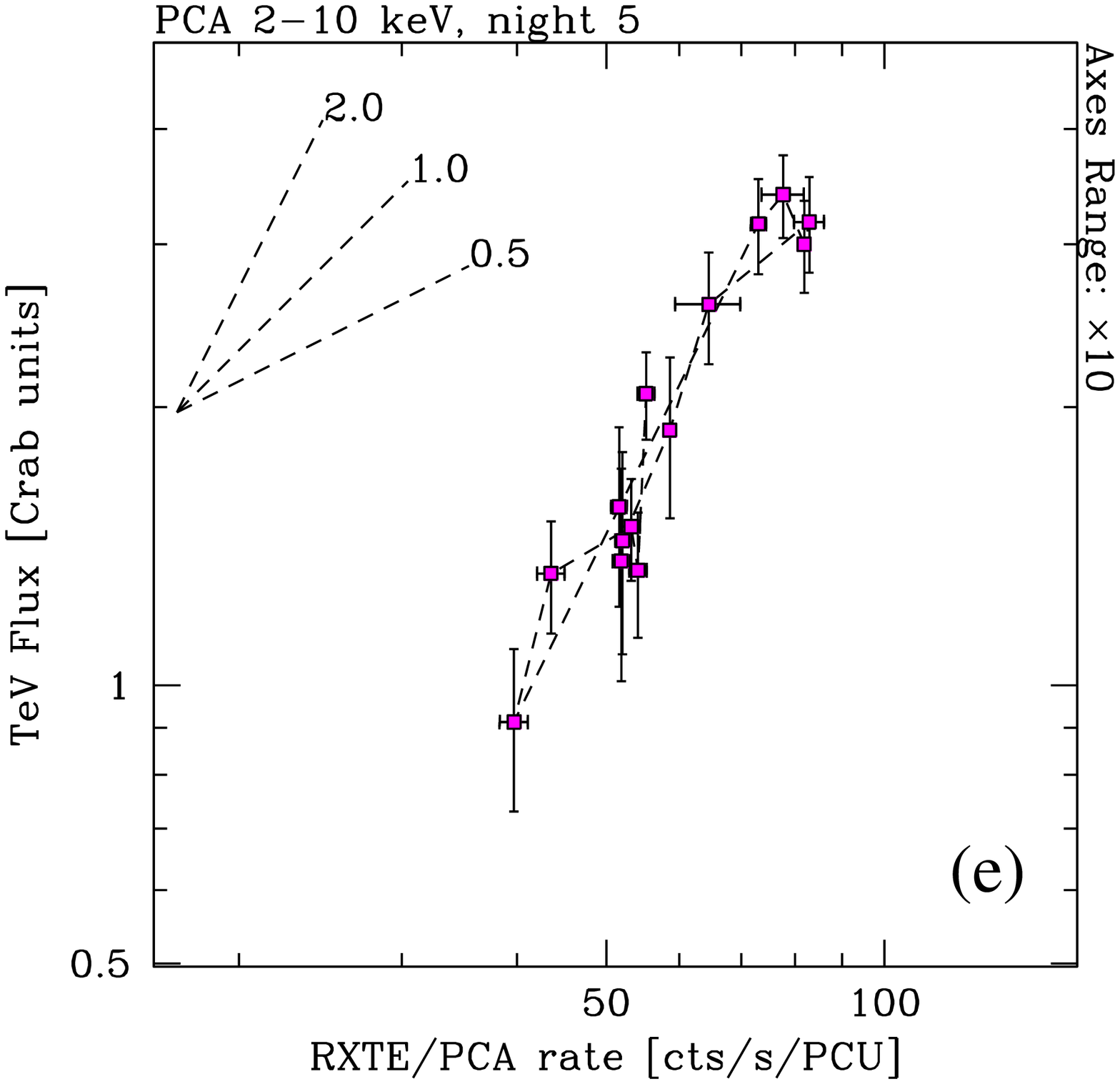}
\hfill
\includegraphics[width=0.33\linewidth]{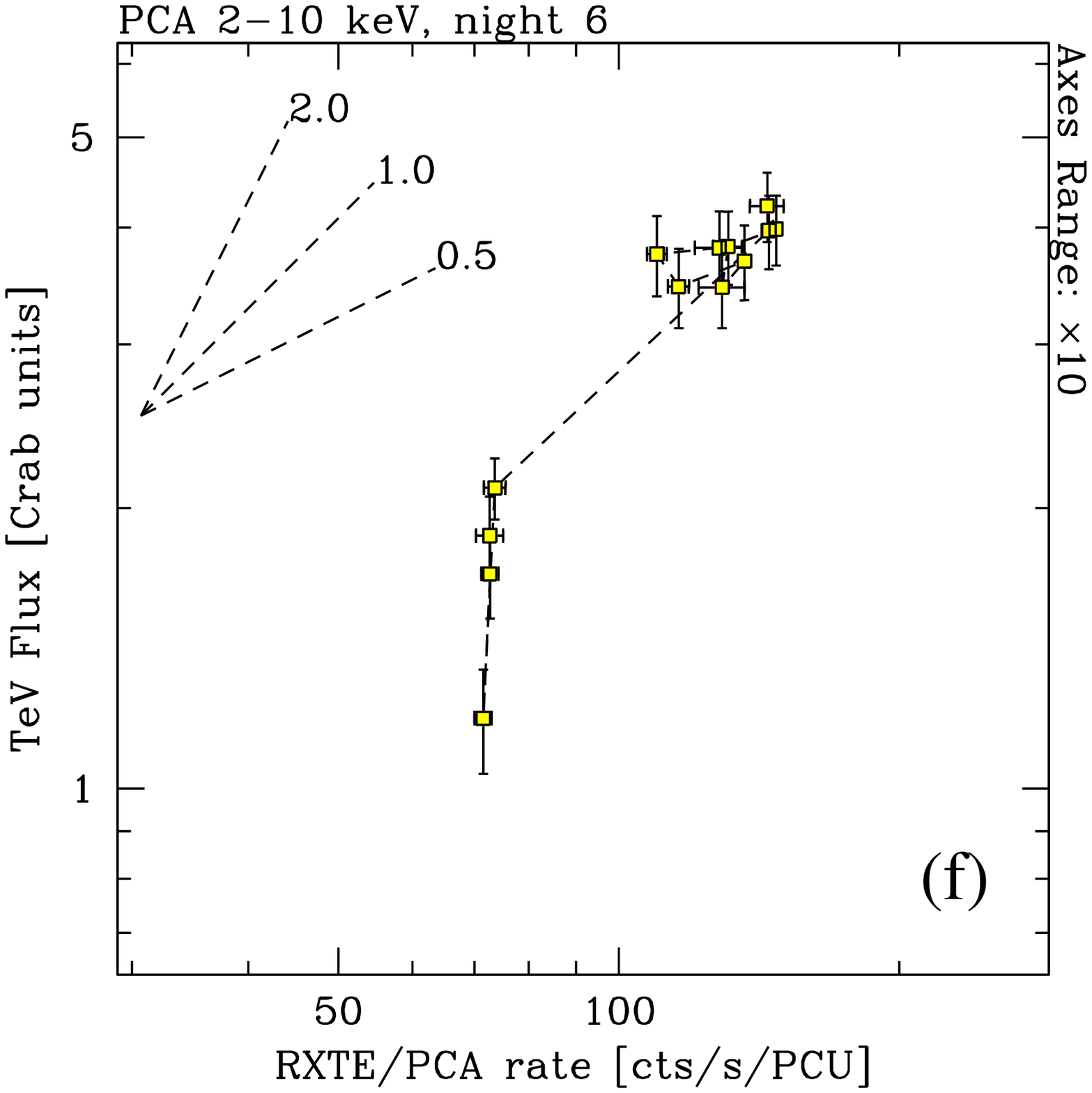}
}
\centerline{%
\includegraphics[width=0.33\linewidth]{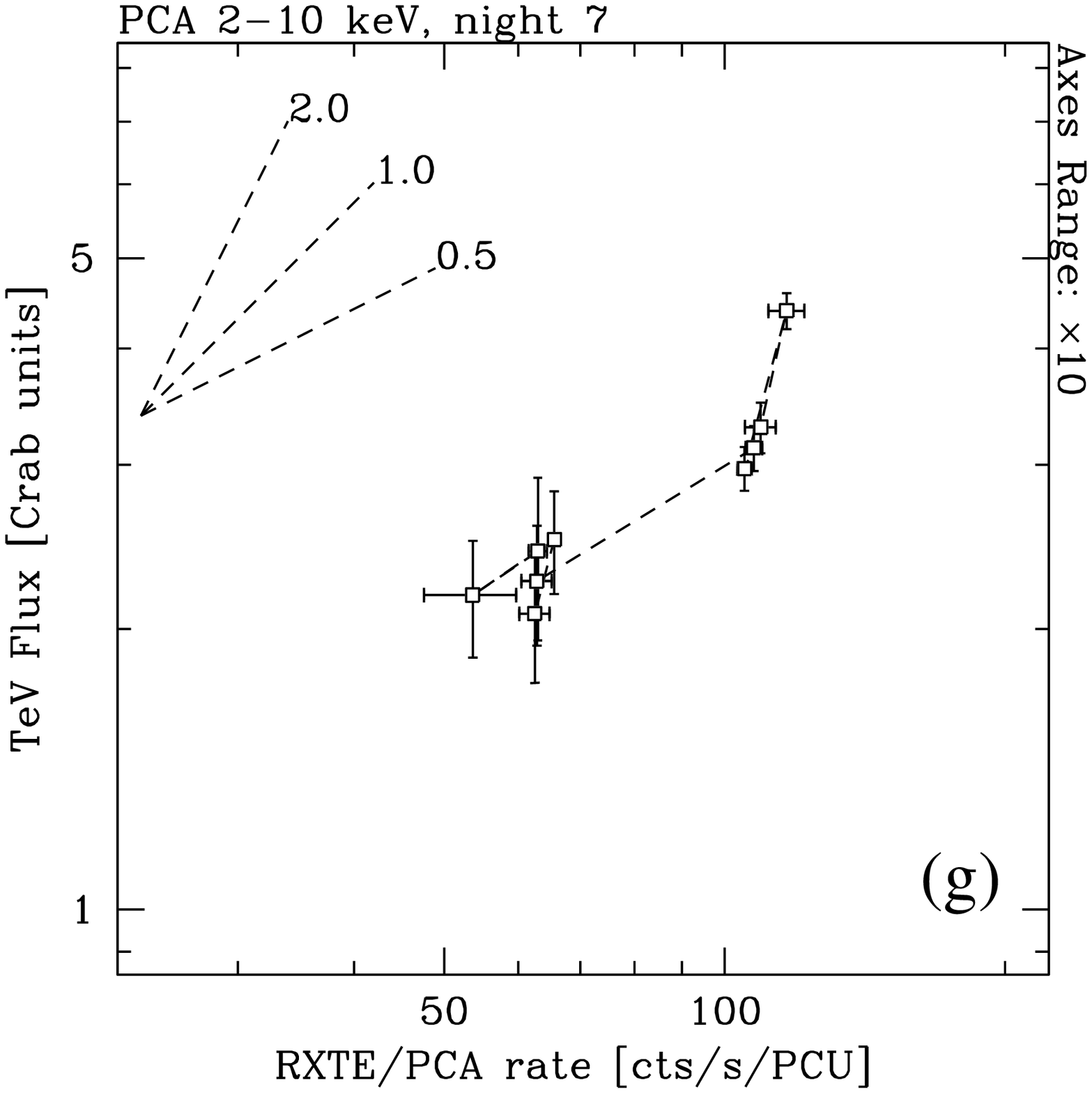}
\hfill
\includegraphics[width=0.33\linewidth]{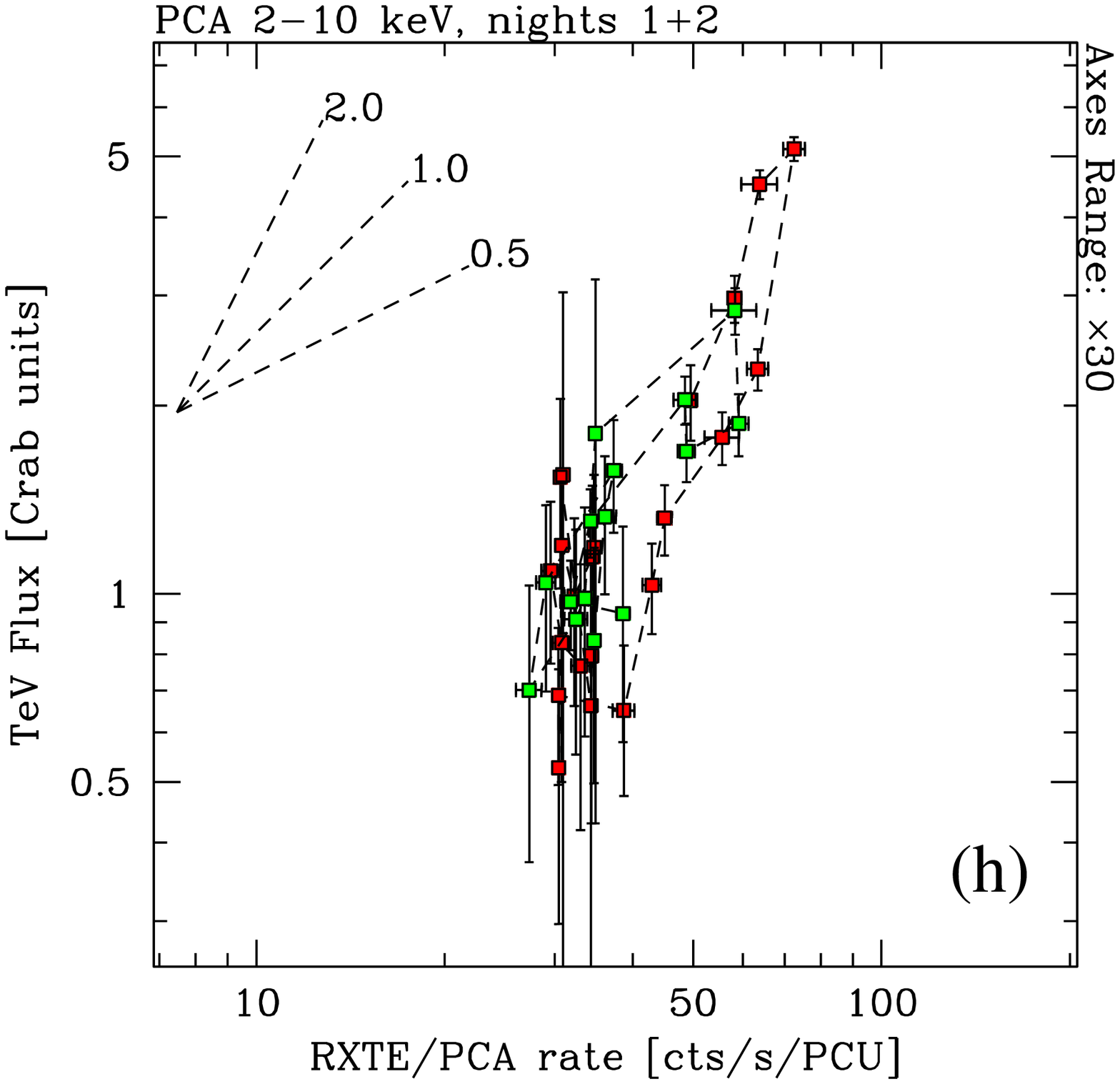}
\hfill
\includegraphics[width=0.33\linewidth]{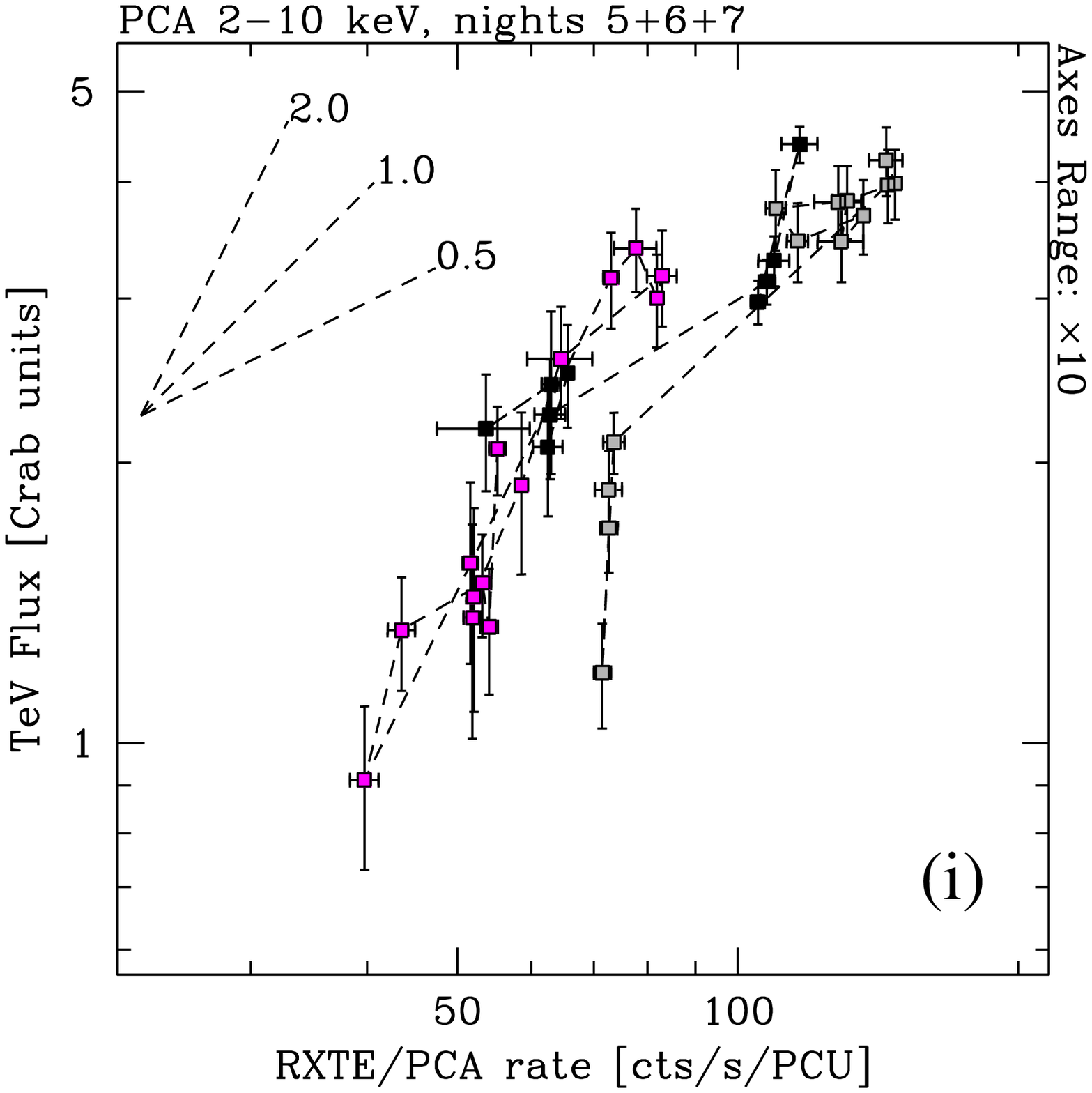}
}
\caption{%
\small
Plot of the Whipple$+$HEGRA TeV flux \vs \xray 2--10\,keV count rate for each
individual observation night, and the combination of nights $1+2$ and $5+6+7$.
Axes range is $\times 10$ in all panels except for those involving the
March 19 (night $1$) data, whose variation range is larger ($\times 30$).
Some of the symbols are shaded in gray to help recognize different sets.
\label{fig:ff_individual_nights}
}
\end{figure*}

\begin{figure*}[t]
\centerline{%
\hfill%
\includegraphics[width=0.49\linewidth]{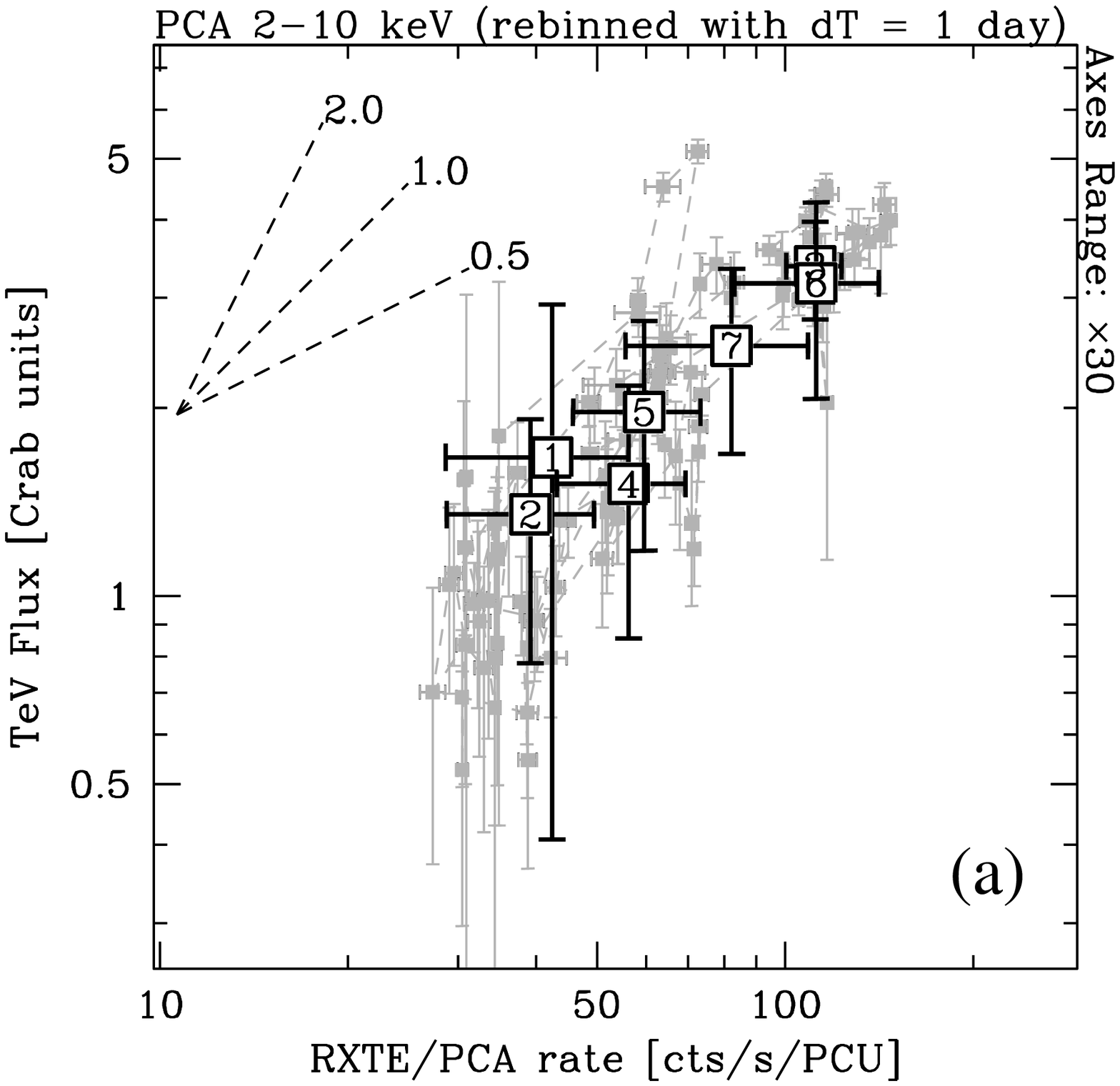}%
\hfill%
\includegraphics[width=0.49\linewidth]{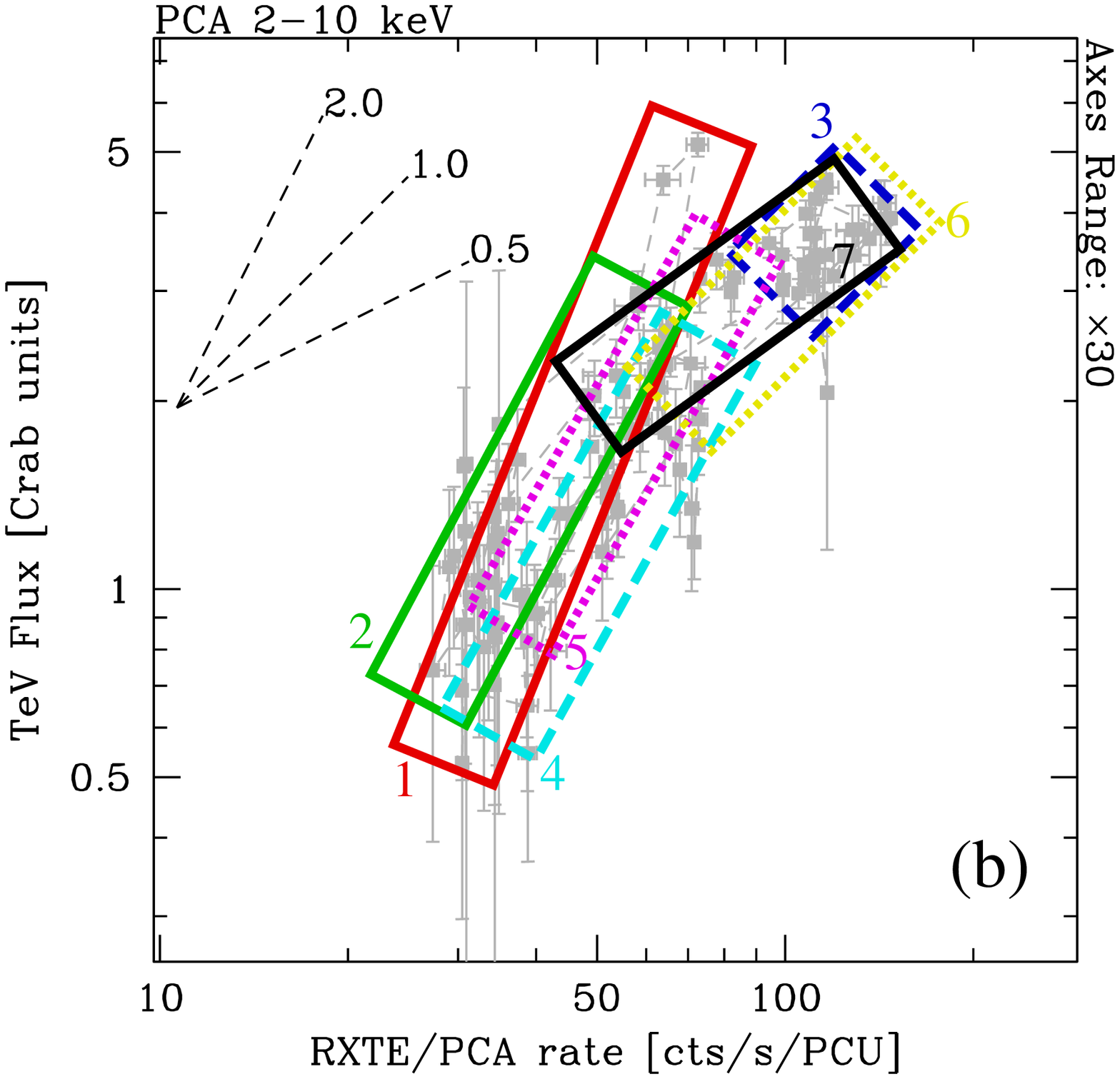}%
\hfill%
}
\caption{%
\small
a): \xray \vs \gray one-day averaged brightnesses.  
The error bars represent the variance during the interval, which can be considerable.
The correlation is approximately linear (see Table\,\ref{tab:slopes}).
Numbers refer to the campaign night sequence.
b): boxes (approximately) representing the regions of the diagram
occupied by the data of each individual night. 
The combination of steep(er) intranight and flat(ter) longer term, due to
shift of the ``barycenters'', correlations is more easily shown.  Both
plots are on the same axes scale and range.
Box contours are solid for nights 1, 2 and 7; long-dashed for 3 and 4;
dotted for 5 and 6.
\label{fig:ellipses}
\label{fig:boxy_contours}
\label{fig:1day_averages}
}
\end{figure*}

\subsection{X--ray \vs TeV flux correlation}
\label{sec:ff_correl}

Comparison of the variability amplitudes (as opposed to phases) offers
different constraints.
As clearly illustrated in Figures~\ref{fig:x_and_g_all_week}
and~\ref{fig:x_and_g_individual_nights}, the source shows stronger variability
in the \grays than in the \xraysnosp: in fact, in all panels the flux scale for
the TeV data spans a range that is the square of that of the rate scale
used for the \rxtenosp/PCA data, and the light curves run in parallel.
This is confirmed in Figure~\ref{fig:ff_all_week} that shows \gray flux
as a function of \xray count rate in different \xray energy bands.  The TeV
data are binned on approximately 28-minute runs.
The \rxte count rates correspond to the average over intervals overlapping
with the TeV observations (as indicated by the shaded boxes in
Figure\,\ref{fig:x_and_g_individual_nights}), and the error bars represent
their variance (height of the shaded boxes).

We fit the log--log data with a linear relationship (\ie $F_\gamma \propto
F_X^\eta$), which provides a satisfactory description in all cases. 
The best fit slopes are reported in Table~\ref{tab:slopes} (top row), along
with their errors. 
We also analyzed the \xraynosp/TeV correlation for different sections of the
campaign, and individually for each night, with the intent of looking for
possible variations.
Individual nights plots are shown in Figure~\ref{fig:ff_individual_nights}.
In Table~\ref{tab:slopes} we report the best fit correlation slopes for the
best single nights, and for a few combinations of consecutive nights (\#1$+$2,
\#4$+$5, \#6$+$7, \#5$+$6$+$7).

It is worth noting that again the March 18/19 (day 1) \xraynosp/\gray 
observations provide the best case study, for the large amplitude of
variability, likely ensuring us that the observed amplitude is close
to the intrinsic one.
The presence, and contribution, of a steady (variable on longer timescale)
emission diluting the flaring one could alter the perceived amplitude of
flares.
This is a long-standing issue that is difficult to address, but in this
respect the March 18/19 flare is a unique event.

The unprecedented quality of this dataset enables us not only to establish
the existence of the correlation between the TeV and \xray luminosities,
but also to start unveiling some of its more detailed characteristics, 
\eg its evolution with time.
The emerging picture is complex.
There are several observational findings that we would like to point out.

\begin{itemize}
\item[$\bullet$]
The first, most direct and general, observation is that the TeV flux shows
a definitive correlation with the \xray rate, for all \xray energy bands
(see Figure\,\ref{fig:ff_all_week}).  Considering the entire week-long
dataset, 105 data pairs, the correlation is approximately linear (see
Table~\ref{tab:slopes}).
The same is apparent when looking at the data binned over 1-day timescale,
Figure~\ref{fig:boxy_contours}a.

\item[$\bullet$] 
A more careful inspection of the flux--flux diagrams suggests however a 
richer phenomenology.
In fact, we may be observing a series of parallel ``flux--flux paths'',
individually obeying a steep (\eg quadratic) trend, but that taken
together produce a rather flat envelope producing the linear trend
emerging for the global cases, because of a drift of their barycenters.
There is indeed a secular increase of the source brightness over the
course of the campaign, and it seems to be more enhanced in \xraynosp.
Its amplitude is of the order of intranight brightness variations,
thus altering the \xraynosp/\gray correlation on longer timescales.
Figure\,\ref{fig:boxy_contours}b shows how the regions covered by nightly
data shift from day to day, while broadly maintaining an approximately
quadratic intranight flux correlation trend in most cases.

There is thus an intriguing hint that there might be a split between the
correlation observed on short (hours) timescales and that apparent on
longer (days) timescales, once faster variations are smoothed out.

\item[$\bullet$] 
There may be two different (luminosity related) regimes for the
\xraynosp/TeV flux correlation.
By splitting the data in two sections of significantly different average
brightness level, days 1$+$2 (with or without the pre-flare noisy HEGRA
data section), and days 6$+$7 (or 3$+$6$+$7), we note that that source seems to
exhibit two different behaviors: the TeV \vs \xray relationship is
significantly steeper for the day-1$+$2 subset, with values of $\eta$ for
all 4 PCA energy bands larger than $\eta=1.82 (\pm0.12)$, versus all values
smaller than $\eta=1.03 (\pm 0.14)$ for days 6$+$7 (Table~\ref{tab:slopes}).

\item[$\bullet$]
For two nights (1 and 5) the flux-flux diagram is very tight, with
all points lying on a very narrow path.
In these cases the TeV flux increases \textit{more than linearly} with respect
to the \xray rate.
For the flare of March 19 the correlation is ``super-quadratic'' at all
energies (Table~\ref{tab:slopes}).
Moreover, for these two nights the light curves encompass a full flaring
cycle, \ie we can follow the complete evolution of an outburst, rising and
decaying.
In both cases the paths of the rising and decaying phases in the flux--flux
diagram overlap perfectly.

\item[$\bullet$]
There is no significant change of the slope of the correlation with the
choice of \xray energy band, except for the case of the full-week dataset.
A flatter correlation slope for harder \xrays would be expected because of 
the intrinsically higher amplitude of the variability of the
synchrotron component towards higher energies (if we are already above the peak
energy) (\egnosp, \citealp{fossati00_sax_mkn421_1}).
In fact, the relative variance, $\sigma_F/\langle F\rangle$, increases with
energy, changing from $\simeq 0.45$, to $0.48$, $0.52$, $0.56$ for 2$-$4,
4$-$6, 6$-$8, 9$-$15\,keV respectively.  This change fully accounts for the
flattening of the \xraynosp/\gray correlation slope.
The effect is not observed for smaller subsets of data probably because
of the lower statistics.

The departure of the 20--60\,keV band from this trend could instead be
justified by considering that the flux in this band may comprise a
contribution from the onset of the inverse Compton, which could be regarded
as constant because it would be varying on much longer timescales.
However, this hypothesis does not seem to be supported by the data,
because, though very noisy and with limited energy leverage, the HEXTE
data are consistent with the extrapolation of the steep PCA power law.
Alternatively it is possible that the difficult background subtraction
of the low-count-rate HEXTE data reduces the intrinsic dynamic range
of the \xray variations, thus steepening the correlation.

\end{itemize}

These observational findings have important implications for the physical
conditions and processes responsible for the variability in the scattering
region, as discussed in \S\ref{sec:conclusions}.

\subsubsection{Comments}

Before we proceed to discuss the observational findings, we would like to
put forward a few additional comments concerning some aspects of the
derivation and interpretation of the flux--flux correlation.

\begin{itemize}

\item[$\bullet$]
For simplicity we performed the brightnesses correlation analysis
using count rates for the \rxte data. 
A proper conversion to flux units requires to fit a model to the data for 
each short sub-interval, and it would introduce a different source of
uncertainty.
We tested the correspondence between count rates in \rxtenosp/PCA bands and
model fluxes for a broken power model, with different spectral
indices, and break energy positions, covering the range of values observed
in March 2001 (for full account of the spectral analysis please refer to F08).
For the 2--10\,keV band, the correlation between count rate and flux is slightly
tilted, in the sense of slightly less than linear increase of the flux with
rate, $\mathrm{Flux} \sim \mathrm{Rate}^{0.9}$.
This would thus further steepen the TeV/\xray flux--flux correlation if computed
with the \xray flux.
The effect is small and it is not present when narrower energy bands are
considered.

\item[$\bullet$]
Rebinning the data alters the variance of the light curves, and if the
effect is different for \xray and \gray (namely if their intrinsic power
spectra are different), it could potentially bias the slope of the
correlation.  The comparison of the change of variance of \xray and
\gray (starting from the 256\,s-binned one when possible) light curves for
different rebinnings, suggests that the effect is at most of the order
of 10\%.  The effect is small in comparison with the overall range spanned
by the data, which is of the order of a factor of at least five for the
week long dataset.
Hence, we deem the effect of the choice of time binning on the
determination of the flux-flux correlation slope not significant.

\item[$\bullet$]
Since we are measuring the fluxes in limited energy bands, the slope of
the relation depends also on the position of the synchrotron and \gray 
peaks with respect to the observed energy bands.  The reason is that as 
the peak moves from lower frequency into the bandpass of a detector, a
small change in the peak position yields a larger variation of the flux.
A simple shift in frequency would be degenerate with a true increase in
luminosity. 
Once the spectral peak falls within the bandpass, and it is shifting within
it, this ``spurious'' effect becomes un-important.
In a very simplified case, taking Mrk\,501 as test SED,
\citet{tavecchio01_mkn501} showed that the \gray \vs \xray flux
relationship predicted for variability simply due to a change in the maximum
particle energy (and in turn synchrotron peak energy), can vary between
flatter-than-linear to steeper-than-quadratic (the effect was however
enhanced by the fact that the authors compared monochromatic fluxes).
\citet{kk_etal_2005} performed a thorough analysis of the effect
of the position of observed energy bands with respect to the synchrotron or
IC peak energies in the context of the \xray \vs \gray brightness correlation,
and found that it can change the slope over a broad range of values,
including linear and quadratic.
This apparent freedom is however lost if data following the full evolution
of a flare are available.
In fact their conclusion with respect to an outburst developing like
that of March 19 is that explaining the observed correlation by means
of specific choices of spectral bands is problematic and it would
require very contrived assumptions.

\end{itemize}

The characteristics of the \xray variability
itself seem to evolve during the campaign.  In particular, it is important
to recall that the spectrum becomes significantly harder over the course
of the week-long campaign, accompanying a gradual brightness increase.
Rather than a caveat this is probably a point in support of the apparent
change of \xraynosp/\gray behavior between the first and second part of the week.
The spectral analysis of the March 2001 Whipple data, reported separately by
\cite{krennrich_2003_icrc}, showed that the TeV spectra also 
significantly hardened between March 19 and 25.
The spectral indices for a power law fit with exponential cutoff (fixed at
4.3\,TeV) shift from $\Gamma\simeq2.3$ to $\Gamma\simeq1.8$ ($\pm0.15$), 
\ie suggesting that the IC peak moved from below to within the Whipple
bandpass (\ie in the latter case the \gray emission would peak at about 1\,TeV).
\rxte spectra present a similar picture of the \xray evolution, namely
that the synchrotron peak shifted into the PCA bandpass.  Broken power law fits
show that the lower energy spectral index becomes harder than $\Gamma\!=\!2$ (F08).
Unfortunately even with the available statistics, because of the limited
energy leverage, it is not possible to pinpoint robustly the energy
of the synchrotron peak and its evolution (as was the case with \saxnosp).

If the peak of one component (synchrotron or IC) moves into the observed
band, we would then be observing the variations of a lower, possibly below peak, 
section of the electron spectrum, instead of the more highly variable
higher energy end.
Depending on whether this happens to both peaks or just one, we expect
to observe a different phenomenology: \eg if this happens only for the
TeV band the correlation with the \xray data should become flatter (smaller
\gray variation for a given \xray one).
This might explain the apparent change of the flux-flux correlation trend
between the beginning and end of the week-long campaign.

However, it is worth noting that \cite{krennrich_2003_icrc} find that
during the ``flare'' of March 25, a flux variation larger than a factor of
2 does not seem to be accompanied by any spectral change. 
Given the characteristics of the spectra, namely the fact that the IC peak
at most moved marginally within the observed band, this achromaticity can
not be convincingly ascribed to the fact that Whipple was observing the
lower energy shoulder of the IC peak.
The possibility that it is intrinsic has to be contemplated.

Therefore the possible change of the \xraynosp/TeV flux correlation may also be
attributed to some intrinsic effect, possibly related to the longer term 
increase of luminosity.

\subsection{X--ray \vs TeV spectra and spectral energy distributions}
\label{sec:x_tev_spectra}

\subsubsection{Intranight \xraynosp/\gray spectra pairs}
\label{sec:x_tev_spectra_march19}

Besides the unprecedented quality of the \xray and \gray light curves
that we have illustrated and discussed in the preceding sections, the
March 2001 dataset affords us a unique opportunity of following the
spectral evolution itself, with a time resolution that allows
meaningful intra-flare analysis.
Detailed SED-snapshot and time dependent modeling analyses are beyond
the scope of this paper and will be presented in a forthcoming
publication.
Here we present the subset of \xraynosp/\gray spectra for the March 19 event (Whipple),
and for the March 21/22, 22/23 flares (HEGRA, presented by \citealp{aharonian02_mrk421_spectral_var}).

\begin{figure*}[t]
\centerline{
      \hfill\includegraphics[width=0.50\linewidth,clip=]{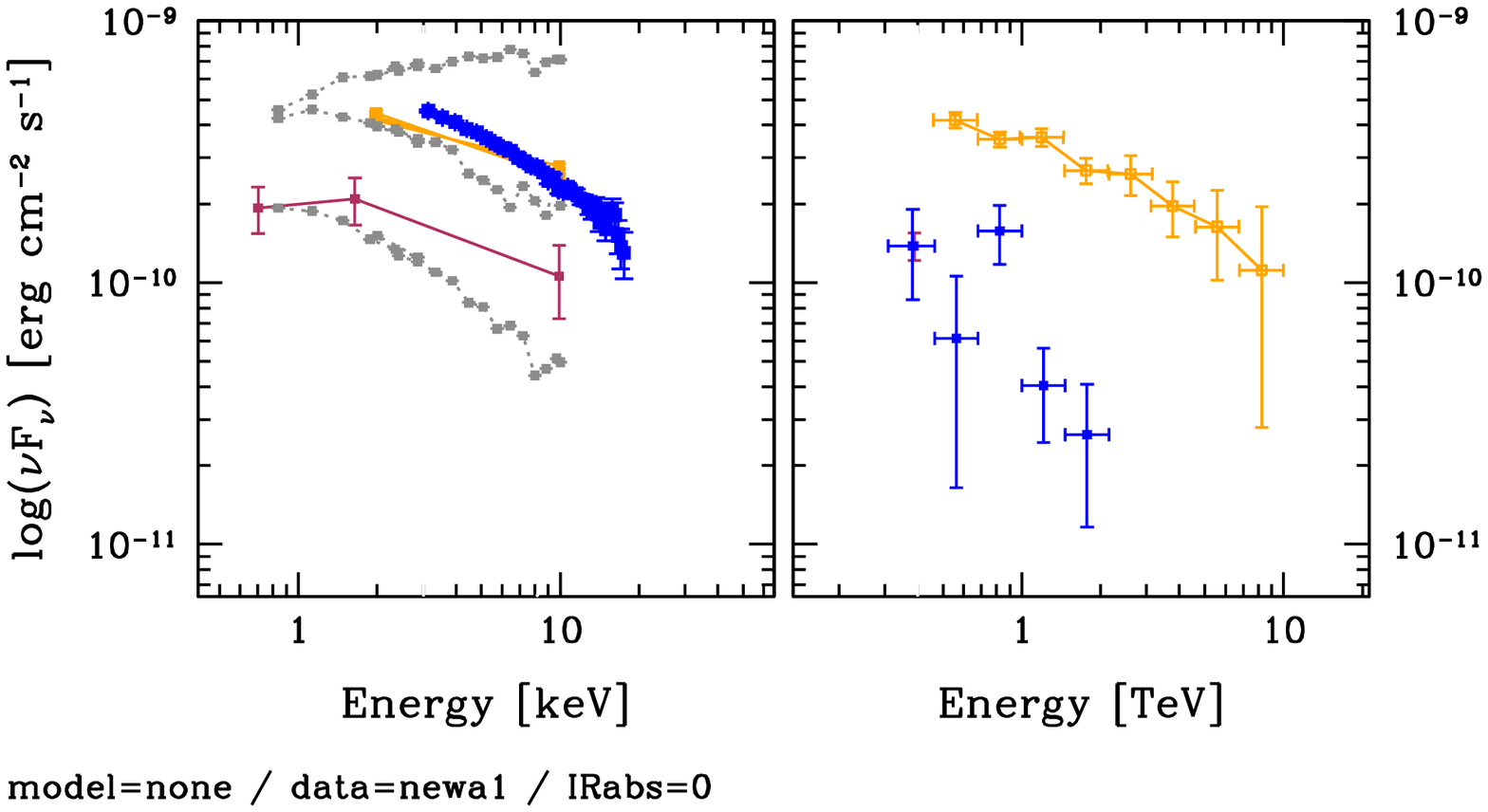}
      \hfill\includegraphics[width=0.50\linewidth,clip=]{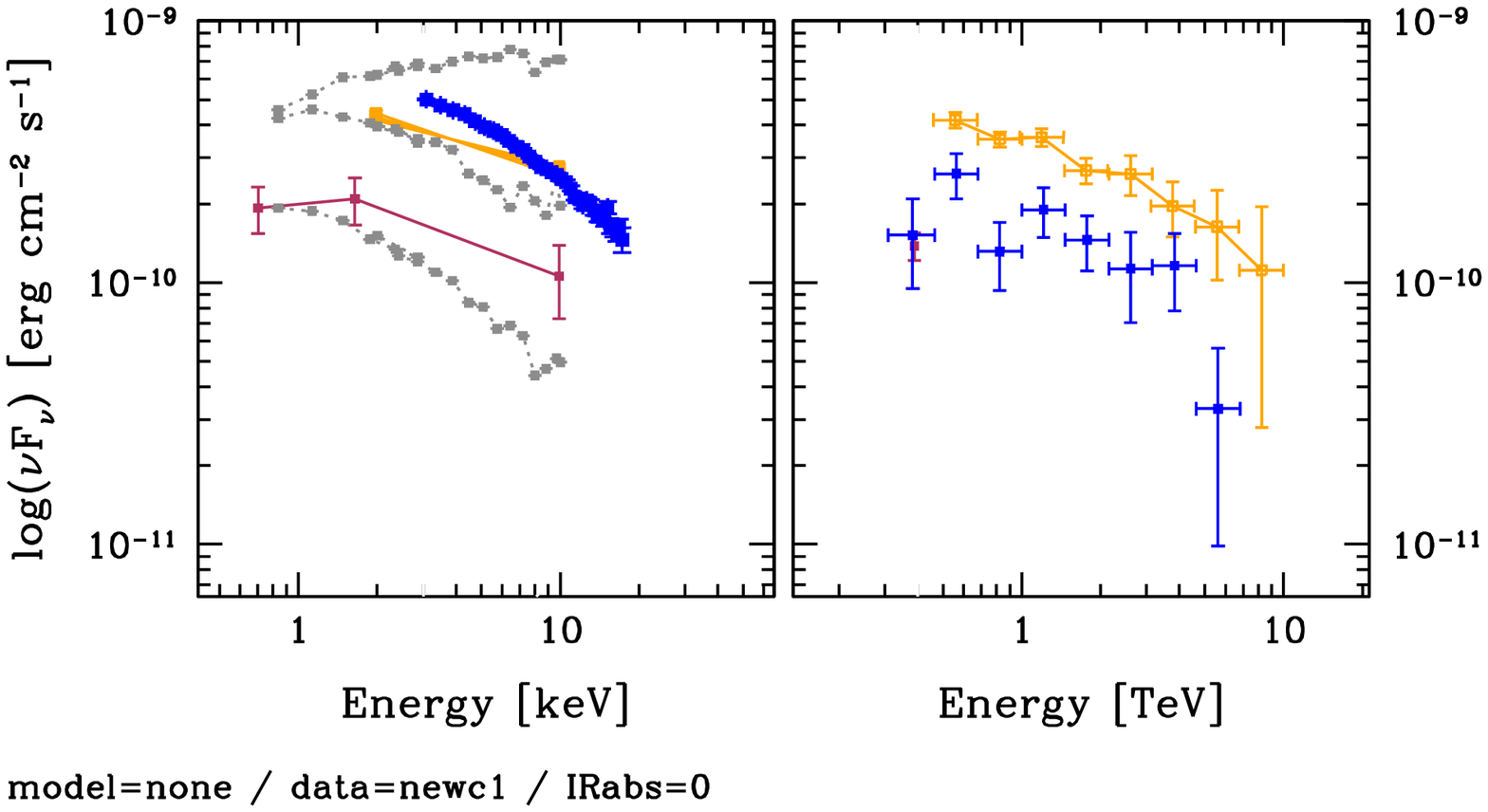}
      \hfill}
\centerline{
      \hfill\includegraphics[width=0.50\linewidth,clip=]{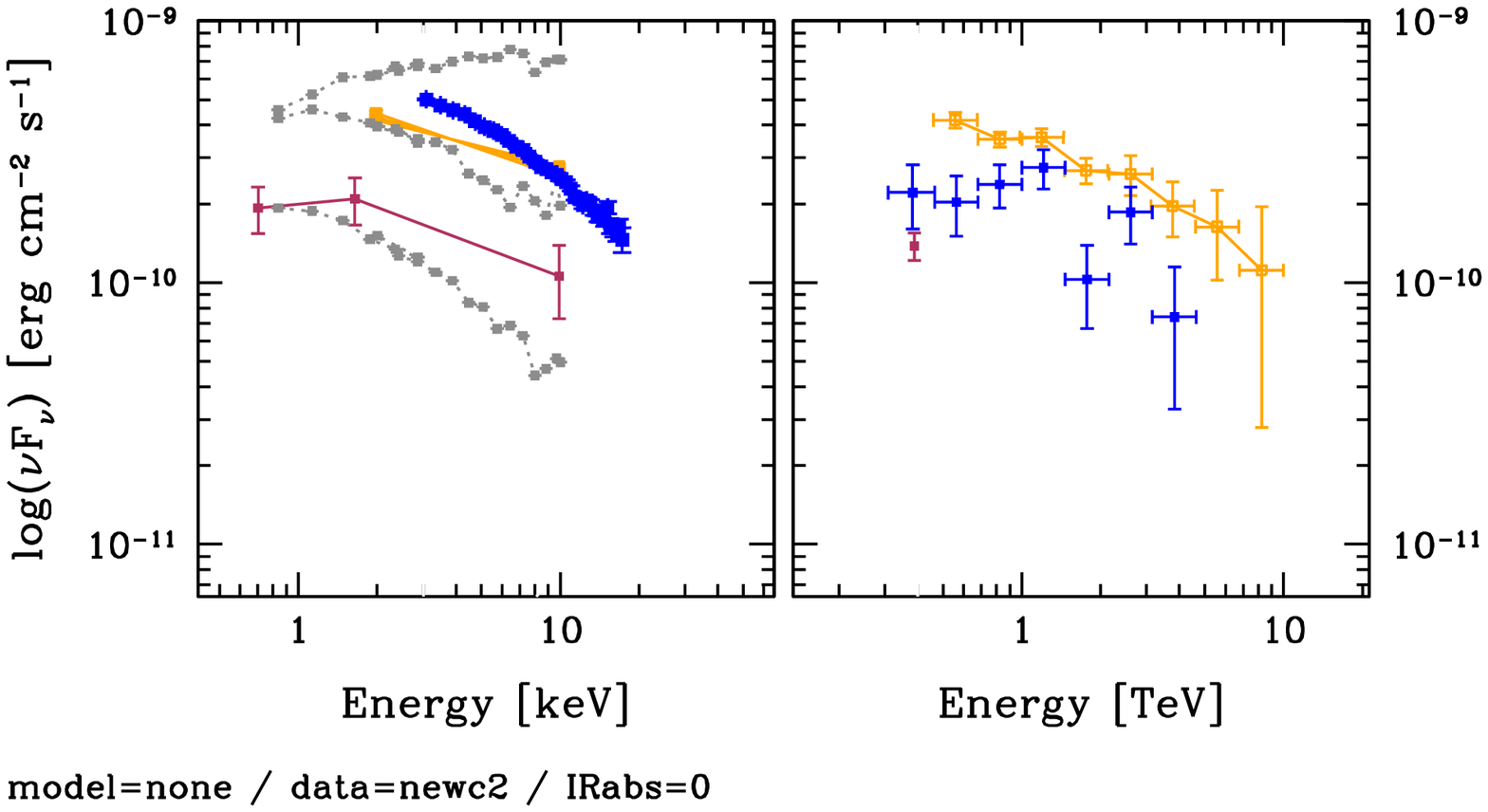}
      \hfill\includegraphics[width=0.50\linewidth,clip=]{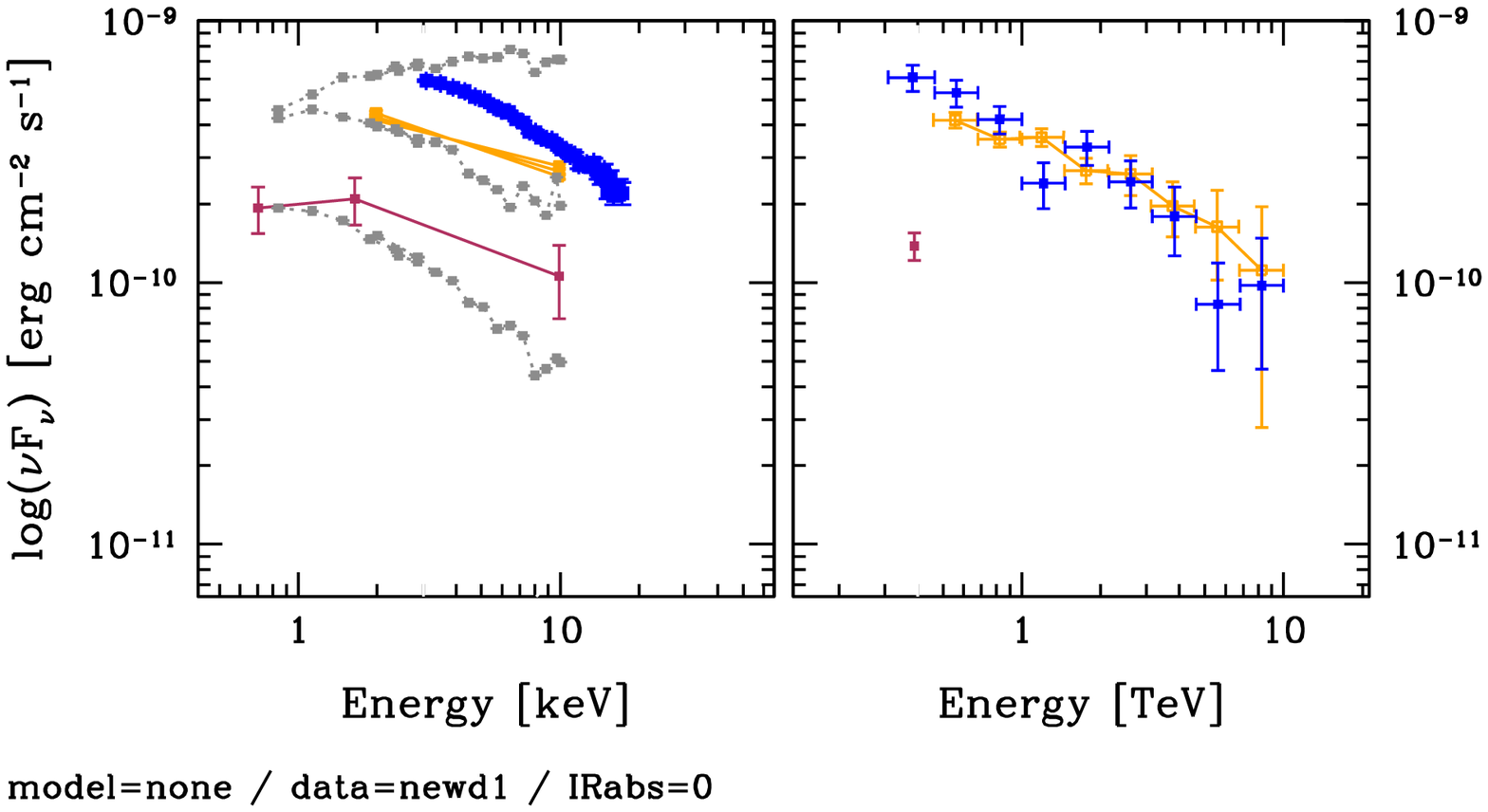}
      \hfill}
\centerline{
      \hfill\includegraphics[width=0.50\linewidth,clip=]{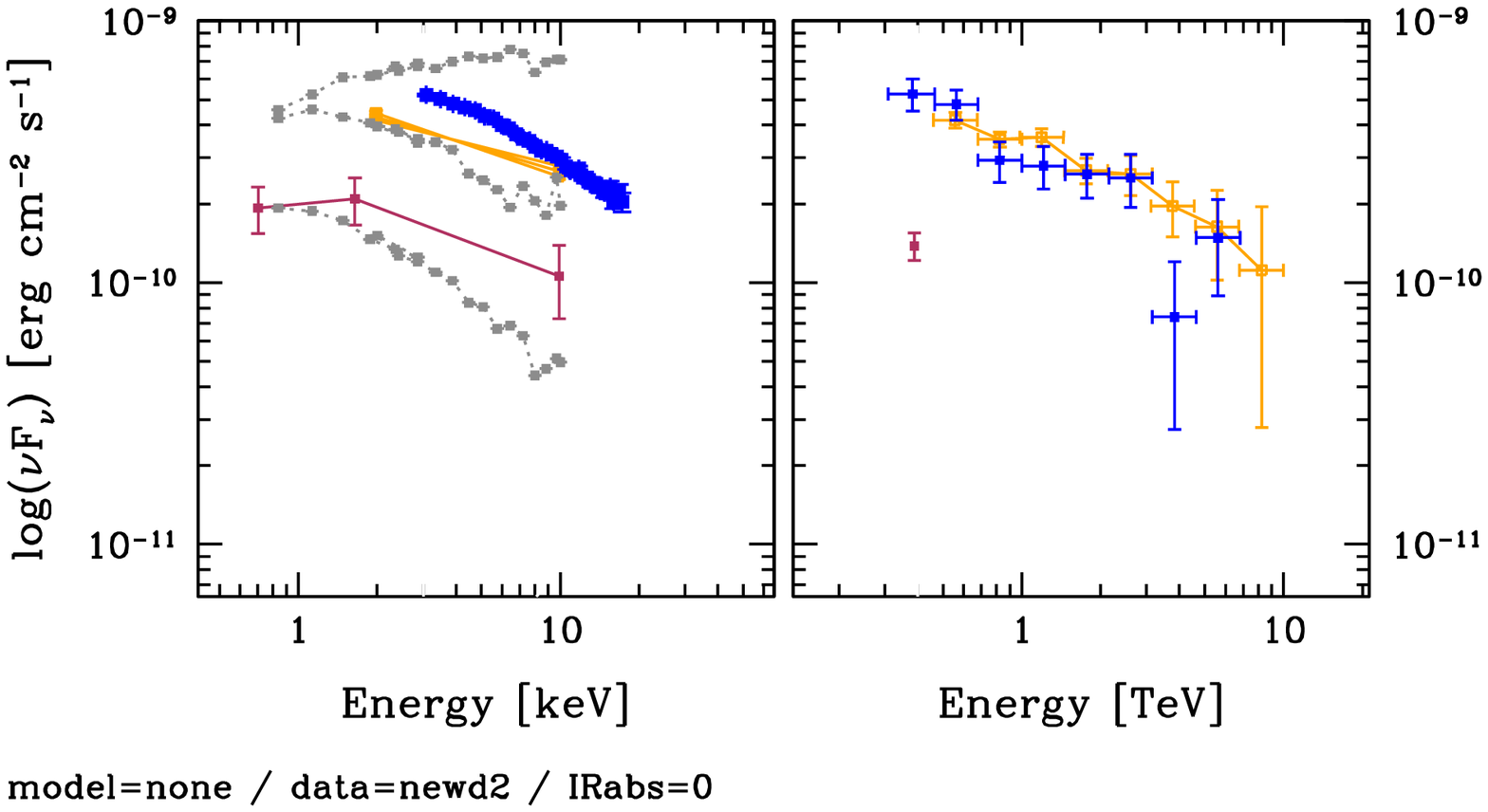}
      \hfill\includegraphics[width=0.50\linewidth,clip=]{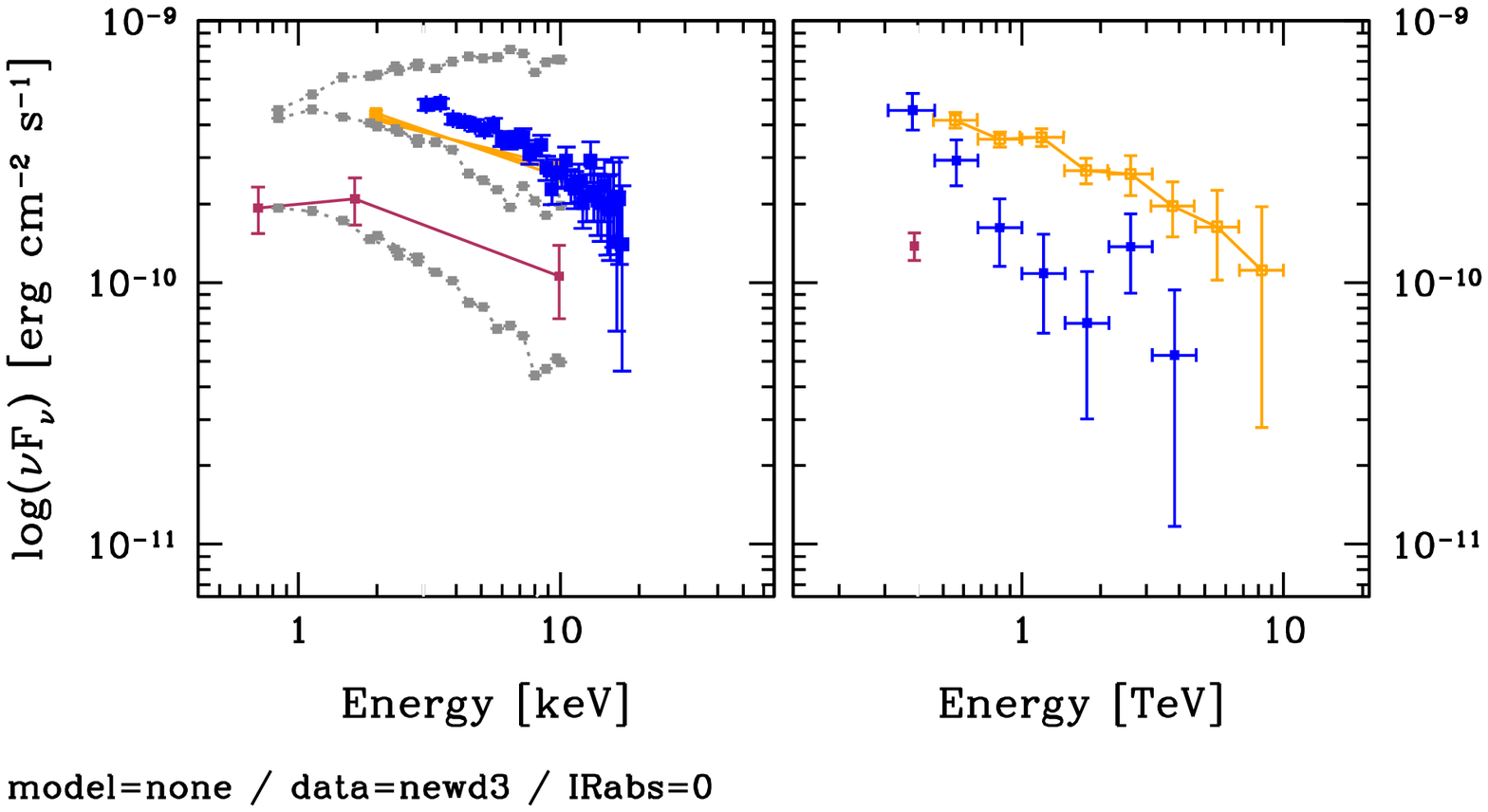}
      \hfill}
\centerline{
      \hfill\includegraphics[width=0.50\linewidth,clip=]{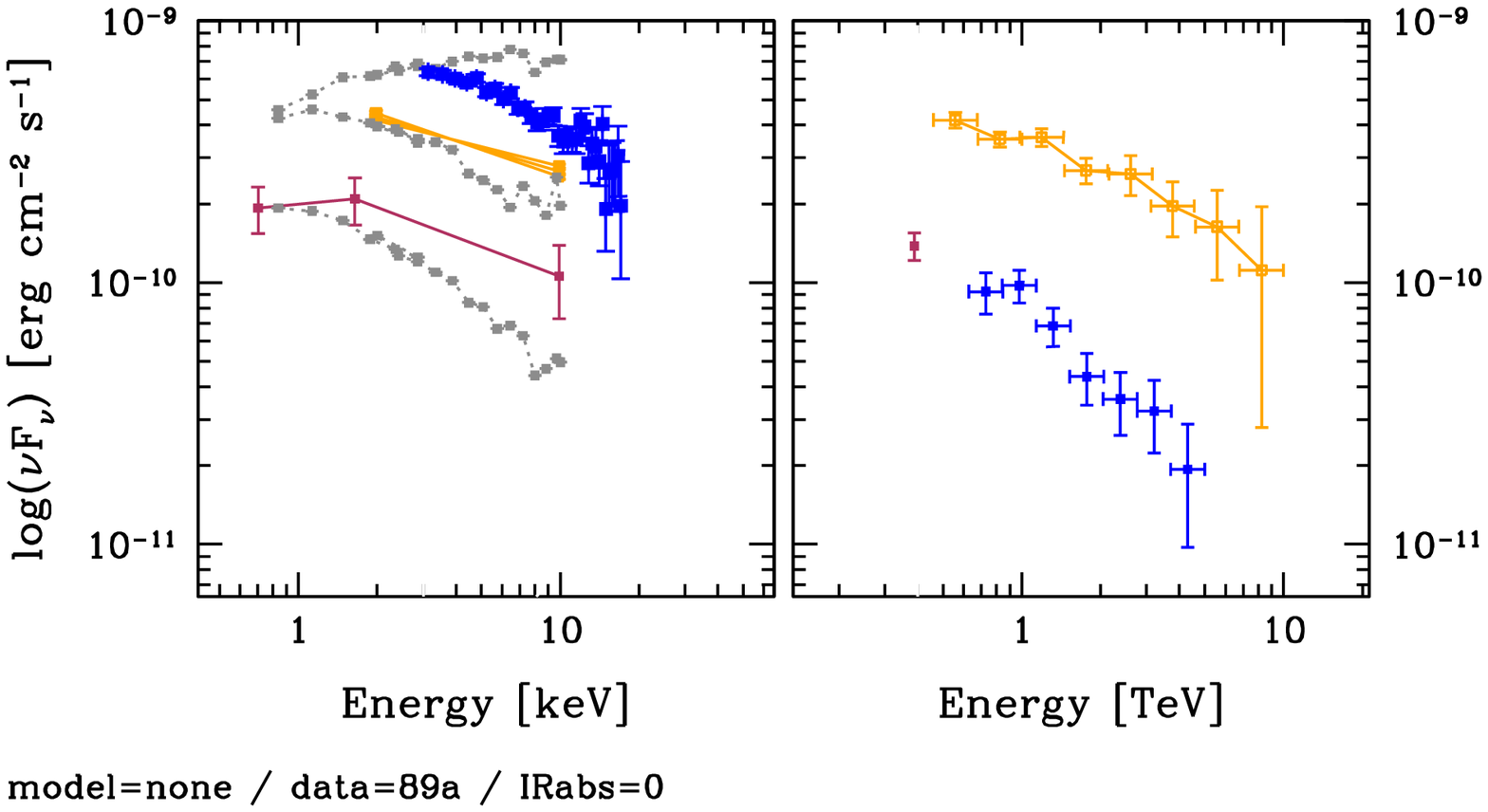}
      \hfill\includegraphics[width=0.50\linewidth,clip=]{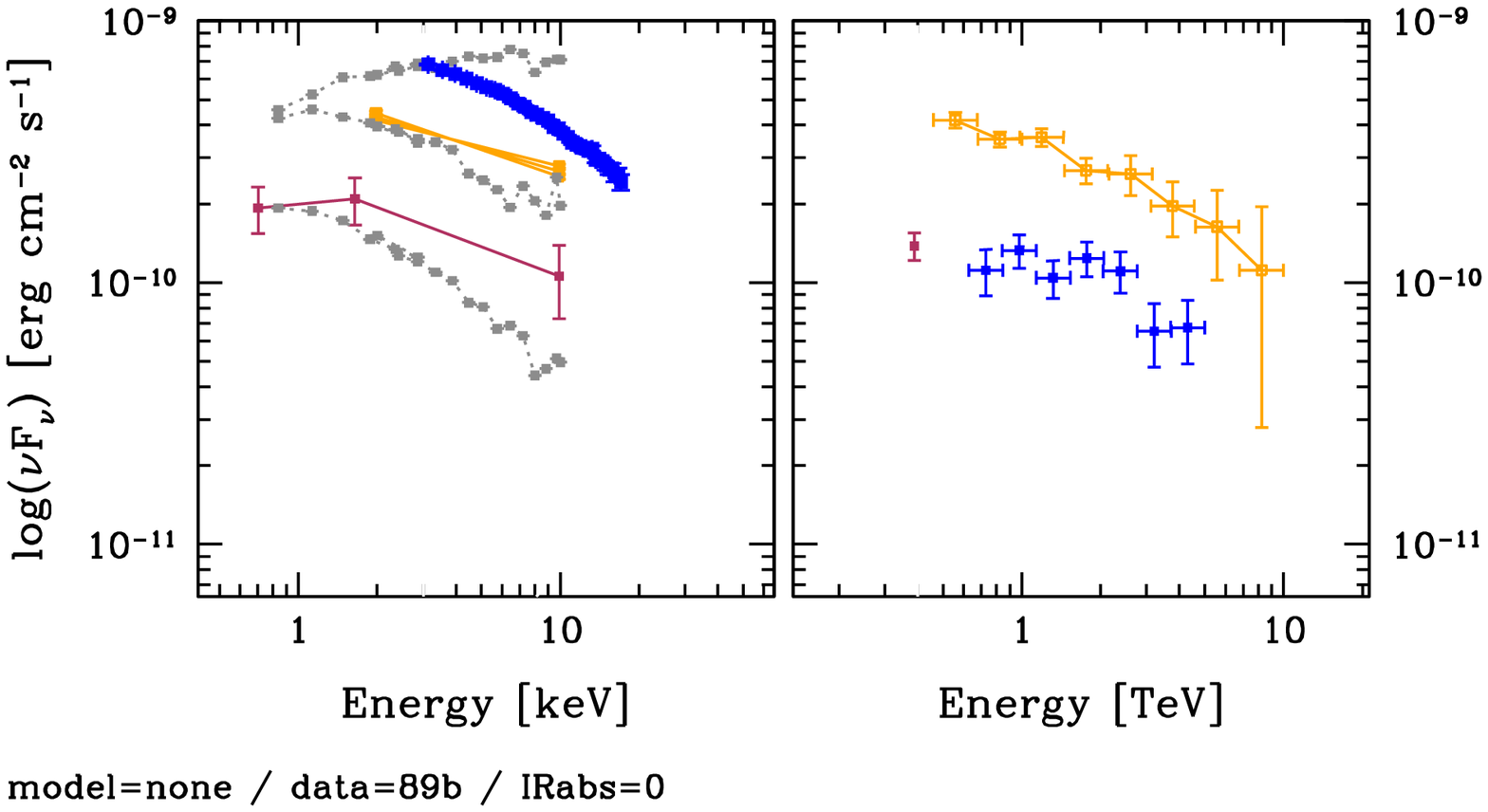}
      \hfill}
\centerline{
      \hfill\includegraphics[width=0.50\linewidth,clip=]{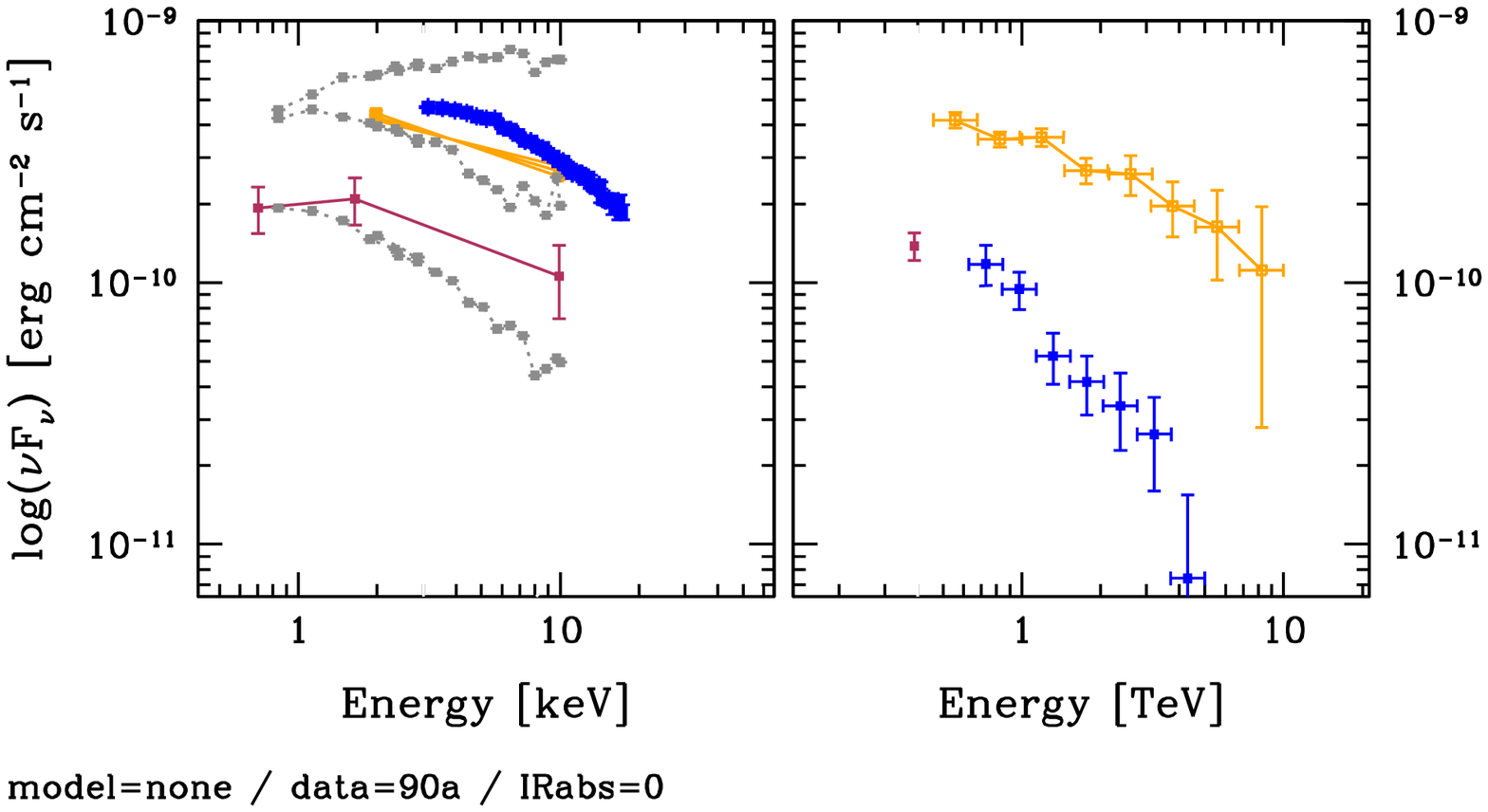}
      \hfill\includegraphics[width=0.50\linewidth,clip=]{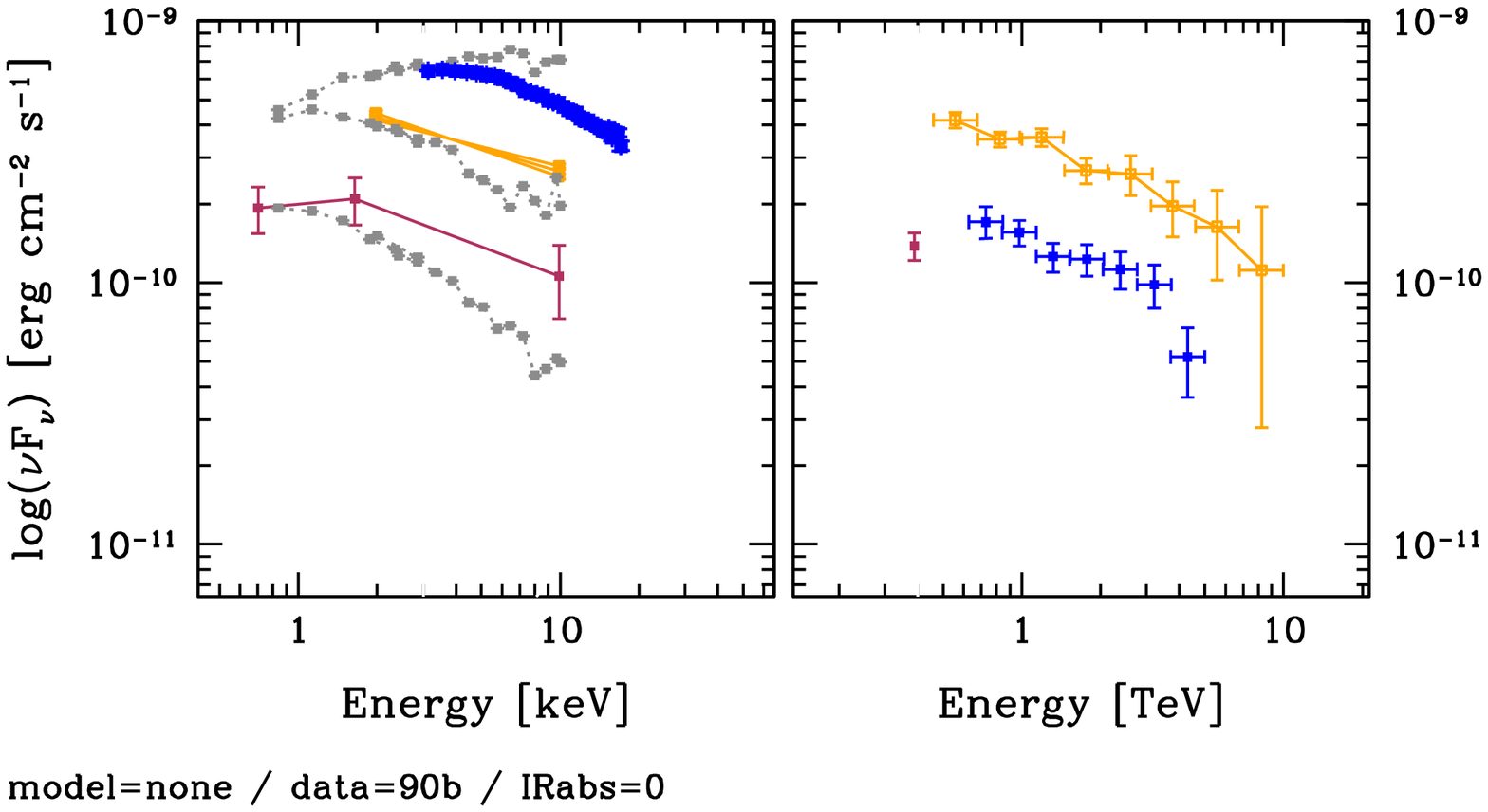}
      \hfill}
\caption{%
\small
Gallery of \rxtenosp, Whipple/HEGRA spectra pairs, in blue symbols.
Time elapses left to right, top to bottom.
The top six panels refer to March 19 (Whipple): approximate times are (UTC)
 05:39, 
 07:04, 
 07:34, 
 08:33 (flare top), 
 08:59, 
 09:30.
The bottom two pairs are for March 21/22 and 22/23, with HEGRA data
(\citealt{aharonian02_mrk421_spectral_var}, preflare and flare).
For reference we also plot: 
i) The simultaneous observations of the May 1994 reported by
\citet{macomb95_mkn421} (maroon, 3-point \xray spectrum and single flux point at
0.4\,TeV, at $\approx 10^{-10}$).
ii) The highest state observed in 1996 \citep{zweerink97_mrk421_flare} 
(orange, \xray power law from ASCA, and 0.4$-$10 TeV spectrum by Whipple).
iii) Denser-points light-gray 0.7$-$10\,keV \xray spectra are (bottom to
top) lowest and highest state during \sax 1998 campaign
\citep{fossati00_sax_mkn421_2}, and the highest \sax 2000 state (Fossati
\etalnosp, in preparation).
\label{fig:xray_gamma_spectra_pairs}
}
\end{figure*}

A summary ``gallery'' of the pairings of \xray and \gray spectra for these
flares is shown in Figure\,\ref{fig:xray_gamma_spectra_pairs}.
For reference we plotted also some historical observations.

It is worth noting that this gallery does not include the highest
luminosity \xray states, nor in general (\ie irrespective of simultaneous
\gray data), neither among the intervals matching TeV observations.
On the other hand, the peak of the March 19 flare does constitute the
most luminous TeV spectrum of the 2001 campaign, and in fact it 
matches the spectrum and luminosity of the most intense flare ever
recorded for \mrknosp, that of May 7, 1996 \citep{zweerink97_mrk421_flare}.

\begin{figure*}[!t]
\centerline{%
\includegraphics[width=0.51\linewidth,clip=]{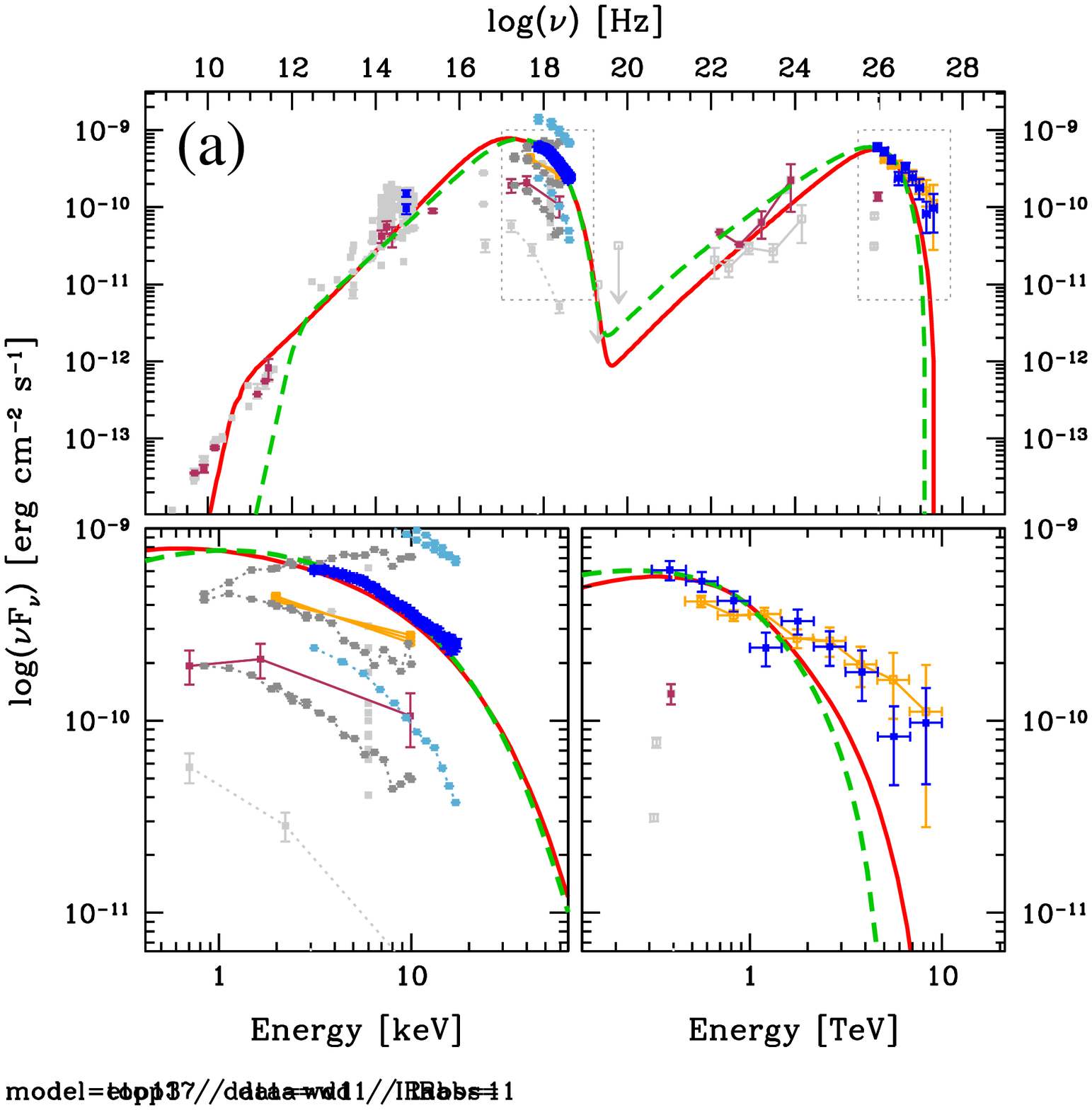}%
\hfill
\includegraphics[width=0.51\linewidth,clip=]{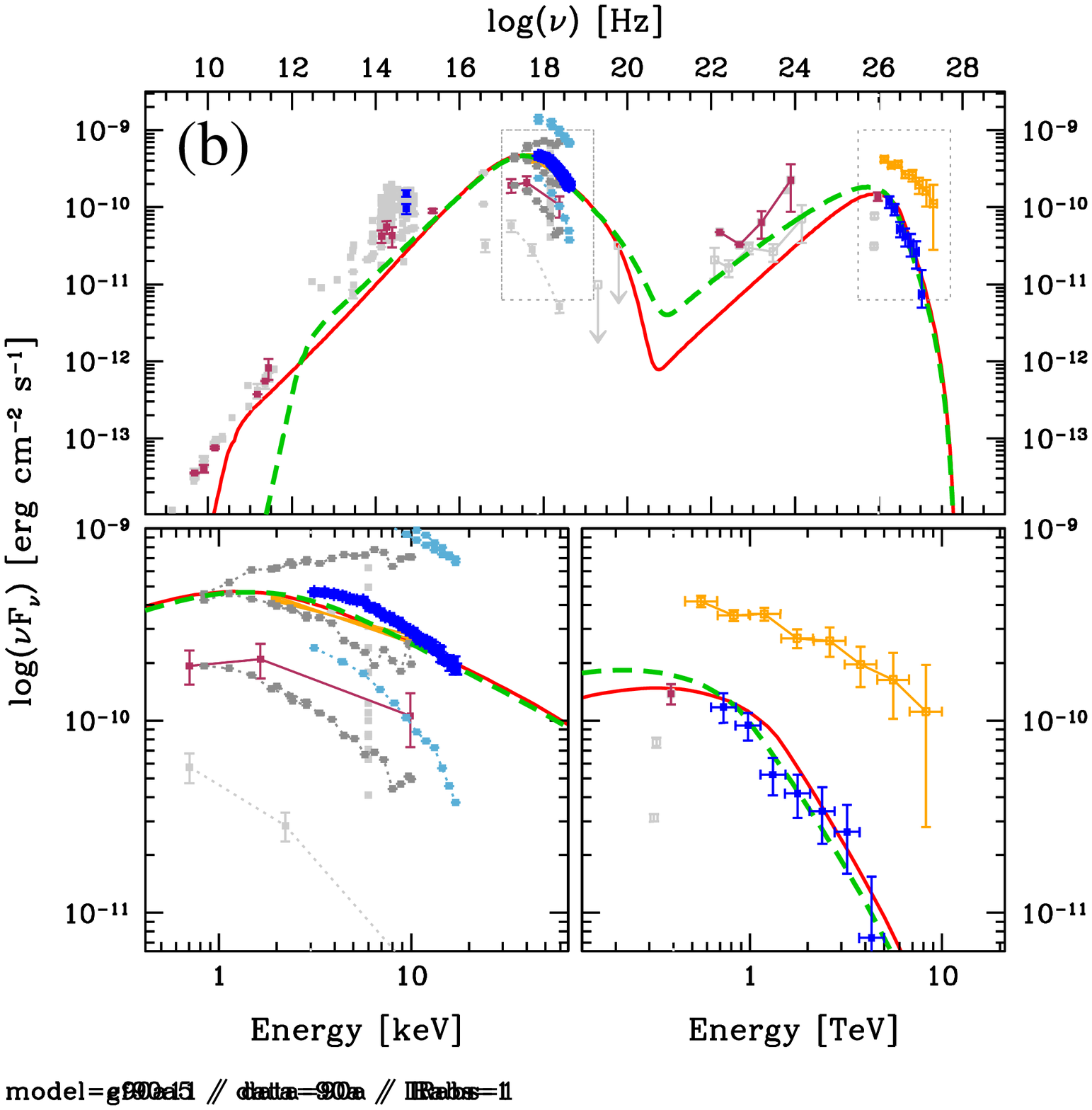}%
}
\caption{%
\small
Spectral Energy Distributions for two epochs during the March 2001 campaign.
Left, time around the peak of the March 19 flare.
Right, the ``low'' --preflare-- state observed by HEGRA on Match 22/23 (HEGRA spectra
from~\citealt{aharonian02_mrk421_spectral_var}).
Simultaneous 2001 data are shown in blue.
For the optical we plot the highest and lowest fluxes observed during the
campaign.
The sparser, connected, light blue, 3$-$20\,keV \xray spectra show the highest
and lowest X--ray observed states.
For reference, we plot in gray multiwavelength data from a collection of
historical data from NED and \citet{macomb95_mkn421}.
The maroon points are the simultaneous observations for May 1994 \citep{macomb95_mkn421}.
For the description of some of the \xray and \gray data see also
Figure~\ref{fig:xray_gamma_spectra_pairs}.
The solid red lines represent ``fits'' with a simple one-zone
homogeneous SSC model with $B\simeq0.1-0.15$\,G, $\delta=20-25$, $R_\mathrm{blob}=10^{16}$\,cm.
The dashed green SED models are for ``extreme'' cases, with $B\simeq1$\,G, $\delta=100$,
$R_\mathrm{blob}=0.5-1\times10^{14}$\,cm.
}
\label{fig:seds}
\end{figure*}

For ease of comparison we prefer to adopt the same axis scales for \xray and \graynosp,
and this makes the variability of the \rxte spectra not as easily
noticeable as that of Whipple/HEGRA spectra.  
Nevertheless the level of variability can be appreciated 
by comparison with the reference historical spectra.

We would like to highlight a few observational findings.
The peaks of the synchrotron and IC components never cross into the
telescopes bandpasses, despite the relatively large luminosity variations.
Increases in brightness are accompanied by significant spectral hardening,
but there is no compelling sign that this is also accompanied/due by a
shift of the SED peak energies.
Among the data presented here, the only instances when the synchrotron
peak might be/is directly detected are the spectra for the March 22/23
flare. 
It indeed seems that the high energy tails move between hard and soft
states as if pivoting with respect to unobserved lower energy parts
of the spectrum, possibly the synchrotron or IC peaks.
This is suggested by the observation that in most cases the lowest
energy data point in successive spectra are approximately at the 
same level, whereas we would expect some ``upward shift'' in both
the case of variations due to a change of energy the SED peak,
and the case of an overall increase of luminosity around the SED peak.

\subsubsection{Spectral Energy Distributions}
\label{sec:seds}

In Figures\,\ref{fig:seds} we show selected simultaneous \xray and \gray  
spectra for the March 2001 campaign, together with a collection of historical
multiwavelength data (see Figure caption for details).

In particular Figure\,\ref{fig:seds}a shows the data for the peak
of the March 19 outburst, and Figure\,\ref{fig:seds}b the pre--flare interval
for March 22/23.
These two are quite representative of a bright and hard, and a fainter and soft cases.
We tried to model this sparse SEDs with a single zone homogeneous SSC model
(\eg \citealp{gg98_sedt}), and example fits are plotted along with the data.

Although the simultaneous data coverage is limited to optical flux and
the \xray and \gray spectra, a coarse search of the parameters space for
a good SSC model fit showed that the constraint are nonetheless very strong.
This is true even though we made no attempt at taking into account self-consistently
the abundant information available ``along the time axis'', such as the 
time resolved spectral variability.
One general difficulty encountered while fitting the SSC model, is that the 
TeV spectra are typically harder than what can be predicted. 
As we illustrate in the section \S\ref{sec:ic_split}, this is in part due
to the effect of the Klein-Nishina (K-N) decrease of the scattering efficiency,
canceling the contribution from the self-Compton of the electrons and
photons emitting/emitted above the synchrotron peak.

\subsubsection{$B-\delta$ diagnostic plane}
\label{sec:bdelta}

Since we can estimate the energies and luminosities of the
synchrotron and IC peaks with reasonable accuracy, in the context of a
single zone SSC model we can draw the locus allowed by a given SED in
the $B-\delta$ parameter space (\eg \citealp{tavecchio_tev_98}).
Besides this primary piece of information, measurements or estimates of
several other quantities (and their combinations) can be exploited to set
additional constraint on the $B-\delta$ relationship.
These include for instance peak luminosities, cooling times, variability
timescales (or source size), intraband time lags.

\begin{figure}[t]
\centerline{%
\includegraphics[width=0.90\linewidth]{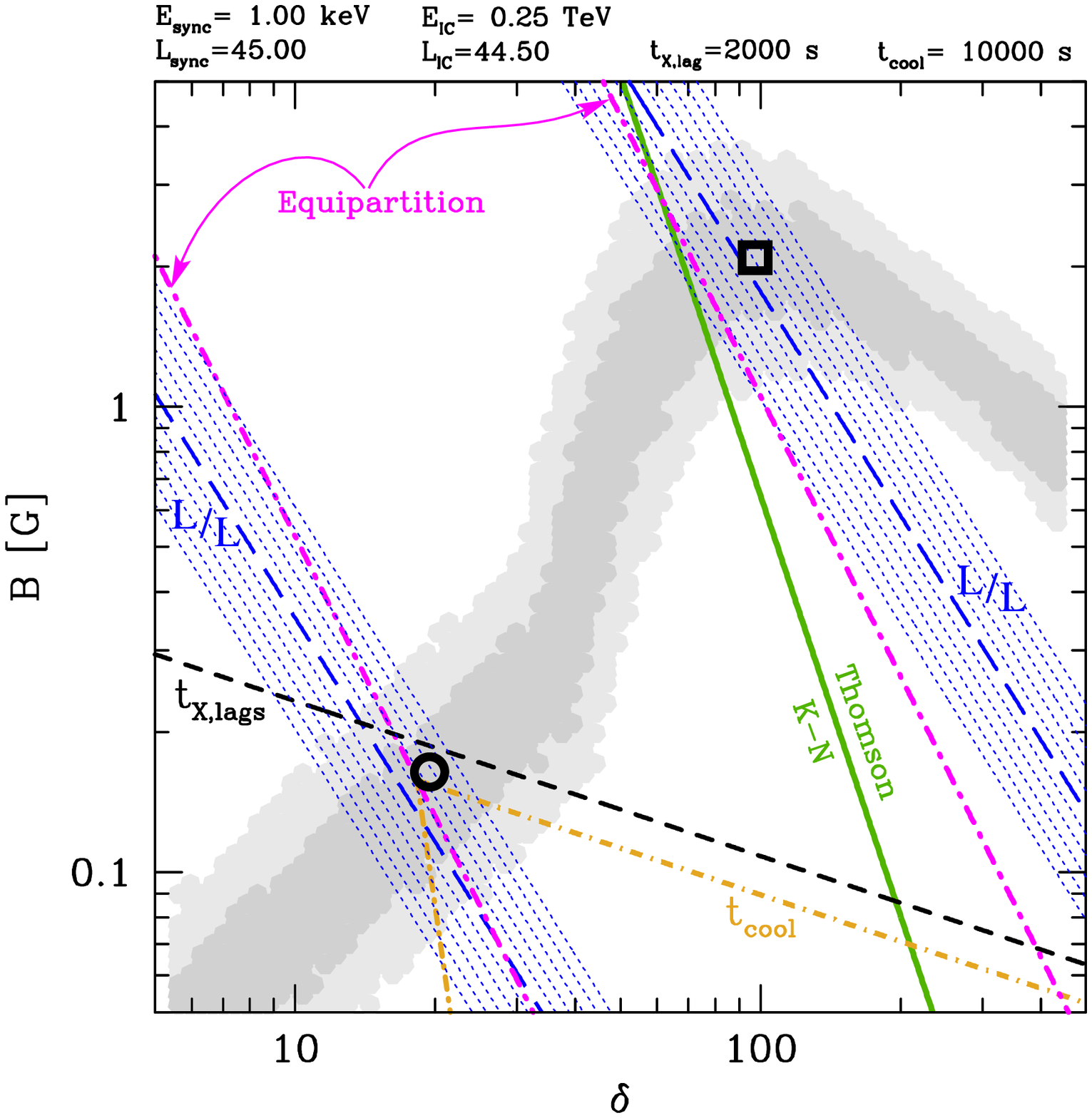}
}
\caption{%
\small
$B-\delta$ plane for a set of parameters ($\nu_\mathrm{peak,X}$, $\nu_\mathrm{peak,\gamma}$, 
$L_\mathrm{peak,X}$, $L_\mathrm{peak,\gamma}$, representative of the 2001 campaign.
The gray locus crossing the plane shows the constraint set by the synchrotron and
IC components peak frequencies (adopted values are shown atop the figure box.)
The long-dash steep lines show the range allowed by the inferred peak
luminosities.
The dot-dashed steep lines marks the equipartition between $U_B$ and $U_\mathrm{rad}$.
There are two sets of long-dash and dot-dashed lines: the left ones correspond
to a standard case of $R=10^{16}$\,cm. 
Those in the right part of the figure refer to the case of $R=5\times10^{13}$\,cm.
The dotted regions are meant to represent the effect of a factor of 3
uncertainty on the main parameters.
The thick black line is the approximate analytical boundary between
Thomson and Klein-Nishina scattering regimes.
The gray dot-dashed lines marks the combination of parameters yielding a
10\,ks cooling time for electrons emitting the synchrotron peak.  
The bottom-right corner wedge delimited by them is non-consistent with
this imposed (putative) limit. 
The short-dash black line is the lower bound allowed by a hypothetical
\xray intraband lag of 2\,ks.  
The black circle marks the parameter choice for the SED model drawn 
solid red in Figures\,\ref{fig:seds}b.
The black square instead marks a possible choice of parameters in the Thomson
regime region of the $B-\delta$ plane, dashed green SED in Figures\,\ref{fig:seds}b.
}
\label{fig:b_delta_plane}
\label{fig:b_delta}
\label{fig:bdelta}
\end{figure}

An example is shown in Figure\,\ref{fig:b_delta_plane}, for March 22/23.
The gray band represents the constraints set by our estimate of the peak
positions.  
The peak luminosities yield a few additional lines in the $B-\delta$
plane, in particular the particle-magnetic field equipartition. 
A requirement on the cooling time of the peak-emitting particles,
for instance to be shorter than the ``typical'' variability timescale
(\eg 10\,ks), translates into an excluded wedge in the lower right part 
of the plane.  The two different lines, meeting at the equipartition
line, correspond to synchrotron or IC dominated cooling regimes.

Detection or an upper limit on the value of intraband \xray lags also sets
a lower boundary to the allowed region in the diagram. 
In Figure\,\ref{fig:b_delta_plane} we draw the limit for a hypothetical
2\,ks lag within the PCA bandpass.  Shorter lags, or upper limits
on them, move this line upwards.
It is worth noting that the detailed cross correlation analysis of the
\xray dataset (F08) does not yield any reliable intraband lag detection.
In particular, no lags have been found in the analysis of all short, single
orbit, subsets with significant variability features (a few dozen), and in
most cases the upper limits is of the order or a few hundred seconds,
$\simeq200-300$\,s (the corresponding line in Figure\,\ref{fig:b_delta_plane}
would be about 0.5~decades, $\times3$, higher).

Finally, we can draw in the $B-\delta$ plane the dividing line between
the Thomson and Klein-Nishina scattering regimes, for the SED peaks.

One of the largest sources of uncertainty for the determination of allowed
region in the $B-\delta$ plane is the position of the IC peak, because 
\begin{equation}
\frac{B}{\delta} \propto \frac{\nu_\mathrm{peak,sync}}{\nu_\mathrm{peak,IC}^2}
\end{equation}
for scattering in Klein-Nishina regime.
This also means that any consideration based on this diagnostic plane
is subject to the uncertainty about the details of the TeV photon 
absorption by the diffuse infrared background.
The models shown here include the effect of the IR absorption, following
the ``low intensity'' model prescription of \citet{stecker_dejager98}.
For our limited modeling purpose the exact choice of IR background
absorption model is not critical.

Figure\,\ref{fig:b_delta_plane} shows the locii and limits obtained for
a SED similar to that on the right panels of Figure\,\ref{fig:seds} (March 22/23), 
for which we obtained a satisfactory SSC model fit.
The relevant observational parameters are reported in the plot.
All constraints are satisfied by a model having a magnetic field 
of $B\simeq0.15$\,G, a Doppler factor $\delta\simeq20$, a blob size
of $R\simeq10^{16}$\,cm, \ie within the range regarded as \textit{standard}
in SSC modeling (see circle in Figure\,\ref{fig:b_delta}, solid line
model in Figure\,\ref{fig:seds}b).

This analysis shows that for this choice of parameters the scattering
producing the TeV emission occurs in the Klein-Nishina regime.
It is in principle possible to shift the ``sweet spot'' in the upper corner
of the diagram, into the Thomson regime region, by adopting a much smaller
source size ($R\simeq5\times10^{13}$\,cm), and in turn $B\simeq2$\,G
and $\delta\simeq100$ (square in Figure\,\ref{fig:b_delta}), and
indeed a similarly satisfactory SSC fit to the snapshot SED can be obtained
(dashed model in Figure\,\ref{fig:seds}b).

A similar analysis was performed for the data of the March 19 flare peak,
shown in Figure\,\ref{fig:seds}a, and also in this case the SED could
be fit both with ``standard'' ($B=0.1$\,G, $\delta\simeq20$,
$R=10^{16}$\,cm) ``extreme'' ($B=1.0$\,G, $\delta=100$,
$R=10^{14}$\,cm) parameters (corresponding SEDs are shown 
as solid and dashed lines in Figure\,\ref{fig:seds}.)

Other considerations can help to discriminate between these scenarios, for
instance arguments concerning time variability properties or the
viability of having such an extreme Doppler factor and blob size.

\subsubsection{TeV spectral decomposition analysis}
\label{sec:ic_split}

In order to try to understand the observed correlated variability between
\xray and TeV fluxes, it is interesting to take a deeper look at the
composition of the emission in the TeV band, in terms of which electrons
and seed photons contribute to the flux at different energies.
This analysis is somewhat model dependent, and we are only showing it
for the \textit{standard} parameter choice introduced before.

The idea is similar to the treatment discussed by \citet{tavecchio_tev_98},
where they split the IC component in four components, produced by the
combinations of electron and synchrotron photons below (L) and above
(H) the synchrotron peak (for electrons the split is done at the energy
mapped to this latter).
For TeV blazars the conditions are such that the component H,H (electrons
and photons both above the peak) is strongly depressed, and becomes
negligible.  The same holds true for the L,H component \citep{tavecchio_tev_98}.

The same approach can be extended to an arbitrary split
of the primary components, aiming at identifying more precisely the
origin of the electrons and photons.

\begin{figure}[!t]
\centerline{%
\includegraphics[width=0.99\linewidth,clip=]{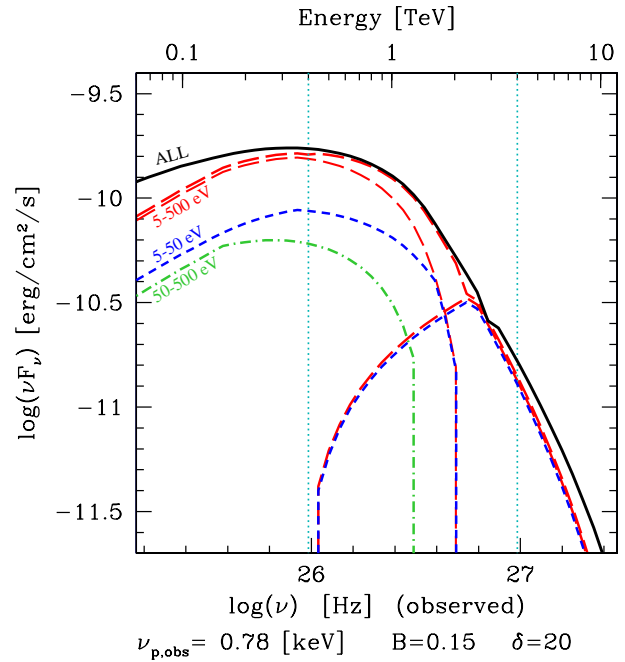}
}
\caption{%
\small
Modeling of the inverse Compton peak for the SSC model shown solid red in
Figure~\ref{fig:seds}b.  The IC emission is split in different
components \textit{\`a la} \citet{tavecchio_tev_98}.
The long-dash black lines show the emission obtained by restricting the
synchrotron seed photons to the 5-500\,eV range.  
It is split in two by considering separately electrons emitting in
synchrotron below and above 2\,keV: the lower energy section is thus coming
from 5--500\,eV photons scattered by electrons up to 2\,keV.
This latter is then further split into the contributions from 5--50\,eV
(light gray short-dashed line) and 50--500\,eV photons (darker gray dot-dashed).
The thick solid line is the full IC component, \ie including all seed
photons contributions.
The vertical dotted lines mark the 0.4--4\,TeV band.
}
\label{fig:ic_split}
\end{figure}

In Figure\,\ref{fig:ic_split} we show the IC peak for the same model
on which we have focused above.  
We adopt as breaking point for the electron spectrum the energy
corresponding to an observed synchrotron emission at 2\,keV, because
this is just below the \rxtenosp/PCA bandpass and this splitting allows
us to divide the electrons between those whose synchrotron
we observe and those we don't.
We consider all electrons, but restrict the allowed seed photons to
those in the range $5-500$\,eV, observer's frame.
Different line types correspond to different ranges of seed photons
within this band.
The long-dash black lines show the full emission by $5-500$\,eV photons
and electrons below and above 2\,keV, with its decomposition in 
the two contributions.
It is clear that most of the emission at around 1\,TeV is
accounted for by these photons scattered by sub-keV electrons.
The gray short-dash and dot-dash lines show a further split of the
photons in the $5-50$\,eV and $50-500$\,eV bands, with the former
contributing about 60\% of the total at around 1\,TeV.

Because of the Klein-Nishina effect, there is no emission by 
self-Compton by electrons observed in synchrotron in the \xray band.
Moreover, the (non-self) IC contribution by \xraynosp-observed electrons is
weak and at high enough energy that we can regard is as irrelevant
to the effect of the observed TeV rate variability.
This latter is dominated by a lower energy section of the bandpass,
around $\simeq1$\,keV.

The K-N depression of the scattering of the higher energy photons
and electrons cuts off the H,H and H,L contributions, and makes
the TeV spectrum always steeper than the \xray one, at least in
the more minimalist scenarios.
Indeed in first approximation the power law of the high-energy tail
of the IC component tends to a value $\alpha_\mathrm{KN} \simeq 2 \alpha_2 - \alpha_1$,
where $\alpha_1$ and $\alpha_2$ are the spectra indices below and above
the synchrotron peak \citep{tavecchio_tev_98}.
This is always steeper than $\alpha_2$, and for fiducial values of
$\alpha_1$ and $\alpha_2$ it is so by $\triangle\alpha\gtrsim0.5$.

In the more extreme scenario contrived to shift the scattering 
into the Thomson regime, the IC peak composition is indeed different.
In particular, as expected, there is a more even contribution by
the three of the four components.  The H,H components is still
negligible.
Photons up to $2-3$\,keV are effectively scattered, and there is
a self-Compton contribution to the emission at $\simeq1$\,keV.
Therefore, a much more direct connection between the two observed
bands is afforded by the more exotic scenario, and in turn the
observed correlated variability would be more readily explained.

\section{Conclusions}
\label{sec:conclusions}

The correlation between the variations in the \xray and TeV bands is
confirmed with unprecedented detail, supporting the idea that the same electron
distribution, in the same physical region, is responsible for the emission
in both energy bands.  However the details of these findings pose a serious
challenge to the emission models.  
Here we would like to sketch a few selected outstanding issues raised by
the correlated variability.  We refer to a forthcoming paper for an 
in-depth analysis comprising more extensive modeling.

\begin{itemize}
\item[$\bullet$] 
If modeled within the realm of standard values for magnetic field,
Doppler factor, source size, the IC scattering responsible for
the observed TeV emission occurs in the Klein-Nishina regime.
This means that with \xray and \gray observations, although
we seem to be observing regions of the spectrum that are
very similar to each other for what concerns their position
with respect to the SED peaks, we are not tracking the 
evolution of the same electrons (and photons).
The extent of the phase and amplitude correlation of the
\xray and \gray variations is however remarkable, and this
sets broad constraints on the characteristics of the
processes responsible/governing the variability, \eg acceleration/injection
of particles, dominant cooling cause.

\item[$\bullet$] 
In this context, the observation in K-N regime of a quadratic relationship
between synchrotron and IC variations (which would be naturally produced in
the Thomson regime because of the effectiveness of self-Compton) constrains
the electron spectrum variations to occur over an energy band broad enough
to affect also the IC seed photons.
Moreover, this variation must be essentially ``achromatic'' (\ie just a
change in normalization), otherwise the extra energy-dependent factor
would produce an observable effect.  

\item[$\bullet$] 
The observation that the flux--flux path of the better observed flares
decay follows closely the bursting path introduces a further complication.
If the flare decay is governed by the cooling of the emitting electrons 
we do not expect the quadratic relationship to hold during the decaying
phase.
In fact, given the energy-dependent nature of synchrotron (and IC) cooling,
with $\tau_{cool}\sim E_{ph}^{-1/2}$, the $50$\,eV seed photons cool
on a longer timescale, \eg $\sim$10 times longer than the timescale for photons
observed at $\gtrsim3$\,keV.  
The possibility that also the electrons contributing to the bulk of
the TeV emission have lower energy than those observed in \xraynosp, 
compounds the problem. 
This means that during the flare decay the \xray and $\gamma$-ray
brightnesses should follow something like a linear relationship, because
the IC (TeV) emission will just reflect the evolution of the electron
spectrum, scattering a ``steady'' seed photon field.

Plain radiative cooling does not seem to match these observations.
A viable mechanisms explaining the flares evolution should allow the
concurrent cooling of a broad portion of the electron distribution.
On the other hand, brightness variations are accompanied by large spectral
changes, and in most cases they are very suggestive of acceleration 
--or injection-- of the higher energy end of the electron population.

\item[$\bullet$] 
Another recurring discrepancy between data and simple one-zone SSC
modeling is that of the TeV spectral shape, which is often harder
than model predictions.
As we illustrated in the previous section the Klein-Nishina effect
plays an important role in this respect.
A more careful analysis is warranted, but it is worth noting that
one alternative option for addressing this problem is that of considering
the effect of additional IC components, off photons external to the blob.
\cite{blazejowski_etal_2005} showed that a multi-component model seems
to be required to fit the 2003$-$2004 observations, and it might mitigate
the discrepancy in the TeV spectrum.
\cite{gg_ft_mc_2005_structure_jets} discuss the effect of including the
effect of the radiation emitted by the putative lower Lorentz factor outer
layer of the jet, namely as source of additional seed photons for IC.

\item[$\bullet$] 
Exotic scenarios could address some of these issues, namely by allowing the
IC scattering to occur in the Thomson regime, and hence the self-Compton to
be effective, thus reinstating the close relationship between the photons
and electrons tracked by \xray and \gray observations.
One such scenario that we discussed briefly would call for very high values
of the Doppler beaming factor, that would reduce the intrinsic energies at
play, higher magnetic field and very small size of the emission region.
In fact recently there has been some interest for less conventional modeling 
of TeV blazars (\eg \citealt{henric01_mrk421_x_tev,rebillot_etal_2006} for \mrknosp).
However, high Doppler factor scenarios raise a series of new issues, 
or re-open some that have been settled for the more traditional model.
First of all, the fact that the beaming cone of the radiation emitted
by such a fast blob is going to be much narrower has to be reconciled
with the population statistics of blazars and radio-galaxies, their
putative parent population (\citealt{up95_review}).
One way of doing it would be to imagine that the jet comprises a very
large number of small high Lorentz factor blobs, fanning out filling
a wider cone, with aperture consistent with the unification statistics.
This would constitute a quite radical change in the jet structure,
from the current one where emission is thought to come from internal shocks.
There are also implications concerning the statistical properties
of the variability, which would likely be due to the combination of
the bursting of different blobs, most likely uncorrelated.

\end{itemize}

The richness and depth of the \xray and \gray data of the March 2001
campaign presented in this paper raise the bar for models.
The aggregate characteristics illustrated here already challenge the simple
traditional SSC model, and the SED-snapshot approach.
In order to answer the questions raised by these observations it is of
paramount importance to exploit fully the time--axis dimension into
the modeling and take a dynamical approach.

The data time density and brightness (and so statistics) are unparalleled, 
enabling time resolved spectroscopy on timescales of the order of the
physically relevant ones, hence allowing to model the phenomenology
self-consistently minimizing the need (freedom) to make assumptions
as to how to connect spectra taken at different times.

It is likely that this dataset is going constitute the best benchmark for
time dependent modeling for some time, despite the great progress
made by ground based TeV atmospheric Cherenkov telescopes in the last
few years, because of the difficulty of securing long uninterrupted 
observations with Chandra and XMM-Newton.

\acknowledgements
GF thanks Jean Swank, Evan Smith and the \rxte GOF for their outstanding
job, in particular for implementing the best \rxte scheduling possible.
GF has been supported by NASA grants NAG5-11796 and NAG5--11853. \\
The Whipple Collaboration is supported by the U.S. Dept. of Energy,
National Science Foundation, the Smithsonian Institution, P.P.A.R.C.
(U.K.), N.S.E.R.C. (Canada), and Enterprise-Ireland. \\
This research has made use of NASA's Astrophysics Data System and 
of the NASA/IPAC Extragalactic Database (NED) that is operated by the Jet
Propulsion Laboratory, California Institute of Technology, under contract
with the National Aeronautics and Space Administration.

{\it Facilities:} \facility{FLWO:10m}, \facility{RXTE (PCA)},
\facility{HEGRA}, \facility{FLWO:1.2m}


\input{gfossati_MS72579_tab1.tex}
\input{gfossati_MS72579_tab2.tex}
\input{gfossati_MS72579_tab3.tex}
\input{gfossati_MS72579_tab4.tex}

\end{document}

%% file: gfossati_MS72579_tab1.tex
\begin{deluxetable}{ccccccccc}
\tabletypesize{\footnotesize}
\tablecolumns{9}
\tablewidth{0pc}
\tablecaption{Journal of \rxte Observations\tablenotemark{a}\label{tab:journal}}
\tablehead{
\colhead{Obs. \#} &
\multicolumn{2}{c}{Date (UTC)} &\phantom{.}&
\multicolumn{2}{c}{PCA} &\phantom{.}&
\multicolumn{2}{c}{HEXTE} \\
\cline{2-3} 
\cline{5-6} 
\cline{8-9} 
\colhead{} &
\colhead{Start} & 
\colhead{Stop} &&
\colhead{T$_\mathrm{exp}$} &
\colhead{\# of GTIs} &&
\colhead{T$_\mathrm{ON}$\tablenotemark{a}} &
\colhead{\# of GTIs}
}
\startdata
{ 1}&{ 18/03/01:21:40 }&{ 19/03/01:05:40 }&&{      16544 }&{  6 }&&{      16544 }&{  6 }\\
{ 2}&{ 19/03/01:05:40 }&{ 19/03/01:06:11 }&&{   \phn1472 }&{  1 }&&{   \phn1824 }&{  1 }\\ 
{ 3}&{ 19/03/01:06:55 }&{ 19/03/01:14:49 }&&{      15056 }&{  5 }&&{      16400 }&{  5 }\\ 
{ 4}&{ 19/03/01:14:49 }&{ 19/03/01:16:17 }&&{   \phn3280 }&{  1 }&&{   \phn3280 }&{  1 }\\ 
{ 5}&{ 19/03/01:18:13 }&{ 19/03/01:20:36 }&&{   \phn5488 }&{  2 }&&{   \phn5984 }&{  2 }\\ 
{ 6}&{ 19/03/01:21:19 }&{ 20/03/01:05:19 }&&{      16464 }&{  6 }&&{      16464 }&{  6 }\\ 
{ 7}&{ 20/03/01:05:19 }&{ 20/03/01:06:05 }&&{   \phn2144 }&{  1 }&&{   \phn2640 }&{  1 }\\ 
{ 8}&{ 20/03/01:06:59 }&{ 20/03/01:08:11 }&&{   \phn1936 }&{  1 }&&{   \phn2576 }&{  1 }\\ 
{ 9}&{ 20/03/01:08:31 }&{ 20/03/01:13:00 }&&{   \phn8432 }&{  3 }&&{   \phn9360 }&{  3 }\\ 
{10}&{ 20/03/01:13:29 }&{ 20/03/01:15:26 }&&{   \phn4560 }&{  2 }&&{   \phn4560 }&{  2 }\\ 
{11}&{ 20/03/01:16:20 }&{ 20/03/01:18:51 }&&{   \phn5984 }&{  2 }&&{   \phn6352 }&{  2 }\\ 
{12}&{ 20/03/01:21:07 }&{ 21/03/01:05:07 }&&{      16544 }&{  6 }&&{      16544 }&{  6 }\\ 
{13}&{ 21/03/01:05:07 }&{ 21/03/01:05:58 }&&{   \phn2304 }&{  1 }&&{   \phn2832 }&{  1 }\\ 
{14}&{ 21/03/01:06:40 }&{ 21/03/01:14:32 }&&{      14720 }&{  5 }&&{      16448 }&{  5 }\\ 
{15}&{ 21/03/01:14:32 }&{ 21/03/01:19:01 }&&{   \phn9440 }&{  3 }&&{   \phn9888 }&{  3 }\\ 
{16}&{ 21/03/01:22:52 }&{ 22/03/01:02:42 }&&{   \phn8784 }&{  4 }&&{   \phn8784 }&{  3 }\\ 
{17}&{ 22/03/01:03:32 }&{ 22/03/01:04:02 }&&{   \phn1760 }&{  1 }&&{   \phn1760 }&{  1 }\\ 
{18}&{ 22/03/01:05:19 }&{ 22/03/01:05:48 }&&{ \phn\phn960 }&{  1 }&&{  \phn1600 }&{  1 }\\ 
{19}&{ 22/03/01:06:41 }&{ 22/03/01:12:45 }&&{      10656 }&{  4 }&&{      12560 }&{  4 }\\ 
{20}&{ 22/03/01:13:13 }&{ 22/03/01:14:20 }&&{   \phn2112 }&{  1 }&&{   \phn2112 }&{  1 }\\ 
{21}&{ 22/03/01:16:07 }&{ 22/03/01:16:46 }&&{   \phn2280 }&{  1 }&&{   \phn2280 }&{  1 }\\ 
{22}&{ 22/03/01:19:19 }&{ 22/03/01:23:48 }&&{   \phn9632 }&{  3 }&&{   \phn9632 }&{  3 }\\ 
{23}&{ 23/03/01:00:07 }&{ 23/03/01:03:50 }&&{   \phn8456 }&{  3 }&&{   \phn8456 }&{  3 }\\ 
{24}&{ 23/03/01:06:24 }&{ 23/03/01:14:15 }&&{      14368 }&{  5 }&&{      16368 }&{  5 }\\ 
{25}&{ 23/03/01:14:15 }&{ 23/03/01:18:25 }&&{   \phn9512 }&{  3 }&&{   \phn9784 }&{  3 }\\ 
{26}&{ 23/03/01:19:06 }&{ 24/03/01:02:21 }&&{      16296 }&{  5 }&&{      16296 }&{  5 }\\ 
{27}&{ 24/03/01:03:02 }&{ 24/03/01:03:51 }&&{   \phn2256 }&{  1 }&&{   \phn2672 }&{  1 }\\ 
{28}&{ 24/03/01:04:41 }&{ 24/03/01:10:54 }&&{   \phn9984 }&{  4 }&&{      12784 }&{  5 }\\ 
{29}&{ 24/03/01:11:21 }&{ 24/03/01:14:54 }&&{   \phn7744 }&{  3 }&&{   \phn7856 }&{  3 }\\ 
{30}&{ 24/03/01:15:45 }&{ 24/03/01:16:37 }&&{   \phn2880 }&{  1 }&&{   \phn2880 }&{  1 }\\ 
{31}&{ 24/03/01:19:01 }&{ 24/03/01:21:06 }&&{   \phn5064 }&{  2 }&&{   \phn5064 }&{  2 }\\ 
{32}&{ 24/03/01:22:26 }&{ 25/03/01:01:52 }&&{   \phn7448 }&{  3 }&&{   \phn7448 }&{  3 }\\ 
{33}&{ 25/03/01:03:12 }&{ 25/03/01:03:42 }&&{   \phn1152 }&{  1 }&&{   \phn1680 }&{  1 }\\ 
{34}&{ 25/03/01:04:55 }&{ 25/03/01:05:49 }&&{\phn\phn832 }&{  1 }&&{   \phn1616 }&{  2 }\\ 
{35}&{ 25/03/01:06:27 }&{ 25/03/01:13:57 }&&{      13024 }&{  5 }&&{      15216 }&{  5 }\\ 
{36}&{ 25/03/01:13:57 }&{ 25/03/01:15:33 }&&{   \phn3296 }&{  1 }&&{   \phn3296 }&{  1 }\\
\hline
{Total}&{ 18/03/01:21:40 }&{ 25/03/01:15:33 }&&{    262864\phn }&{  98\phn }&&{    281840\phn }&{  99\phn }
\enddata
\tablenotetext{a}{\footnotesize%
Non deadtime corrected, and therefore strictly not an exposure time, but an on--source time.
}
\end{deluxetable}

%% file: gfossati_MS72579_tab2.tex
\begin{deluxetable}{ccccccccc}
\tabletypesize{\footnotesize}
\tablecolumns{9}
\tablewidth{0pc}
\tablecaption{\rxte/PCA and Whipple, HEGRA Overlap Statistics\label{tab:overlap}}
\tablehead{
\colhead{Night \#} &
\colhead{Date\tablenotemark{a}} &
\colhead{Date (MJD)} &
\colhead{Whipple} &
\multicolumn{2}{c}{Overlap Fraction} &
\colhead{HEGRA} &
\multicolumn{2}{c}{Overlap Fraction} \\
\cline{5-6}
\cline{8-9}
\colhead{}&
\colhead{}&
\colhead{}&
\colhead{Exp. Time}&
\colhead{detailed\tablenotemark{b}}&
\colhead{run-by-run\tablenotemark{c}}&
\colhead{Exp. Time}&
\colhead{detailed\tablenotemark{b}}&
\colhead{run-by-run\tablenotemark{c}}\\
\colhead{(1)}&
\colhead{(2)}&
\colhead{(3)}&
\colhead{(4)}&
\colhead{(5)}&
\colhead{(6)}&
\colhead{(7)}&
\colhead{(8)}&
\colhead{(9)}
}
\startdata
{All} & {March 18--25}& {51986/51993} & {$   29^\mathrm{h} 52^\mathrm{m}$} & {44~\%} & {   47/    64 \phn(73~\%)}  & {$   32^\mathrm{h} 14^\mathrm{m}$} & {54~\%} & {   58/    69 \phn(84~\%)} \\
\hline
{ 1 } & {March 18/19} & {51986/51987} & {$\phn6^\mathrm{h} 04^\mathrm{m}$} & {42~\%} & {   11/    13 \phn(85~\%)}  & {$\phn5^\mathrm{h} 14^\mathrm{m}$} & {55~\%} & {   11/    11    (100~\%)} \\[0.05in]
{ 2 } & {March 19/20} & {51987/51988} & {$\phn3^\mathrm{h} 44^\mathrm{m}$} & {67~\%} & {\phn7/ \phn8 \phn(87~\%)}  & {$\phn4^\mathrm{h} 33^\mathrm{m}$} & {60~\%} & {\phn8/    10 \phn(80~\%)} \\[0.05in]
{ 3 } & {March 20/21} & {51988/51989} & {$\phn5^\mathrm{h} 36^\mathrm{m}$} & {45~\%} & {\phn9/    12 \phn(75~\%)}  & {$\phn5^\mathrm{h} 30^\mathrm{m}$} & {54~\%} & {   10/    11 \phn(91~\%)} \\[0.05in]
{ 4 } & {March 21/22} & {51989/51990} & {$\phn4^\mathrm{h} 12^\mathrm{m}$} & {44~\%} & {\phn7/ \phn9 \phn(78~\%)}  & {$\phn4^\mathrm{h} 32^\mathrm{m}$} & {54~\%} & {\phn7/    10 \phn(70~\%)} \\[0.05in]
{ 5 } & {March 22/23} & {51990/51991} & {$\phn4^\mathrm{h} 40^\mathrm{m}$} & {30~\%} & {\phn5/    10 \phn(50~\%)}  & {$\phn4^\mathrm{h} 37^\mathrm{m}$} & {57~\%} & {\phn8/    10 \phn(80~\%)} \\[0.05in]
{ 6 } & {March 23/24} & {51991/51992} & {$\phn2^\mathrm{h} 20^\mathrm{m}$} & {55~\%} & {\phn4/ \phn5 \phn(80~\%)}  & {$\phn4^\mathrm{h} 35^\mathrm{m}$} & {58~\%} & {\phn9/    10 \phn(90~\%)} \\[0.05in]
{ 7 } & {March 24/25} & {51992/51993} & {$\phn3^\mathrm{h} 16^\mathrm{m}$} & {31~\%} & {\phn4/ \phn7 \phn(57~\%)}  & {$\phn3^\mathrm{h} 10^\mathrm{m}$} & {37~\%} & {\phn5/ \phn7 \phn(71~\%)} \\
\enddata
\tablenotetext{a}{
HEGRA observing windows typically extend across UTC midnight.
All the Whipple observing windows occur after the UTC midnight.}
\tablenotetext{b}{
Ratio between the actual \rxte/PCA on--source time during Whipple (HEGRA) runs,
and the actual total Whipple (HEGRA) observing time.}
\tablenotetext{c}{
Fraction of Whipple (HEGRA) $\approx$0.5 hrs runs with some overlapping
\rxte/PCA data.  The overlap fraction range between 7\% and 100\%, averaging to
the number reported in columns (5) and (8).}
\end{deluxetable}

%% file: gfossati_MS72579_tab3.tex
\begin{deluxetable}{cccccc}
\tabletypesize{\small}
\tablecolumns{6}
\tablewidth{0pc}
\tablecaption{\rxte/PCA and Whipple+HEGRA Overlap Statistics\label{tab:overlap_all}}
\tablehead{
\colhead{Night \#} &
\colhead{Date\tablenotemark{a}} &
\colhead{Date (MJD)} &
\colhead{TeV} &
\multicolumn{2}{c}{Overlap Fraction} \\
\cline{5-6}
\colhead{}&
\colhead{}&
\colhead{}&
\colhead{Exp. Time}&
\colhead{detailed\tablenotemark{b}}&
\colhead{run-by-run\tablenotemark{c}}\\
\colhead{(1)}&
\colhead{(2)}&
\colhead{(3)}&
\colhead{(4)}&
\colhead{(5)}&
\colhead{(6)}
}
\startdata
{All} & {March 18--25}& {51986/51993} & {$   62^\mathrm{h} 06^\mathrm{m}$} & {49~\%} & {  104/   133 \phn(78~\%)} \\[0.05in]
\hline
{ 1 } & {March 18/19} & {51986/51987} & {$   11^\mathrm{h} 18^\mathrm{m}$} & {48~\%} & {   22/    24 \phn(92~\%)} \\[0.05in]
{ 2 } & {March 19/20} & {51987/51988} & {$\phn8^\mathrm{h} 17^\mathrm{m}$} & {63~\%} & {   15/    18 \phn(83~\%)} \\[0.05in]
{ 3 } & {March 20/21} & {51988/51989} & {$   11^\mathrm{h} 06^\mathrm{m}$} & {48~\%} & {   18/    23 \phn(78~\%)} \\[0.05in]
{ 4 } & {March 21/22} & {51989/51990} & {$\phn8^\mathrm{h} 44^\mathrm{m}$} & {49~\%} & {   14/    19 \phn(74~\%)} \\[0.05in]
{ 5 } & {March 22/23} & {51990/51991} & {$\phn9^\mathrm{h} 17^\mathrm{m}$} & {43~\%} & {   13/    20 \phn(65~\%)} \\[0.05in]
{ 6 } & {March 23/24} & {51991/51992} & {$\phn6^\mathrm{h} 55^\mathrm{m}$} & {57~\%} & {   13/    15 \phn(87~\%)} \\[0.05in]
{ 7 } & {March 24/25} & {51992/51993} & {$\phn6^\mathrm{h} 26^\mathrm{m}$} & {34~\%} & {\phn9/    14 \phn(64~\%)} \\
\enddata
\tablenotetext{a}{
HEGRA observing windows typically extend across UTC midnight.}
\tablenotetext{b}{
Ratio between the actual \rxte/PCA on--source time during Whipple, HEGRA runs,
$\approx0.5$\,hrs, and the actual total observing time.}
\tablenotetext{c}{
Fraction of Whipple, HEGRA runs with some overlapping \rxte/PCA data.
The overlap fraction ranges between 7\% and 100\%, averaging to the number
reported in column (5).}
\end{deluxetable}

%% file: gfossati_MS72579_tab4.tex
\begin{deluxetable}{lc|c|ccccc}
\tabletypesize{\footnotesize}
\tablecolumns{8}
\tablewidth{0pc}
\tablecaption{Slope of the TeV/X--ray correlation.\label{tab:slopes}}
\tablehead{%
\colhead{Night(s)} &
\colhead{\# data} &
\colhead{PCA 2--10 keV} &
\colhead{PCA 2--4 keV} &
\colhead{PCA 4--6 keV} &
\colhead{PCA 6--8 keV} &
\colhead{PCA 9--15 keV} &
\colhead{HEXTE 20--60 keV} 
}
\startdata
{ all }              & {105} & {$0.88 \pm 0.07$} & {$0.93 \pm 0.09$}  & {$0.89 \pm 0.07$} & {$0.81 \pm 0.07$} & {$0.71 \pm 0.06$} & {$1.02 \pm 0.14$} \\
{ all (dT\,=\,1d) }  & {  7} & {$0.83 \pm 0.14$} & {$0.89 \pm 0.16$}  & {$0.82 \pm 0.14$} & {$0.79 \pm 0.13$} & {$0.73 \pm 0.12$} & {$1.96 \pm 0.35$} \\
\cutinhead{Individual Nights}
{ 1$^\mathrm{a}$ }   & {11} & {$2.84 \pm 0.41$} & {$2.29 \pm 1.00$}  & {$2.68 \pm 0.28$} & {$2.51 \pm 0.21$} & {$2.20 \pm 0.12$} & {$2.47 \pm 0.96$} \\[0.01in]
{ 1 }                & {22} & {$2.26 \pm 0.25$} & {$1.98 \pm 0.45$}  & {$2.26 \pm 0.13$} & {$2.12 \pm 0.16$} & {$1.90 \pm 0.12$} & {$2.73 \pm 0.66$} \\[0.01in]
{ 4 }                & {14} & {$1.56 \pm 0.25$} & {$1.31 \pm 0.23$}  & {$1.73 \pm 0.27$} & {$1.65 \pm 0.25$} & {$1.61 \pm 0.22$} & {$1.44 \pm 0.52$} \\[0.01in]
{ 5 }                & {14} & {$1.67 \pm 0.16$} & {$1.81 \pm 0.17$}  & {$1.65 \pm 0.17$} & {$1.61 \pm 0.15$} & {$1.44 \pm 0.14$} & {$1.72 \pm 0.63$} \\
\cutinhead{Selected subsets of Consecutive Nights}
{ $1^\mathrm{a}+2$ } & {26} & {$2.16 \pm 0.21$} & {$1.97 \pm 0.38$}  & {$2.12 \pm 0.17$} & {$2.00 \pm 0.15$} & {$1.82 \pm 0.12$} & {$2.00 \pm 0.61$} \\[0.01in]
{ $1+2$            } & {37} & {$2.04 \pm 0.16$} & {$1.88 \pm 0.28$}  & {$2.03 \pm 0.14$} & {$1.92 \pm 0.12$} & {$1.73 \pm 0.10$} & {$2.12 \pm 0.52$} \\[0.01in]
{ $4+5$ }            & {28} & {$1.74 \pm 0.15$} & {$1.57 \pm 0.23$}  & {$1.78 \pm 0.15$} & {$1.68 \pm 0.12$} & {$1.45 \pm 0.10$} & {$1.77 \pm 0.38$} \\[0.01in]
{ $6+7$ }            & {22} & {$1.00 \pm 0.14$} & {$0.81 \pm 0.15$}  & {$1.03 \pm 0.14$} & {$1.01 \pm 0.13$} & {$0.98 \pm 0.12$} & {$1.12 \pm 0.19$} \\[0.01in]
{ $5+6+7$ }          & {36} & {$0.94 \pm 0.09$} & {$0.84 \pm 0.10$}  & {$0.98 \pm 0.09$} & {$0.95 \pm 0.09$} & {$0.91 \pm 0.08$} & {$1.21 \pm 0.16$} \\
\cutinhead{Subsets by brightness level}
{ $1^\mathrm{a}+2$ } & {26} & {$2.16 \pm 0.21$} & {$1.97 \pm 0.38$}  & {$2.12 \pm 0.17$} & {$2.00 \pm 0.15$} & {$1.82 \pm 0.12$} & {$2.00 \pm 0.61$} \\[0.01in]
{ $6+7$ }            & {22} & {$1.00 \pm 0.14$} & {$0.81 \pm 0.15$}  & {$1.03 \pm 0.14$} & {$1.01 \pm 0.13$} & {$0.98 \pm 0.12$} & {$1.12 \pm 0.19$} \\[0.01in]
{ $3+6+7$ }          & {40} & {$0.99 \pm 0.14$} & {$0.82 \pm 0.14$}  & {$1.04 \pm 0.14$} & {$0.98 \pm 0.14$} & {$0.90 \pm 0.13$} & {$0.74 \pm 0.19$} \\
\enddata
\tablenotetext{a}{
Whipple (flare) only.
}
\end{deluxetable}